\documentclass[12pt,hidelinks]{article}
\usepackage{geometry}
\RequirePackage{amsthm,amsmath,amsfonts,amssymb, amsthm}
\RequirePackage[authoryear]{natbib}
\usepackage{amsmath}

\usepackage{lmodern}
\usepackage[T1]{fontenc}
\usepackage{bm}
\usepackage{natbib}
\usepackage{tikz}
\usetikzlibrary{fit, positioning}
\newcommand\sdots{\hbox to 0.5em{.\hss.\hss.}}
\usepackage[labelformat=simple]{subcaption}

\usepackage{bbm}
\usepackage{color,xcolor,soul}
\usepackage{graphicx}
\usepackage{diagbox}
\usepackage{pdflscape}
\usepackage{rotating}

\usepackage[ruled]{algorithm2e}
\usepackage{caption}
\usepackage{booktabs}
\usepackage{multirow}
\usepackage{hyperref}
\usepackage{authblk}

\usepackage[capitalize]{cleveref}

\usepackage{mathtools} 

\title{Bayesian Dynamic Generalized Additive Model for Mortality during COVID-19 Pandemic}
	
\author[1]{Wei Zhang\footnote{{wei.zhang@usi.ch}}}
\author[1]{Antonietta Mira\footnote{{antonietta.mira@usi.ch}}}
\author[1]{Ernst C. Wit\footnote{{ernst.jan.camiel.wit@usi.ch}}}

\affil[1]{Università della Svizzera Italiana}

\begin{document}
\maketitle

\begin{abstract}
While COVID-19 has resulted in a significant increase in global mortality rates, the impact of the pandemic on mortality from other causes remains uncertain. To gain insight into the broader effects of COVID-19 on various causes of death, we analyze an Italian dataset that includes monthly mortality counts for different causes from January 2015 to December 2020. Our approach involves a generalized additive model enhanced with correlated random effects. The generalized additive model component effectively captures non-linear relationships between various covariates and mortality rates, while the random effects are multivariate time series observations recorded in various locations, and they embody information on the dependence structure present among geographical locations and different causes of mortality. Adopting a Bayesian framework, we impose suitable priors on the model parameters. For efficient posterior computation, we employ variational inference, specifically for fixed effect coefficients and random effects, Gaussian variational approximation is assumed, which streamlines the analysis process. The optimisation is performed using a coordinate ascent variational inference algorithm and several computational strategies are implemented along the way to address the issues arising from the high dimensional nature of the data, providing accelerated and stabilised parameter estimation and statistical inference.
\end{abstract}


\section{Introduction}
\label{Introduction}

The COVID-19 pandemic has had profound impacts across the globe. Researches focus on various aspects such as health disparities linked to racial and socio-economic factors and the healthcare system's adaptation in terms of testing, contact tracing, and vaccine rollouts \citep{kretzschmar2020impact, peretti2020future, abedi2021racial}. A key area of investigation is excess mortality, which provides an overarching view of the pandemic's impact on human health. This includes factors such as government lockdown measures and disruptions to non-COVID healthcare services \citep{wang2022estimating, msemburi2023estimates}. While excess mortality offers a general perspective, examining the consequences of specific causes of death as a result of these factors during the pandemic is crucial for developing more targeted future mitigation strategies. For example, the pandemic has contributed to increases in deaths from chronic conditions as observed by \citet{shiels2022leading}. There also has been a rise in accidental deaths, homicides or suicides \citep{pell2020coronial, mitchell2021state, dmetrichuk2022retrospective}. Contributing to this research field, we analyze Italian monthly death counts recorded in 21 Italian regions from 2015 to 2020, categorized according to the International Classification of Diseases, 10th Revision (ICD-10) \citep{world1992icd}, to understand the effects on specific human mortality during the pandemic better.

We apply the generalized additive model (GAM) \citep{hastie1987generalized, hastie2017generalized} to study non-linear relationships between response, cause-specific mortality rate, and continuous covariates, including lockdown intensity level and age, in terms of smooth functions. GAMs have demonstrated considerable potential to study COVID-19 mortality \citep{koum2021excess, zhu2021association, basellini2022explaining}. In the more general context of spatial-temporal data modeling, GAMs are well-suited for analyzing changes over time and across different geographic regions, crucial in understanding the differentiating consequences. One way of employing GAMs to model the change of spatial pattern over time is via a tensor product smoother \citep{wood2017generalized, feng2022spatial}. Alternatively, \citet{clark2023dynamic} introduce dynamic GAMs where extra dynamic spatial random effects are incorporated into the mean of response variable, offering a solution to forecasting discrete time series while estimating relevant nonlinear predictor associations that conventional generalized linear models (GLM) plus spatial temporal random effects are not able to account for \citep{nazia2022identifying, yin2023spatio}. In our study, we follow \citet{clark2023dynamic}'s approach and assume that the mortality rate of specific cause in a region for a certain month combines a GAM regression component and random effects. The random effects are further assumed to be correlated as opposed to being independent, facilitating statistical inference on dependence structure between geographical regions and various causes of death \citep{rao2015small, smith2015restricted}. We adopt a Bayesian approach and assign suitable priors on model parameters to better infer correlation structure from the data.

The model framework is a special case of non-Gaussian state space models where the observation equation is Poisson distribution. Inference with state space models includes filtering to estimate the current state given past observations and smoothing to estimate past states given all observations up to the current time \citep{petris2009dynamic, durbin2012time}. Filtering and smoothing is exact in linear Gaussian state space models where Kalman Filter can be applied. When the model is non-Gaussian, particle filter approximates the states using a set of weighted particles \citep{doucet2001sequential}. The posterior samples in \citet{clark2023dynamic} are drawn either in the Gibbs sampling software JAGS \citep{plummer2003jags} or with the Hamiltonian Monte Carlo in Stan \citep{carpenter2017stan}. However, the approximation for the target deteriorates as the dimension increases \citep{10.1214/13-AAP951, van2019particle}. Due to the high dimensionality inherent in the data and model, these sampling algorithms are impractical. To address the issue, we use a variational inference algorithm for fast approximation \citep{jordan1999introduction, blei2017variational}; more specifically, the joint variational density of fixed effect coefficients and random effects takes the Gaussian form. Various methods exist for optimizing the mean and variance of this approximating Gaussian distribution. For instance, \citet{titsias2014doubly} propose to parameterize the density in terms of its mean and a lower triangular scale matrix whereas \citet{tan2018gaussian} incorporate sparsity in the Cholesky factors of the precision matrix. Both developed stochastic gradient methods for optimization. Our approach instead involves Newton's method and fixed point iteration as in \citet{arridge2018variational} to achieve faster convergence in fewer iterations. This aligns with recent advancements with recent development in Gaussian variational approximations for high-dimensional state space models \citep{quiroz2023gaussian}, but distincts in that we do not define structure of the proposed variational approximation thanks to our chosen optimization technique.
  
The rest of the paper is organized as follows. In Section \ref{Model}, we formulate the model and specify the priors imposed on parameters. In Section \ref{Posterior inference}, we demonstrate how to derive the ELBO in this model setup and present the variational algorithm for posterior inferences. We then apply the model and the algorithm to the Italian monthly mortality data in Section \ref{real data application}. Finally, Section \ref{Summary and Future Work} gives some concluding remarks and points to future work.

\section{Bayesian dynamic GAM}
\label{Model}

Let $Y_{n,t}$ be the mortality count of instance $n$ at time $t$, $n=1,\dots,N, t=1,\dots,T$. For each $n$, covariates such as ordinal $a_n$ for age categorical $g_n$ for gender, $l_n$ for $L$ geographical locations and $k_n$ for $K$ different causes of death are available. Additionally, we are interested in the lockdown effect on outcome variable $Y_{n,t}$, therefore we include a stringency index that quantifies the intensity of government restriction policies for each geographical location over time and we denote it by $r_{n,t}$, which depends on $n$ through $l_n$. We assume that $Y_{n,t}$ is Poisson random variable with rate equal to
\begin{equation*}
    \epsilon_{n,t}\exp\left[\mathbf{x}^{\boldsymbol{\beta}}_n\boldsymbol{\beta} + f^r(r_{n,t}) + f^a(a_n) +  f^{k,r}(k_n, r_{n,t}) + f^{k,a}(k_n, a_n) + f^{g,a}(g_n, a_n) + z^*_{n,t} \right]
\end{equation*}
where $\epsilon_{n,t}$ is the offset, $\mathbf{x}^{\boldsymbol{\beta}}_n\boldsymbol{\beta}$ are parametric terms which include gender effect. $f^{r}(r_{n,t})$ and $f^{a}(a_n)$ are natural cubic splines for government intervention effect and age effect respectively and they take the form 
\begin{equation*}
    f^r\left(r\right) = \sum_{j=1}^J u^r_j\left(r\right)\beta^r_j, \quad f^a\left(a\right) = \sum_{j=1}^J u^a_j\left(a\right)\beta^a_j,
\end{equation*}
where $J$ is the number of knots, $u^r_j, u^a_j$ are natural cubic spline bases, $\beta^r_j$ and $\beta^a_j$ are coefficients to be estimated. The smoothness penalization terms associated with the bases are defined as
\begin{equation*}
    \lambda^r_1(\boldsymbol{\beta}^r)' S^r_1\boldsymbol{\beta}^r + \lambda^r_2(\boldsymbol{\beta}^r)' S^r_2\boldsymbol{\beta}^r, \quad 
    \lambda^a_1(\boldsymbol{\beta}^a)' S^a_1\boldsymbol{\beta}^a + \lambda^a_2(\boldsymbol{\beta}^a)' S^a_2\boldsymbol{\beta}^a.
\end{equation*}
Here $S^r_1, S^r_1, S^a_1, S^a_2$ contain known coefficients. $\boldsymbol{\beta}^r=\left(\beta^r_1, \dots, \beta^r_J\right)'$ and $\boldsymbol{\beta}^a=\left(\beta^a_1, \dots, \beta^a_J\right)'$ are vectors of spline coefficients. From a Bayesian perspective, this is equivalent to imposing the following multivariate normal priors on $\boldsymbol{\beta}_r$ and $\boldsymbol{\beta}_a$
\begin{gather*}
    \boldsymbol{\beta}^r \sim \mathcal{N}\left(\mathbf{0}, \left(\lambda^r_1S^r_1+\lambda^r_2 S^r_2\right)^{-1}\right) \\
        \boldsymbol{\beta}^a \sim \mathcal{N}\left(\mathbf{0}, \left(\lambda^a_1 S^a_1+\lambda^a_2 S^a_2\right)^{-1}\right).
\end{gather*}
$\lambda^r_1, \lambda^r_2, \lambda^a_1$ and $\lambda^a_2$ control the level of roughness.

As for the two-way interaction terms, $f^{k,r}(k, r)$ models interactions between cause of death and stringency index in a non-parametric manner. $f^{k,a}(k, a)$ accounts for interactions between causes of death and age while $f^{g,a}(g, a)$ captures interactions between gender and age. The three interactions terms are constructed in the following way. For each level $k$ of causes of death or each $g$ of gender, a unique spline is specified \citep{wood2017generalized}. We formulate $f^{k,r}(k, r)$ as an example,  $f^{k,a}(k, a)$ and $f^{g,a}(g, a)$ are modeled in a similar way. For each $k$ other than the baseline, $f^{k,r}(k, r)$ is assumed to be natural cubic spline such that
\begin{equation*}
    f^{k,r}(k, r) = \sum_{j=1}^J u^{k,r}_{k,j}\left(r\right)\beta^{k,r}_{k,j}.
\end{equation*}
The $J(K-1)$ dimensional vector $\boldsymbol{\beta}^{k,r}=\left(\beta^{k,r}_{2,1},\dots,\beta^{k,r}_{2,J},\beta^{k,r}_{3,1}\dots,\beta^{k,r}_{K,J}\right)'$ is jointly penalised by $S^{k,r}_1$ and $S^{k,r}_2$ with smoothing parameters $\lambda^{k,r}_1$ and $\lambda^{k,r}_2$ and the penalisation term is $\lambda^{k,r}_1(\boldsymbol{\beta}^{k,r})' S^r_1\boldsymbol{\beta}^{k,r} + \lambda^{k,r}_2(\boldsymbol{\beta}^{k,r})' S^r_2\boldsymbol{\beta}^{k,r},$
which, in terms of Bayesian prior, translates to
\begin{equation*}
    \boldsymbol{\beta}^{k,r} \sim \mathcal{N}\left(\mathbf{0}, \left(\lambda^{k,r}_1S^{k,r}_1+\lambda^{k,r}_2 S^{k,r}_2\right)^{-1}\right).
\end{equation*}

The last term $z^*_{n,t}$ embeds the spatial-temporal structure in the data; in fact, $z^*_{n,t}$ depends on $n$ through $l_n$ and $k_n$, therefore in total, we have $LK$ time series of length $T$. $z^*_{n,t}$ can be modeled with great flexibility. We make the following latent state assumption. Let $\mathbf{z}^*_t=(z^*_{1,t}, \dots, z^*_{N,t})'$ and $\mathbf{z}_t=(z_{1,t}, \dots, z_{LK,t})'$ such that
\begin{equation*}
    \mathbf{z}^*_t = \left(I_{LK}\bigotimes \mathbf{1}\right)\mathbf{z}_t,
\end{equation*}
where $\mathbf{1}$ stands for a vector whose elements are all equal to 1. The dimension of $\mathbf{1}$ is determined by the number of age groups and gender. The assumption implies that residual mortality rates $z^*_{n,t}$ of the same causes of death in the same region are identical regardless of age and gender, which is reasonable since we have already taken into account age and gender effect in GAM component. We further assume that marginally the latent state $\mathbf{z}_t\sim \mathcal{N}\left(\boldsymbol{\mu},\Sigma\right)$ and $P\left(\mathbf{z}_t - \boldsymbol{\mu}\right)$ follows a simple autoregressive (AR) process such that 
\begin{equation*}
    P\left(\mathbf{z}_t - \boldsymbol{\mu}\right) = \Phi P\left(\mathbf{z}_{t-1} - \boldsymbol{\mu}\right) + \boldsymbol{\epsilon}_t, \quad \boldsymbol{\epsilon}_t\sim \mathcal{N}\left(\mathbf{0}, I_{LK}-\Phi\Phi'\right),
\end{equation*}
where $P$ is an upper triangular matrix such that $P'P=\Omega=\Sigma^{-1}$, i.e. P are the Cholesky factors of the precision matrix $\Omega$.
$\Phi$ contains multivariate autoregressive coefficients. For simplicity, we require $\Phi$ to be diagonal with diagonal entries $\boldsymbol{\phi}=c(\phi_1, \dots, \phi_{LK})'$ such that $-1<\phi_1, \dots, \phi_{LK}<1$. The pre-multiplying of $\mathbf{z}_t$ by $P$ is to ensure that marginally $\mathbf{z}_t \sim \mathcal{N}\left(\mathbf{0}, \Sigma\right), t=0, \dots, T$ where $\Sigma$ incorporates dependence structure of geographical locations as well as mortality causes. Reorganizing $\mathbf{z}_t$ as matrices $Z_t$ of size $L\times K$, we assume that 
\begin{equation*}
    Z_t \sim \text{MN}\left(\boldsymbol{\mu}, \Sigma^k, \Sigma^l\right),
\end{equation*}
which is a matrix normal distribution equivalent to 
\begin{equation*}
    \mathbf{z}_t \sim \mathcal{N}\left(\boldsymbol{\mu}, \Sigma^k \bigotimes \Sigma^l\right),
\end{equation*}
where $\bigotimes$ stands for the Kronecker product between $\Sigma^k$ and $\Sigma^l$, the covariance matrices of the causes of death dependence and regional dependence respectively. We impose the following priors on the model parameters. Firstly, $\boldsymbol{\mu}$ is assigned a normal prior $\mathcal{N}\left(\mathbf{0}, \sigma^2_\mu I_{LK}\right)$. $\Omega^k = \left(\Sigma^k\right)^{-1}$ is the precision matrix that underlines the conditional independence of mortality causes. For simplicity, we stick to the usual Wishart distribution
\begin{equation*}
    \Omega^k \sim \text{Wishart}\left(\delta^k, \theta^k I_K\right).
\end{equation*}
The same reasoning applies to the precision matrix $\Omega^l=\left(\Sigma^l\right)^{-1}$ and it follows that
\begin{equation*}
    \Omega^l \sim \text{Wishart}\left(\delta^l, \theta^l I_L\right).
\end{equation*}


We complete the model with a hierarchy of prior specification on the linear regression coefficients, on penalization parameters that control smoothness of splines and on the autoregressive coefficients, on the cause specific mean. Denoting by $\mathbf{x}_n$ the vector containing $\mathbf{x}^{\boldsymbol{\beta}}_n$ and all bases in GAMs for $n$, the hierarchical model can be summarized as
\begin{gather*}
    Y_{n,t} \sim \text{Poisson}\left[ \epsilon_{n,t}\exp\left(\mathbf{x}_n\boldsymbol{\beta}^* + z^*_{n,t} \right)\right], \quad \boldsymbol{\beta}^* = \left(\boldsymbol{\beta}', \left(\boldsymbol{\beta}^r\right)', \left(\boldsymbol{\beta}^a\right)', \left(\boldsymbol{\beta}^{k,r}\right)', \left(\boldsymbol{\beta}^{k,a}\right)', \left(\boldsymbol{\beta}^{g,a}\right)'\right)', \\
\boldsymbol{\beta} \sim \mathcal{N}\left(\mathbf{0}, \sigma^2_\beta I\right), \quad \boldsymbol{\beta}^r \sim \mathcal{N}\left(\mathbf{0}, \left(\lambda^r_1S^r_1+\lambda^r_2 S^r_2\right)^{-1}\right), \quad
        \boldsymbol{\beta}^a \sim \mathcal{N}\left(\mathbf{0}, \left(\lambda^a_1 S^a_1+\lambda^a_2 S^a_2\right)^{-1}\right), \\
    \boldsymbol{\beta}^{k,r} \sim \mathcal{N}\left(\mathbf{0}, \left(\lambda^{k,r}_1S^{k,r}_1+\lambda^{k,r}_2 S^{k,r}_2\right)^{-1}\right), \quad \boldsymbol{\beta}^{k,a} \sim \mathcal{N}\left(\mathbf{0}, \left(\lambda^{k,a}_1S^{k,a}_1+\lambda^{k,a}_2 S^{k,a}_2\right)^{-1}\right), \\
    \boldsymbol{\beta}^{g,a} \sim \mathcal{N}\left(\mathbf{0}, \left(\lambda^{g,a}_1S^{g,a}_1+\lambda^{g,a}_2 S^{g,a}_2\right)^{-1}\right), \\
   \lambda^r_1, \lambda^r_2, \lambda^a_1, \lambda^a_2, \lambda^{k,r}_1, \lambda^{k,r}_2, \lambda^{k,a}_1, \lambda^{k,a}_2, \lambda^{g,a}_1, \lambda^{g,a}_2 \sim \text{Gamma}\left(\alpha_\lambda, \beta_\lambda\right), \\
   \mathbf{z}^*_t = \left(I_{LK}\bigotimes \mathbf{1}\right)\mathbf{z}_t, \\
       P\left(\mathbf{z}_t-\boldsymbol{\mu}\right) = \Phi P\left(\mathbf{z}_{t-1}-\boldsymbol{\mu}\right) + \boldsymbol{\epsilon}_t, \quad \boldsymbol{\epsilon}_t\sim \mathcal{N}\left(\mathbf{0}, I_{LK}-\Phi\Phi'\right), \quad
       \boldsymbol{\mu}\sim\mathcal{N}\left(\mathbf{0},\sigma^2_\mu I_{LK}\right), \\
       \Phi = \text{diag}\left(\phi_1,\dots,\phi_{LK}\right), \quad \frac{\phi_1+1}{2}, \dots, \frac{\phi_{LK}+1}{2} \overset{i.i,d}{\sim} \text{Beta}(\alpha_\phi, \beta_\phi), \\
       \quad P'P = \Omega = \Omega^k \bigotimes \Omega^l,\quad
        \Omega^k \sim \text{Wishart}\left(\delta^k, \theta^k I_K\right), \quad \Omega^l \sim \text{Wishart}\left(\delta^l, \theta^lI_L\right).   
\end{gather*}

\section{Posterior inference via variational approximation}
\label{Posterior inference}

To make posterior inference, one may devise suitable Markov Chain Monte Carlo (MCMC) algorithms to obtain posterior samples. However, the algorithm may fail to deliver desirable convergent output within reasonable time when it is difficult to explore the geometry of the target distribution due to the sheer dimension of the data. Therefore we resort to variational inference approach for fast approximation. The target posterior distribution is 
\begin{equation*}
\begin{aligned}
        p\left(\right. & \boldsymbol{\beta}^*,\boldsymbol{\lambda},\mathbf{z}_{0:T}, \left.\boldsymbol{\phi}, \boldsymbol{\mu}, \Omega^k, \Omega^l \mid \mathbf{y}\right) \\
        &\propto p\left(\mathbf{y}\mid \boldsymbol{\beta}^*, \mathbf{z}_{0,T}\right)p\left(\boldsymbol{\beta}^* \mid \boldsymbol{\lambda}\right)p\left(\boldsymbol{\lambda}\right)p\left(\mathbf{z}_{0:T} \mid \boldsymbol{\phi}, \boldsymbol{\mu}, \Omega^k, \Omega^l\right)p\left(\boldsymbol{\mu}\right)p\left(\boldsymbol{\phi}\right)p\left(\Omega^k\right)p\left(\Omega^l\right),
\end{aligned}
\end{equation*}
and the goal is to find the optimal $q^* \left(\boldsymbol{\beta}^*,\boldsymbol{\lambda},\mathbf{z}_{0:T}, \boldsymbol{\mu},\boldsymbol{\phi}, \Omega^k, \Omega^l \right)$ from a pre-specified family of distributions such that the Kullback–Leibler (KL) divergence, defined as
\begin{equation*}
    \text{KL}\left[q \left(\cdot \right) || p\left(\cdot \mid \mathbf{y} \right)\right] = \text{E}_q\left[\log\frac{q\left(\cdot\right)}{p\left(\cdot \mid \mathbf{y} \right)} \right],
\end{equation*}
with expectation taken with respect to $q\left(\cdot\right)$, is minimized. This is equivalent to maximizing the evidence lower bound (ELBO) given by
\begin{equation*}
\resizebox{\textwidth}{!}{$
    \begin{aligned}
    &\text{ELBO}\left[ p\left(\cdot \mid \mathbf{y} \right)\right] \\
    =&\text{E}_q \left[\log\frac{p\left(\mathbf{y}\mid \boldsymbol{\beta}^*, \mathbf{z}_{0,T}\right)p\left(\boldsymbol{\beta}^* \mid \boldsymbol{\lambda}\right)p\left(\mathbf{z}_{0:T} \mid \boldsymbol{\phi}, \boldsymbol{\mu}, \Omega^k, \Omega^l\right)p\left(\boldsymbol{\lambda}\right)p\left(\boldsymbol{\mu}\right)p\left(\boldsymbol{\phi}\right)p\left(\Omega^k\right)p\left(\Omega^l\right)}{q \left(\boldsymbol{\beta}^*,\mathbf{z}_{0:T}, \boldsymbol{\lambda},\boldsymbol{\mu},\boldsymbol{\phi}, \Omega^k, \Omega^l \right)} \right].
\end{aligned}
$}
\end{equation*}
We further assume that the family of candidate  approximation $q(\cdot)$ can be factorized as
\begin{equation}
    q \left(\boldsymbol{\beta}^*,\mathbf{z}_{0:T}, \boldsymbol{\lambda},\boldsymbol{\mu},\boldsymbol{\phi}, \Omega^k, \Omega^l \right) = q \left(\boldsymbol{\beta}^*,\mathbf{z}_{0:T}\right)q\left(\boldsymbol{\lambda}\right)q\left(\boldsymbol{\mu}\right)q\left(\boldsymbol{\phi}\right)q\left(\Omega^k\right)q\left(\Omega^l\right),
\label{eq:factorization}
\end{equation}
where $q \left(\boldsymbol{\beta}^*,\mathbf{z}_{0:T}\right)$ is multivariate normal distribution with mean $\mathbf{m}$ and covariance matrix $M$. This is essentially the variational Gaussian approximation (VGA) that has been widely implemented in literature and its theoretical properties when applied to Poisson data are studied 
by \citet{arridge2018variational}. $M$ can be full or it can be sparse block diagonal matrix so that there is no correlation between $\boldsymbol{\beta}^*$ and $\mathbf{z}_{0:T}$, which means that $q \left(\boldsymbol{\beta}^*,\mathbf{z}_{0:T}\right)$ can be further factorized as the product of two multivariate normal densities, $q \left(\boldsymbol{\beta}^*\right)$ and $q \left(\mathbf{z}_{0:T}\right)$. The sparse version of $M$ greatly reduces algorithm complexity. However, in our application, since the dimension of $\mathbf{z}_{0:T}$ is overwhelmingly larger than $\boldsymbol{\beta}^*$, the block diagonal assumption does not produce much gain; therefore we stick to the full matrix characterisation of $M$. Furthermore, we choose the following variational densities implied by mean field approximation: for each element in $\boldsymbol{\lambda}$, it is a point mass at $\lambda^{q,r}_1, \lambda^{q,r}_2, \lambda^{q,a}_1, \lambda^{q,a}_2, \lambda^{q,k,r}_1, \lambda^{q,k,r}_2, \lambda^{q,k,a}_1, \lambda^{q,k,a}_2, \lambda^{q,g,a}_1, \lambda^{q,g,a}_2$. Together they are $\boldsymbol{\lambda}^q$. The Dirac measure choice as a variational density avoids evaluating expectation with respect to otherwise non-trivial variational densities. For $q\left(\boldsymbol{\mu}\right)$, we assume independent normal approximation densities $q\left(\boldsymbol{\mu}\right)=\mathcal{N}\left(\boldsymbol{\mu}^q, \text{diag}\left[\left(\boldsymbol{\sigma}^q\right)^2\right]\right)$. For $q\left(\boldsymbol{\phi}\right)$, even though it is analytically possible to compute the expectation when assuming that $(\phi_1+1)/2\overset{i.i,d}{\sim} \text{Beta}(\alpha^{q,\phi}_1, \beta^{q,\phi}_1), \dots, (\phi_{LK}+1)/2 \overset{i.i,d}{\sim} \text{Beta}(\alpha^{q,\phi}_{LK}, \beta^{q,\phi}_{LK})$, the optimization scheme we adapt diverges as only the means of beta variational distributions is identifiable, therefore we use Dirac measures on $q\left(\boldsymbol{\phi}\right)$ so they are $\boldsymbol{\phi}^q$. Finally, we set  $q(\Omega^k) = \text{Wishart}(\delta^{q,k}, D^{q,k})$ and $q(\Omega^l) = \text{Wishart}(\delta^{q,l}, D^{q,l})$.

\subsection{ELBO calculation}
With \eqref{eq:factorization}, the ELBO can be written as the sum of expectations
\begin{align*}
    &\text{ELBO}\left[ p\left(\cdot \mid \mathbf{y} \right)\right] \\
    =&\text{E}_q \left[\log\frac{p\left(\mathbf{y}\mid \boldsymbol{\beta}^*, \mathbf{z}_{0,T}\right)p\left(\boldsymbol{\beta}^* \mid \boldsymbol{\lambda}\right)p\left(\mathbf{z}_{0:T} \mid \boldsymbol{\phi}, \boldsymbol{\mu}, \Omega^k, \Omega^l\right)}{q\left(\boldsymbol{\beta}^*,\mathbf{z}_{0:T}\right)} \right] + \text{E}_{q(\boldsymbol{\lambda)}} \left[\log\frac{p\left(\boldsymbol{\lambda)}\right)}{q\left(\boldsymbol{\lambda)}\right)} \right] \\
    & + \text{E}_{q(\boldsymbol{\mu}) } \left[\log\frac{p\left(\boldsymbol{\mu}\right)}{q\left(\boldsymbol{\mu}\right)} \right] + \text{E}_{q(\boldsymbol{\phi}) } \left[\log\frac{p\left(\boldsymbol{\phi}\right)}{q\left(\boldsymbol{\phi}\right)} \right] + \text{E}_{q(\Omega^k) }  \left[\log\frac{p\left(\Omega^k\right)}{q\left(\Omega^k\right)} \right] + \text{E}_{q(\Omega^l ) } \left[\log\frac{p\left(\Omega^l \right)}{q\left(\Omega^l \right)} \right].
\end{align*}
The first term on the right hand side of the equation bears most importance as it connects likelihood with prior and it is also the most computationally heavy part to optimize due to the dimension of $\mathbf{z}_{0,T}$. Denote the mean and covariance matrix of joint multivariate normal prior on $\boldsymbol{\beta}^* $ and $\mathbf{z}_{0:T}$ by $\mathbf{m}_0$ and $M_0$.
$M_0$ is block diagonal, whose upper diagonal block corresponds to the covariance of $\boldsymbol{\beta}^*$ and lower diagonal block is the vector autoregressive matrix that derives from the assumed $\mathbf{z}_t$ dynamics. The lower diagonal block is

\begin{equation*}
    \left(I_T\bigotimes P^{-1}\right)\begin{pmatrix}
I_{LK} & \Phi & \Phi^2 & \cdots & \Phi^{T-1} \\
\Phi & I_{LK} & \Phi & \cdots & \Phi^{T-2} \\
\Phi^2 & \Phi & I_{LK} & \cdots & \Phi^{T-3} \\
\vdots & \vdots & \vdots & \ddots & \vdots \\
\Phi^{T-1} & \Phi^{T-2} & \Phi^{T-3} & \cdots & I_{LK} \\
\end{pmatrix}\left(I_T\bigotimes \left(P^{-1}\right)'\right).
\end{equation*}
It is more convenient to use the precision matrix as the term repeatedly appears in the object function that we aim to optimize. The precision matrix can be expressed as $\left(I_T\bigotimes P'\right)R\left(I_T\bigotimes P\right)$ with
\begin{equation*}
\resizebox{\textwidth}{!}{$
    R= \begin{pmatrix}
(I_{LK} - \Phi^2)^{-1} & -(I_{LK} - \Phi^2)^{-1}\Phi & \mathbf{0} & \cdots & \mathbf{0} \\
-(I_{LK} - \Phi^2)^{-1}\Phi & (I_{LK} - \Phi^2)^{-1}(I_{LK} + \Phi^2) & -(I_{LK} - \Phi^2)^{-1}\Phi & \cdots & \mathbf{0} \\
\mathbf{0} & -(I_{LK} - \Phi^2)^{-1}\Phi & (I_{LK} - \Phi^2)^{-1}(I_{LK} + \Phi^2) & \cdots & \mathbf{0} \\
\vdots & \vdots & \vdots & \ddots & \vdots \\
\mathbf{0} & \mathbf{0} & \mathbf{0} & \cdots & (I_{LK} - \Phi^2)^{-1} \\
\end{pmatrix}.
$}
\end{equation*}
The first terms is hence
\begin{align*}
& \text{E}_q \left[\log\frac{p\left(\mathbf{y}\mid \boldsymbol{\beta}^*, \mathbf{z}_{0,T}\right)p\left(\boldsymbol{\beta}^* \mid \boldsymbol{\lambda}\right)p\left(\mathbf{z}_{0:T} \mid \boldsymbol{\mu}, \boldsymbol{\phi}, \Omega^k, \Omega^l\right)}{q\left(\boldsymbol{\beta}^*,\mathbf{z}_{0:T}\right)} \right] \\
= & \mathbf{y}'X\mathbf{m}-\sum_{i=1}^{NT}\epsilon_i\exp\left(\mathbf{x}'_i\mathbf{m}+\frac{1}{2}\mathbf{x}'_i M \mathbf{x}_i\right) -\frac{1}{2}\left[\mathbf{m}-\text{E}_q(\mathbf{m}_0)\right]'\text{E}_q\left(M_0^{-1}\right)\left[\mathbf{m}-\text{E}_q(\mathbf{m}_0)\right] \\
& -\frac{1}{2}\text{tr}\left[\text{E}_q(M_0^{-1})M\right] + \frac{1}{2}\log|M| - \frac{1}{2}\text{tr}\left[\text{E}_q(M_0^{-1}))\text{COV}_q(\mathbf{m}_0)\right] + \frac{1}{2}\text{E}_q\left(\log|M_0^{-1}|\right) \\
& + constant
\end{align*}
Here, $X$ is a two-block design matrix. The left block matrix consists of all regressors while the right block is the kronecker product $I_{LKT}\bigotimes \mathbf{1}$. The length of vector $\mathbf{1}$ relies on specific data configuration; in our case, it equals the product between number of age groups and gender. $\mathbf{x}_i$ is the column vector of $i$-th row of the design matrix $X$. The expected value $\text{E}_q(\mathbf{m}_0)=(\mathbf{0}', (\mathbf{1}_T\bigotimes \boldsymbol{\mu}^q)')'$ and $\text{E}_q\left(M_0^{-1}\right)$ is
\begin{equation*}
\resizebox{\textwidth}{!}{$
    \text{E}_q\left(M_0^{-1}\right)=\begin{pmatrix}
\frac{1}{\sigma_\beta^2}I & \mathbf{0} & \cdots & \mathbf{0} & \mathbf{0} \\
\mathbf{0} & \lambda_1^{q,r} S_1^r+\lambda_2^{q,r} S_2^r & \cdots & \mathbf{0} & \mathbf{0} \\
\vdots  & \vdots & \ddots & \mathbf{0} & \mathbf{0} \\
\mathbf{0} & \mathbf{0} & \mathbf{0} & \lambda_1^{q,g,a} S_1^{g,a}+\lambda_2^{q,g,a} S_2^{g,a} & \mathbf{0} \\
\mathbf{0} & \mathbf{0} & \mathbf{0} & \mathbf{0} & \text{E}_q\left[\left(I_T\bigotimes P'\right)\text{E}_q(R)\left(I_T\bigotimes P\right)\right]
\end{pmatrix},$
}
\end{equation*}
where 
\begin{equation*}
\text{E}_q\left( R \right) =\begin{pmatrix}  \text{E}^1_q\left( R \right) & \text{E}^2_q\left( R \right)  & \mathbf{0} & \cdots & \mathbf{0} \\
 \text{E}^2_q\left( R \right)  &  \text{E}^3_q\left( R \right)  &  \text{E}^2_q\left( R \right)  & \cdots & \mathbf{0} \\
\mathbf{0} &  \text{E}^2_q\left( R \right)  &  \text{E}^3_q\left( R \right)  & \cdots & \mathbf{0} \\
\vdots & \vdots & \vdots & \ddots & \vdots \\
\mathbf{0} & \mathbf{0} & \mathbf{0} & \cdots & \text{E}^1_q\left( R \right) 
\end{pmatrix},
\end{equation*}
with diagonal matrices
\begin{equation*}
    \text{E}^1_q\left( R \right) = \text{diag}\left(\frac{1}{1-(\boldsymbol{\phi}^q)^2}\right), \quad \text{E}^2_q\left( R \right) = \text{diag}\left(-\frac{\boldsymbol{\phi}^q}{1-(\boldsymbol{\phi}^q)^2}\right), \quad \text{E}^3_q\left( R \right) = \text{diag}\left(\frac{1+(\boldsymbol{\phi}^q)^2}{1-(\boldsymbol{\phi}^q)^2}\right).
\end{equation*}
Here we vectorize the function by writing its arguments in terms of vectors $\boldsymbol{\phi}^q$. The expectation is therefore
\begin{equation*}
\resizebox{\textwidth}{!}{$
\text{E}_q\left[\left(I_T\bigotimes P'\right)\text{E}_q(R)\left(I_T\bigotimes P\right)\right] = \begin{pmatrix}  \text{E}_q\left[P'\text{E}^1_q\left( R \right)P\right] & \text{E}_q\left[P'\text{E}^2_q\left( R \right)P\right]  & \mathbf{0} & \cdots & \mathbf{0} \\
 \text{E}_q\left[P'\text{E}^2_q\left( R \right)P\right]  &  \text{E}_q\left[P'\text{E}^3_q\left( R \right)P\right]  &  \text{E}_q\left[P'\text{E}^2_q\left( R \right)P\right]  & \cdots & \mathbf{0} \\
\mathbf{0} &  \text{E}_q\left[P'\text{E}^2_q\left( R \right)P\right]  &  \text{E}_q\left[P'\text{E}^3_q\left( R \right)P\right]  & \cdots & \mathbf{0} \\
\vdots & \vdots & \vdots & \ddots & \vdots \\
\mathbf{0} & \mathbf{0} & \mathbf{0} & \cdots & \text{E}_q\left[P'\text{E}^1_q\left( R \right)P\right]
\end{pmatrix}.
$}
\end{equation*}
Recall that $P'P=\Omega = \Omega^k \bigotimes \Omega^l$, when the variational densities of precision matrices $\Omega^k$ and $\Omega^l$ are $\text{Wishart}(\delta^{q,k}, D^{q,k})$ and $\text{Wishart}(\delta^{q,l}, D^{q,l})$, the derived Cholesky upper triangular matrix $P$ can be written as $P = \left(A^kV^{q,k}\right) \bigotimes \left(A^lV^{q,l}\right)$ with $(V^{q,k}A^k)'A^kV^{q,k} = \Omega^k, (V^{q,l}A^l)'A^lV^{q,l} = \Omega^l$. $V^{q,k}, V^{q,l}$ are Cholesky factors of $D^{q,k}$ and $D^{q,l}$ and 
\begin{equation*}
A^k = \begin{pmatrix}
    c^k_1 & n^k_{1,2} & n^k_{1,3} & \cdots & n^k_{1,K} \\
    0 & c^k_2 & n^k_{2,3} & \cdots & n^k_{2,K} \\
    0 & 0 & c^k_3 &  \cdots & n^k_{3,K} \\
    \vdots & \vdots & \vdots & \ddots & \vdots \\
    0 & 0 & 0 & \cdots & c^k_K 
\end{pmatrix}, \quad 
A^l = \begin{pmatrix}
    c^l_1 & n^l_{1,2} & n^l_{1,3} & \cdots & n^l_{1,K} \\
    0 & c^l_2 & n^l_{2,3} & \cdots & n^l_{2,K} \\
    0 & 0 & c^l_3 &  \cdots & n^l_{3,K} \\
    \vdots & \vdots & \vdots & \ddots & \vdots \\
    0 & 0 & 0 & \cdots & c^l_K 
\end{pmatrix},
\end{equation*}
with $c^k_i\sim \chi^2_{\delta^{q,k}-i+1}, c^l_i\sim \chi^2_{\delta^{q,l}-i+1}, n^k_{i,j}, n^l_{i,j} \sim \mathcal{N}\left(0,1\right)$ independently. This is known as the Bartlett decomposition \citep{anderson1958introduction, smith1972algorithm}. Thanks to the decomposition and diagonal assumption on $\Phi$, we are able to work out the expectation of the quadratic forms. In the following, we demonstrate the details of deriving $\text{E}_q\left[P'\text{E}^1_q\left( R \right)P\right]$. $\text{E}_q\left[P'\text{E}^2_q\left( R \right)P\right]$ and $\text{E}_q\left[P'\text{E}^3_q\left( R \right)P\right]$ have similar structures.
\begin{align*}
    \text{E}_q\left[P'\text{E}^1_q\left( R \right)P\right] & = \text{E}_q\left[\left(V^{q,k}\bigotimes V^{q,l}\right)'\left(A^k\bigotimes A^l\right)'\text{E}^1_q\left( R \right)\left(A^k\bigotimes A^l\right)\left(V^{q,k} \bigotimes V^{q,l}\right)\right] \\
    & = \left(V^{q,k}\bigotimes V^{q,l}\right)'\text{E}_q\left[\left(A^k\bigotimes A^l\right)'\text{E}^1_q\left( R \right)\left(A^k\bigotimes A^l\right)\right]\left(V^{q,k} \bigotimes V^{q,l}\right).
\end{align*}
Now the problem simplifies to take the expectation of $\left(A^k\bigotimes A^l\right)'\text{E}^1_q\left( R \right)\left(A^k\bigotimes A^l\right)$ with
\begin{equation*}
    A^k\bigotimes A^l = \begin{pmatrix}
    c^k_1 A^l  & n^k_{1,2}A^l  & n^k_{1,3}A^l  & \cdots & n^k_{1,K}A^l  \\
    \mathbf{0} & c^k_2 A^l  & n^k_{2,3}A^l  & \cdots & n^k_{2,K}A^l  \\
    \mathbf{0} & \mathbf{0} & c^k_3 A^l &  \cdots & n^k_{3,K}A^l  \\
    \vdots & \vdots & \vdots & \ddots & \vdots \\
    \mathbf{0} & \mathbf{0} & \mathbf{0} & \cdots & c^k_K A^l  
    \end{pmatrix}.
\end{equation*}
Since $c^k_i,n_{i,j}^k$ are independent, the expected value is a block diagonal matrix with diagonal entries $\text{E}\left[\left(c_1^k\right)^2\right]\text{E}\left\{\left(A^l\right)'\left[\text{E}^1_q\left( R \right)\right]_{1:L}A^l\right\}$, $\text{E}\left[\left(n_{1,2}^k\right)^2\right]\text{E}\left\{\left(A^l\right)'\left[\text{E}^1_q\left( R \right)\right]_{1:L}A^l\right\} +\text{E}\left[\left(c_2^k\right)^2\right]\text{E}\left\{\left(A^l\right)'\left[\text{E}^1_q\left( R \right)\right]_{(L+1):2L}A^l\right\}, \dots, \text{E}\left[\left(n_{1,K}^k\right)^2\right]\text{E}\left\{\left(A^l\right)'\left[\text{E}^1_q\left( R \right)\right]_{1:L}A^l\right\} +\cdots+$ 

\noindent $\text{E}\left[\left(c_K^k\right)^2\right]\text{E}\left\{\left(A^l\right)'\left[\text{E}^1_q\left( R \right)\right]_{((K-1)L+1):KL}A^l\right\}$
where $\left[\text{E}^1_q\left( R \right)\right]_{1:L}$ denotes the diagonal blocks indexed from 1 to $L$ and so on. By the same reasoning, $\text{E}\left\{\left(A^l\right)'\left[\text{E}^1_q\left( R \right)\right]_{1:L}A^l\right\}$ is a diagonal matrix whose diagonal entries are $\text{E}\left[(c_1^l)^2\right]\left[\text{E}^1_q\left( R \right)\right]_1$, $\text{E}\left[(n_{1,2}^l)^2\right]\left[\text{E}^1_q\left( R \right)\right]_1+\text{E}\left[(c_2^l)^2\right]\left[\text{E}^1_q\left( R \right)\right]_2, \dots, \text{E}\left[(n_{1,L}^l)^2\right]\left[\text{E}^1_q\left( R \right)\right]_1+\cdots+\text{E}\left[(c_L^l)^2\right]\left[\text{E}^1_q\left( R \right)\right]_L$. The same structure also holds for the remaining matrix expectations. The last two quantities in the expectation $\text{E}_q \left[\log\frac{p\left(\mathbf{y}\mid \boldsymbol{\beta}^*, \mathbf{z}_{0,T}\right)p\left(\boldsymbol{\beta}^* \mid \boldsymbol{\lambda}\right)p\left(\mathbf{z}_{0:T} \mid \boldsymbol{\mu}, \boldsymbol{\phi}, \Omega^k, \Omega^l\right)}{q\left(\boldsymbol{\beta}^*,\mathbf{z}_{0:T}\right)} \right]$ are
\begin{equation*}
    \text{COV}_q\left(\mathbf{m}_0\right)= \begin{pmatrix}
        \mathbf{0} & \mathbf{0} \\
        \mathbf{0} & I_T \bigotimes \text{diag}\left[\left(\boldsymbol{\sigma}^q\right)^2\right]
    \end{pmatrix},
\end{equation*}
and
\begin{align*}
    \text{E}_q\left(\log|M_0^{-1}|\right) = & 
    \log\left|\lambda_1^{q,r}S_1^r+\lambda_2^{q,r}S_2^r\right| + 
\log\left|\lambda_1^{q,a}S_1^a+\lambda_2^aS_2^{q,a}\right| + \log\left|\lambda_1^{q,k,r}S_1^{k,r}+\lambda_2^{q,k,r}S_2^{k,r}\right| \\
& + \log\left|\lambda_1^{q,k,a}S_1^{k,a}+\lambda_2^{q,k,a}S_2^{k,a}\right| + \log\left|\lambda_1^{q,g,a}S_1^{g,a}+\lambda_2^{q,g,a}S_2^{g,a}\right| \\
&+\left(T-1\right)\sum_{i=1}^{LK}\log\left(\frac{1}{1-(\phi_i^q)^2}\right)+LT\left[\log|D^{q,k}|+\sum_{k=1}^K \psi\left(\frac{\delta^{q,k}-k+1}{2}\right)\right] \\
& + KT\left[\log|D^{q,l}|+\sum_{l=1}^L \psi\left(\frac{\delta^{q,l}-l+1}{2}\right)\right] + constant,
\end{align*}
with $\psi(\cdot)$ representing the digamma function.
The remaining terms in the ELBO are
\begin{align*}
    \text{E}_{q({\boldsymbol{\lambda}}) } \left[\log\frac{p\left({\boldsymbol{\lambda}}\right)}{q\left({\boldsymbol{\lambda}}\right)} \right] = & \left(\alpha - 1\right)\log\boldsymbol{\lambda}^q - \beta\boldsymbol{\lambda}^q + constant \\
    \text{E}_{q(\boldsymbol{\mu}) } \left[\log\frac{p\left(\boldsymbol{\mu}\right)}{q\left(\boldsymbol{\mu}\right)} \right] = & -\frac{1}{2\sigma_\mu^2}{\boldsymbol{\mu}^q}'\boldsymbol{\mu}^q - \frac{1}{2\sigma_\mu^2} \sum \left({\boldsymbol{\sigma}^q}\right)^2
    + \frac{1}{2}\sum\log\left[\left({\boldsymbol{\sigma}^q}\right)^2\right] + constant \\
    \text{E}_{q({\boldsymbol{\phi}}) } \left[\log\frac{p\left({\boldsymbol{\phi}}\right)}{q\left({\boldsymbol{\phi}}\right)} \right] = & \sum_{i=1}^{LK} (\alpha_\phi-1)\log\left(1+\phi_i^2\right) + (\beta_\phi-1)\log\left(1-\phi_i^2\right) + constant \\
    \text{E}_{q(\Omega^k) }  \left[\log\frac{p\left(\Omega^k\right)}{q\left(\Omega^k\right)} \right] = &\frac{\delta^k}{2}\log|D^{q,k}| + \frac{\delta^k-\delta^{q,k}}{2}\sum_{i=1}^K \psi\left(\frac{\delta^{q,k}-i+1}{2}\right) - \frac{\delta^{q,k}}{2\theta^k}\text{tr}(D^{q,k}) \\
    & +\frac{\delta^{q,k}K}{2}+\log\Gamma_K\left(\frac{\delta^{q,k}}{2}\right) + constant \\
    \text{E}_{q(\Omega^l) }  \left[\log\frac{p\left(\Omega^l\right)}{q\left(\Omega^l\right)} \right] = &\frac{\delta^l}{2}\log|D^{q,l}| + \frac{\delta^l-\delta^{q,l}}{2}\sum_{i=1}^L \psi\left(\frac{\delta^{q,l}-i+1}{2}\right) - \frac{\delta^{q,l}}{2\theta^l}\text{tr}(D^{q,l}) \\
    & +\frac{\delta^{q,l}L}{2}+\log\Gamma_L\left(\frac{\delta^{q,l}}{2}\right) + constant 
\end{align*}
Now that we have formulated the ELBO, a coordinate ascent variational inference (CAVI) algorithm is devised and employed to maximize the target; that is we optimize the ELBO with respect to $\mathbf{m}, M, \boldsymbol{\lambda}^q, \boldsymbol{\phi}^q, \delta^{q,k}, D^{q,k}, \delta^{q,l}, D^{q,l}$ sequentially while keeping the other parameters fixed. This is outlined in Algorithm \ref{alg:CAVI}.
               \begin{algorithm}
\caption{CAVI for Bayesian dynamic GAM}
\label{alg:CAVI}
\KwIn{data set $\mathbf{y}, X, \boldsymbol{\epsilon}$ \newline
GAM regularization matrices $S_1^r, S_2^r, S_1^a, S_2^a,S_1^{k,r}, S_2^{k,r}, S_1^{k,a}, S_2^{k,a}, S_1^{s,a}, S_2^{s,a}$ \newline
priors $p(\boldsymbol{\beta}), p(\boldsymbol{\lambda}), p(\boldsymbol{\mu}),p(\boldsymbol{\phi}), q(\Omega^l), q(\Omega^k)$}
\KwOut{variational densities $q(\boldsymbol{\beta}^*, \mathbf{z}_{0:T}), q(\boldsymbol{\lambda}), q(\boldsymbol{\mu}), q(\boldsymbol{\phi}), q(\Omega^l), q(\Omega^k)$}
\SetKwInput{kwInit}{Initialization}
\kwInit{variational densities $q(\boldsymbol{\beta}^*, \mathbf{z}_{0:T}), q(\boldsymbol{\lambda}), q(\boldsymbol{\mu}), q(\boldsymbol{\phi}), q(\Omega^l), q(\Omega^k)$}

\While{the algorithm has not converged}{
set $q(\boldsymbol{\beta}^*, \mathbf{z}_{0:T})$ according to \ref{subsection:m_M} \\
set $q(\boldsymbol{\lambda})$ according to \ref{subsection:lambda} \\
set $q(\boldsymbol{\mu})$ according to \ref{subsection:mu_sigma} \\
set $q(\boldsymbol{\phi})$ according to \ref{subsection:phi} \\
set $q(\Omega^l)$ according to \ref{subsection:deltal_Vl} \\
set $q(\Omega^k)$ according to \ref{subsection:deltak_Vk} \\
}
\end{algorithm}

\subsection{optimizing with respect to \texorpdfstring{$\mathbf{m}$}{m} and \texorpdfstring{$M$}{M}}
\label{subsection:m_M}

The first order conditions of optimal $\mathbf{m}, M$ are
\begin{align*}
    \frac{\partial \text{ELBO}}{\partial \mathbf{m}} = X' & \mathbf{y} - \sum_{i=1}^{NT} \epsilon_i \exp\left(\mathbf{x}'_i\mathbf{m}+ \frac{1}{2}\mathbf{x}'_i M\mathbf{x}_i\right)\mathbf{x}_i - \text{E}_q\left(M_0^{-1}\right)\left[\mathbf{m}-\text{E}_q\left(\mathbf{m}_0\right)\right] \\
    \frac{\partial \text{ELBO}}{\partial M} = \frac{1}{2} \Biggl\{ &-X'\text{diag}\biggl[\epsilon_1\exp\left(\mathbf{x}'_1\mathbf{m}+\frac{1}{2}\mathbf{x}'_1 M \mathbf{x}_1\right), \dots, \\
&\epsilon_{NT}\exp\left(\mathbf{x}'_{NT}\mathbf{m}+\frac{1}{2}\mathbf{x}'_{NT} M \mathbf{x}_{NT} \right)\biggl]X 
    -\text{E}_q\left(M_0^{-1}\right)+M^{-1} \Biggl\}
\end{align*}
We follow the numeric algorithm proposed by \citet{arridge2018variational} and update $\mathbf{m}$ using Newton's method, which requires the computation of Hessian matrix
\begin{align*}
    H^{\text{ELBO}}_{\mathbf{m}} = -X'\text{diag} &\left[\epsilon_1\exp\left(\mathbf{x}'_1\mathbf{m}+\frac{1}{2}\mathbf{x}'_1 M \mathbf{x}_1\right), \dots, \right.\\
    &  \left. \epsilon_{NT}\exp\left(\mathbf{x}'_{NT}\mathbf{m}+\frac{1}{2}\mathbf{x}'_{NT} M \mathbf{x}_{NT} \right)\right]X  - \text{E}_q\left(M_0^{-1}\right),
\end{align*}
and the new $\mathbf{m}^{(h+1)}$ at iteration $h+1$ is updated based on the previous value $\mathbf{m}^{(h)}$
\begin{equation*}
    \mathbf{m}^{(h+1)} = \mathbf{m}^{(h)} -\left(H^{\text{ELBO}}_{\mathbf{m}}\right)^{-1}\frac{\partial \text{ELBO}}{\partial \mathbf{m}}.
\end{equation*}
To obtain the optimal $M$, a fixed-point method is employed and at iteration $h+1$, it updates $M$ according to
\begin{align*}
\begin{split}
        M^{(h+1)} = g\left(M^{(h)}\right) = \Bigl(X'\text{diag} &\left[\epsilon_1\exp\left(\mathbf{x}'_1\mathbf{m}+\frac{1}{2}\mathbf{x}'_1 M^{(h)} \mathbf{x}_1\right), \dots, \right.\\
    &  \left. \epsilon_{NT}\exp\left(\mathbf{x}'_{NT}\mathbf{m}+\frac{1}{2}\mathbf{x}'_{NT} M^{(h)} \mathbf{x}_{NT} \right)\right]X +\text{E}_q\left(M_0^{-1}\right)\Bigl)^{-1}, 
\end{split}
\end{align*}
where $M^{(h)}$ stands for the value at iteration $h$.

In practice, it is possible that the fixed-point iteration algorithm takes many iterations to converge or it oscillates between two values. Since to compute $g(M^{(h)})$ is the most expensive step in the whole algorithm due to the matrix inversion operation of the potentially high dimensional matrix, we wish to avoid such situations from happening, therefore we employ the Anderson acceleration technique as outlined in \citet{walker2011anderson}. Essentially, the Anderson acceleration algorithm uses a combination of $S$ previous iterates to form the next iterate, improving convergence rate. The pseudo code of the algorithm is given in Algorithm \ref{alg:AA}.

\begin{algorithm}[H]
\DontPrintSemicolon
\KwIn{function $g$, memory $S$}
\KwOut{fixed point $M^*$}
\SetKwInput{kwInit}{Initialization}
\kwInit{$M^{(0)}$}
set $h=1$ and $M^{(1)} = g(M^{(0)})$

\While{the algorithm has not converged}{
set $s_h = \min \{S,h\}$ \\
set $F_h = (f_{h - s_h}, \dots, f_h)$, where $ f_i = \text{vec}[g(M^{(i)}) - M^{(i)}]$ \\
determine $\gamma^{(h)} = \left(\gamma^{(h)}_0, \dots, \gamma^{(h)}_{s_h}\right)'$ that solves
\begin{equation*}
    \min_{\gamma = \left(\gamma_0, \dots, \gamma_{s_h}\right)'} ||F_h\gamma||_2 \quad \text{s.t.} \quad \sum_{i=0}^{s_h}\gamma_i=1 
\end{equation*} \\
set $M^{(h+1)} = \sum_{i=0}^{s_h}\gamma^{(h)}_i g\left(M^{(h-s_h+1)}\right)$
}
\caption{Anderson acceleration for fixed-point iteration solving for $M$}
\label{alg:AA}
\end{algorithm}


\subsection{optimizing with respect to \texorpdfstring{$\boldsymbol{\lambda}^q$}{lambda}}
\label{subsection:lambda}

We optimize $\boldsymbol{\lambda}^q$ jointly through Newton's method as well. The steps to obtain first order conditions and Hessian matrices with respect to $\lambda_1^{q,r}$ and $\lambda_2^{q,r}$ are illustrated as examples and the rest of the smoothing parameters have similar formulas. The first order conditions of $\lambda^{q,r}_1, \lambda^{q,r}_2$ are
\begin{align*}
    \frac{\partial \text{ELBO}}{\partial \lambda_1^{q,r}} = & -\frac{1}{2}\left(\mathbf{m}^r\right)'S_1^r\mathbf{m}^r-\frac{1}{2}\text{tr}\left(S_1^rM^r\right)+\frac{1}{2}\text{tr}\left[\left(\lambda_1^{q,r}S_1^r+\lambda_2^{q,r}S_2^r\right)^{-1}S_1^r\right]+ \frac{\alpha-1}{\lambda_1^{q,r}} - \beta, \\
    \frac{\partial \text{ELBO}}{\partial \lambda_2^{q,r}} = & -\frac{1}{2}\left(\mathbf{m}^r\right)'S_2^r\mathbf{m}^r-\frac{1}{2}\text{tr}\left(S_2^rM^r\right) +\frac{1}{2}\text{tr}\left[\left(\lambda_1^{q,r}S_1^r+\lambda_2^{q,r}S_2^r\right)^{-1}S_2^r\right]+ \frac{\alpha-1}{\lambda_2^{q,r}} - \beta,
\end{align*}
where $\mathbf{m}^r$ and $M^r$ take corresponding entries in $\mathbf{m}$ and $M$ that are associated with the spline of stringency index. The Hessian matrix block associated with $\lambda^{q,r}_1, \lambda^{q,r}_2$ is
\begin{equation*}
\resizebox{\textwidth}{!}{$
    H_{\lambda_1^{q,r},\lambda_2^{q,r}}^{\text{ELBO}} = \begin{pmatrix}
        -\frac{1}{2}\text{tr}\left[\left(\lambda_1^{q,r}S_1^r+\lambda_2^{q,r}S_2^r\right)^{-1}S_1^r\left(\lambda_1^{q,r}S_1^r+\lambda_2^{q,r}S_2^r\right)^{-1}S_1^r\right] - \frac{\alpha-1}{\left(\lambda_1^{q,r}\right)^2} & -\frac{1}{2}\text{tr}\left[\left(\lambda_1^{q,r}S_1^r+\lambda_2^{q,r}S_2^r\right)^{-1}S_1^r\left(\lambda_1^{q,r}S_1^r+\lambda_2^{q,r}S_2^r\right)^{-1}S_2^r\right]\\
        -\frac{1}{2}\text{tr}\left[\left(\lambda_1^{q,r}S_1^r+\lambda_2^{q,r}S_2^r\right)^{-1}S_1^r\left(\lambda_1^{q,r}S_1^r+\lambda_2^{q,r}S_2^r\right)^{-1}S_2^r\right] & -\frac{1}{2}\text{tr}\left[\left(\lambda_1^{q,r}S_1^r+\lambda_2^{q,r}S_2^r\right)^{-1}S_2^r\left(\lambda_1^{q,r}S_1^r+\lambda_2^{q,r}S_2^r\right)^{-1}S_2^r\right]  - \frac{\alpha-1}{\left(\lambda_2^{q,r}\right)^2}
    \end{pmatrix}.
$}
\end{equation*}
Similarly we can derive first order conditions and Hessian matrices with respect to $\lambda_1^{q,a}, \lambda_2^{q,a}$, $\lambda_1^{q,a}, \lambda_2^{q,a}, \lambda_1^{q,k,r}, \lambda_2^{q,k,r}, \lambda_1^{q,k,a}, \lambda_2^{q,k,a}, \lambda_1^{q,g,a}, \lambda_2^{q,g,a}$.
The joint updating of smoothing parameters $\boldsymbol{\lambda}^q$ at iteration $h+1$ is then
\begin{equation}
    \left(\boldsymbol{\lambda}^q\right)^{(h+1)} = \left(\boldsymbol{\lambda}^q\right)^{(h)} - \left(H^{\text{ELBO}}_{\boldsymbol{\lambda}^q}\right)^{-1}\frac{\partial \text{ELBO}}{\partial \boldsymbol{\lambda}^q}.
\end{equation}
Here $H^{\text{ELBO}}_{\boldsymbol{\lambda}^q}$ is a block diagonal matrix with block entries $H_{\lambda_1^{q,r},\lambda_2^{q,r}}^{\text{ELBO}}, H_{\lambda_1^{q,a},\lambda_2^{q,a}}^{\text{ELBO}},
H_{\lambda_1^{q,k,r},\lambda_2^{q,k,r}}^{\text{ELBO}}$,
$H_{\lambda_1^{q,k,a},\lambda_2^{q,k,a}}^{\text{ELBO}}$ and
$H_{\lambda_1^{q,g,a},\lambda_2^{q,g,a}}^{\text{ELBO}}$. Note that $\boldsymbol{\lambda}^q$ is subject to positive constraints, therefore in each step, we apply a projected Newton's method with backtracking line search to find the optimal solution \citep{bertsekas1982projected, tibshirani2015general}.

\subsection{optimizing with respect to \texorpdfstring{$\boldsymbol{\mu}^q$}{mu} and \texorpdfstring{$\left(\boldsymbol{\sigma}^q\right)^2$}{sigma}}
\label{subsection:mu_sigma}
First order conditions with respect to $\boldsymbol{\mu}_i^q$, the $i$-th element of $\boldsymbol{\mu}^q$ is
\begin{equation*}
    \frac{\partial \text{ELBO}}{\partial \boldsymbol{\mu}_i^q} = \text{E}_q\left(M_0^{-1}\right)\left[\mathbf{m}-\text{E}_q(\mathbf{m}_0)\right]\frac{\partial \text{E}_q(\mathbf{m}_0)}{\partial \boldsymbol{\mu}_i^q}-\frac{1}{\sigma^2_\mu}\boldsymbol{\mu}_i^q.
\end{equation*}
The second derivatives are
\begin{equation*}
    \frac{\partial^2 \text{ELBO}}{\partial \boldsymbol{\mu}_i^q\partial \boldsymbol{\mu}_j^q} = \left(\frac{\partial \text{E}_q(\mathbf{m}_0)}{\partial \boldsymbol{\mu}_i^q}\right)'\text{E}_q\left(M_0^{-1}\right)\frac{\partial \text{E}_q(\mathbf{m}_0)}{\partial \boldsymbol{\mu}_j^q}-\frac{1}{\sigma^2_\mu}
\end{equation*}
when $i=j$ and 
\begin{equation*}
    \frac{\partial^2 \text{ELBO}}{\partial \boldsymbol{\mu}_i^q\partial \boldsymbol{\mu}_j^q} = \left(\frac{\partial \text{E}_q(\mathbf{m}_0)}{\partial \boldsymbol{\mu}_i^q}\right)'\text{E}_q\left(M_0^{-1}\right)\frac{\partial \text{E}_q(\mathbf{m}_0)}{\partial \boldsymbol{\mu}_j^q}
\end{equation*}
when $i\neq j$. Newton's method is then directly applicable to obtain optimal $\boldsymbol{\mu}^q$.

First order conditions to update $\left(\boldsymbol{\sigma}^q_i\right)^2$, the $i$-th element of $\left(\boldsymbol{\sigma}^q\right)^2$ are
\begin{equation*}
    \frac{\partial \text{ELBO}}{\partial \left(\boldsymbol{\sigma}^q_i\right)^2} = - \frac{1}{2}\text{tr}\left[\text{E}_q(M_0^{-1}) \frac{\text{COV}_q(\mathbf{m}_0)}{\partial \left(\boldsymbol{\sigma}^q_i\right)^2}\right] - \frac{1}{2\sigma_\mu^2}
    + \frac{1}{2\left(\boldsymbol{\sigma}^q_i\right)^2}.
\end{equation*}
$\left(\boldsymbol{\sigma}^q\right)^2$ can be updated by setting first order conditions equal to 0 and then solve the linear system. The positivity constraint is automatically satisfied here. 

\subsection{optimizing with respect to \texorpdfstring{$\boldsymbol{\phi}^q$}{phi}}
\label{subsection:phi}

To update a specific $\phi^q_i$ in the $l$-th position of sequence $1,\dots,L$ and $k$-th position of sequence $1,\dots,K$, first note that the term containing $\phi^q_i$ in the expectation of quantity $\left[\left(A^k\right)'\bigotimes \left(A^l\right)'\right]\text{E}^1_q\left( R \right)\left(A^k\bigotimes A^l\right)$ is 
\begin{equation*}
    1/\left(1-(\phi^q_i)^2\right)\text{diag}\left[\left(0,\dots,0,\delta^{q,k}-k+1,1,\dots,1\right)'\bigotimes\left(0,\dots,0,\delta^{q,l}-l+1,1,\dots,1\right)'\right],
\end{equation*}
with $l$-th entry of vector $\left(0,\dots,0,\delta^{q,k}-k+1,1,\dots,1\right)'$ equal to $\delta^{q,k}-k+1$ and $k$-th entry of vector $\left(0,\dots,0,\delta^{q,l}-l+1,1,\dots,1\right)'$ equal to $\delta^{q,l}-l+1$. Similarly, in the expectation of $\left[\left(A^k\right)'\bigotimes \left(A^l\right)'\right]\text{E}^2_q\left( R \right)\left(A^k\bigotimes A^l\right)$ and $\left[\left(A^k\right)'\bigotimes \left(A^l\right)'\right]\text{E}^3_q\left( R \right)\left(A^k\bigotimes A^l\right)$, we change $1/\left(1-(\phi^q_i)^2\right)$ to $-\phi^q_i/\left(1-(\phi^q_i)^2\right)$ and $\left(1+(\phi^q_i)^2\right)/\left(1-(\phi^q_i)^2\right)$ accordingly. The first order condition with respect to $\phi^q_i$ is therefore
\begin{align*}
    \frac{\partial \text{ELBO}}{\partial \phi^q_i} = -\frac{1}{2}\text{tr} \Biggl\{&\left(I_T \bigotimes V^{q,k}\bigotimes V^{q,l}\right) \left[\mathbf{m}-\text{E}_q\left(\mathbf{m}_0\right)\right] \left[\mathbf{m}-\text{E}_q\left(\mathbf{m}_0\right)\right]'  \\
    &\left(I_T \bigotimes V^{q,k}\bigotimes V^{q,l}\right)'\frac{\partial \text{E}^*\left(\boldsymbol{\phi},\delta^{q,l}, \delta^{q,k}\right)}{\partial \phi^q_i} \Biggl\} \\
    -\frac{1}{2}\text{tr}\Biggl\{&\left(I_T \bigotimes V^{q,k}\bigotimes V^{q,l}\right)M\left(I_T \bigotimes V^{q,k}\bigotimes V^{q,l}\right)'\frac{\partial \text{E}^*\left(\boldsymbol{\phi},\delta^{q,l}, \delta^{q,k}\right)}{\partial \phi^q_i} \Biggl\} \\
    + &\frac{T-1}{2}\left(\frac{1}{1-\phi^q_i} - \frac{1}{1+\phi^q_i}\right) + \frac{\alpha_\phi-1}{1+\phi^q_i} - \frac{\beta_\phi-1}{1-\phi^q_i}
\end{align*}
with $ \text{E}^*\left(\boldsymbol{\phi},\delta^{q,l}, \delta^{q,k}\right) = \text{E}_q\left[\left(I_T \bigotimes A^k\bigotimes A^l\right)'\text{E}_q\left( R \right)\left(I_T \bigotimes A^k\bigotimes A^l\right) \right]$.

We can further derive the second order condition
\begin{align*}
    \frac{\partial^2 \text{ELBO}}{\partial (\phi^q_i)^2} = -\frac{1}{2}\text{tr} \Biggl\{ &\left(I_T \bigotimes V^{q,k}\bigotimes V^{q,l}\right) \left[\mathbf{m}-\text{E}_q\left(\mathbf{m}_0\right)\right] \left[\mathbf{m}-\text{E}_q\left(\mathbf{m}_0\right)\right]' \\
    & \left(I_T \bigotimes V^{q,k}\bigotimes V^{q,l}\right)'\frac{\partial^2 \text{E}^*\left(\boldsymbol{\phi},\delta^{q,l}, \delta^{q,k}\right)}{\partial (\phi^q_i)^2} \Biggl\} \\
    -\frac{1}{2}\text{tr} \Biggl\{ &\left(I_T \bigotimes V^{q,k}\bigotimes V^{q,l}\right)M\left(I_T \bigotimes V^{q,k}\bigotimes V^{q,l}\right)'\frac{\partial^2 \text{E}^*\left(\boldsymbol{\phi},\delta^{q,l}, \delta^{q,k}\right)}{\partial (\phi^q_i)^2} \Biggl\} \\
    + &\frac{T-1}{2}\left(\frac{1}{\left(1-\phi^q_i\right)^2} + \frac{1}{\left(1+\phi^q_i\right)^2}\right) - \frac{\alpha_\phi-1}{\left(1+\phi^q_i\right)^2} - \frac{\beta_\phi-1}{\left(1-\phi^q_i\right)^2}
\end{align*}
Newton's method is then applied to update $\phi^q_i$ for $i=1,\dots,LK$ sequentially. If the number $LK$ is large in the application, $\phi^q_i$ can also be updated in parallel as the second order condition involves only $\phi^q_i$.

\subsection{optimizing with respect to \texorpdfstring{$\delta^{q,l}$}{deltal} and \texorpdfstring{$V^{q,l}$}{Vl}}
\label{subsection:deltal_Vl}

$\delta^{q,l}$ and $V^{q,l}$ are updated jointly. First order conditions are
\begin{align*}
    \frac{\partial \text{ELBO}}{\partial \delta^{q,l}} = -\frac{1}{2}\text{tr} \Biggl\{&\left(I_T \bigotimes V^{q,k}\bigotimes V^{q,l}\right) \left[\mathbf{m}-\text{E}_q\left(\mathbf{m}_0\right)\right] \left[\mathbf{m}-\text{E}_q\left(\mathbf{m}_0\right)\right]' \\
    & \left(I_T \bigotimes V^{q,k}\bigotimes V^{q,l}\right)' \frac{\partial \text{E}^*\left(\boldsymbol{\phi},\delta^{q,l}, \delta^{q,k}\right)}{\partial \delta^{q,l}} \Biggl\} \\
    -\frac{1}{2}\text{tr}\Biggl\{& \left(I_T \bigotimes V^{q,k}\bigotimes V^{q,l}\right)M\left(I_T \bigotimes V^{q,k}\bigotimes V^{q,l}\right)'\frac{\partial \text{E}^*\left(\boldsymbol{\phi},\delta^{q,l}, \delta^{q,k}\right)}{\partial \delta^{q,l}} \Biggl\} \\
    +\frac{KT}{4}&\sum_{l=1}^L \psi_1\left(\frac{\delta^{q,l}-l+1}{2}\right) + \frac{\delta^l - \delta^{q,l}}{4}\sum_{l=1}^L\psi_1\left(\frac{\delta^{q,l}-l+1}{2}\right)-\frac{1}{2\theta^l}\text{tr}\left(D^{q,l}\right) + \frac{L}{2}
\end{align*}
for degree of freedom $\delta^{q,l}$,
\begin{align*}
    \frac{\partial \text{ELBO}}{\partial V^{q,l}_{i,i}} = -\text{tr} \Biggl\{ & \left[\mathbf{m}-\text{E}_q\left(\mathbf{m}_0\right)\right] \left[\mathbf{m}-\text{E}_q\left(\mathbf{m}_0\right)\right]' \left(I_T \bigotimes V^{q,k}\bigotimes V^{q,l}\right)' \\
&\text{E}^*\left(\boldsymbol{\phi},\delta^{q,l}, \delta^{q,k}\right)\frac{\partial \left(I_T \bigotimes V^{q,k}\bigotimes V^{q,l}\right)}{\partial V^{q,l}_{i,i}} \Biggl\} \\
    -\text{tr} \Biggl\{ &M \left(I_T \bigotimes V^{q,k}\bigotimes V^{q,l}\right)'\text{E}^*\left(\boldsymbol{\phi},\delta^{q,l}, \delta^{q,k}\right)\frac{\partial \left(I_T \bigotimes V^{q,k}\bigotimes V^{q,l}\right)}{\partial V^{q,l}_{i,i}} \Biggl\} 
    + \frac{KT+\delta^l}{V^{q,l}_{i,i}} - \frac{\delta^{q,l}V^{q,l}_{i,i}}{\theta^l}
\end{align*}
for diagonal elements $V^{q,l}_{i,i}$, and
\begin{align*}
    \frac{\partial \text{ELBO}}{\partial V^{q,l}_{i,j}} = -\text{tr} \Biggl\{& \left[\mathbf{m}-\text{E}_q\left(\mathbf{m}_0\right)\right] \left[\mathbf{m}-\text{E}_q\left(\mathbf{m}_0\right)\right]' \left(I_T \bigotimes V^{q,k}\bigotimes V^{q,l}\right)' \\
    &\text{E}^*\left(\boldsymbol{\phi},\delta^{q,l}, \delta^{q,k}\right)\frac{\partial \left(I_T \bigotimes V^{q,k}\bigotimes V^{q,l}\right)}{\partial V^{q,l}_{i,j}} \Biggl\} \\
    -\text{tr}\Biggl\{ &M \left(I_T \bigotimes V^{q,k}\bigotimes V^{q,l}\right)'\text{E}^*\left(\boldsymbol{\phi},\delta^{q,l}, \delta^{q,k}\right)\frac{\partial \left(I_T \bigotimes V^{q,k}\bigotimes V^{q,l}\right)}{\partial V^{q,l}_{i,j}} \Biggl\}-\frac{\delta^{q,l}V^{q,l}_{i,j}}{\theta^l},
\end{align*}
for off-diagonal elements $V^{q,l}_{i,j}$ with $i<j$.
In the Hessian matrix, the second derivatives are
\begin{align*}
    \frac{\partial^2 \text{ELBO}}{\left(\partial \delta^{q,l}\right)^2} = &\frac{KT}{8}\sum_{l=1}^L \psi_2\left(\frac{\delta^{q,l}-l+1}{2}\right) - \frac{1}{4}\sum_{l=1}^L \psi_1\left(\frac{\delta^{q,l}-l+1}{2}\right)+\frac{\delta^l - \delta^{q,l}}{8}\sum_{l=1}^L \psi_2\left(\frac{\delta^{q,l}-l+1}{2}\right),
\end{align*}
\begin{align*}
    \frac{\partial^2 \text{ELBO}}{\left(\partial V^{q,l}_{i,i}\right)^2} = -\text{tr} \Biggl\{& \left[\mathbf{m}-\text{E}_q\left(\mathbf{m}_0\right)\right] \left[\mathbf{m}-\text{E}_q\left(\mathbf{m}_0\right)\right]' \frac{\left(I_T \bigotimes V^{q,k}\bigotimes V^{q,l}\right)'}{\partial V^{q,l}_{i,i}} \\
    &\text{E}^*\left(\boldsymbol{\phi},\delta^{q,l}, \delta^{q,k}\right)\frac{\partial \left(I_T \bigotimes V^{q,k}\bigotimes V^{q,l}\right)}{\partial V^{q,l}_{i,i}} \Biggl\} \\
    -\text{tr} \Biggl\{ &M \frac{\left(I_T \bigotimes V^{q,k}\bigotimes V^{q,l}\right)'}{\partial V^{q,l}_{i,i}} \text{E}^*\left(\boldsymbol{\phi},\delta^{q,l}, \delta^{q,k}\right)\frac{\partial \left(I_T \bigotimes V^{q,k}\bigotimes V^{q,l}\right)}{\partial V^{q,l}_{i,i}} \Biggl\}-\frac{KT+\delta^l}{\left(V^{q,l}_{i,i}\right)^2} - \frac{\delta^{q,l}}{\theta^l},
\end{align*} and
\begin{align*}
    \frac{\partial^2 \text{ELBO}}{\left(\partial V^{q,l}_{i,j}\right)^2} = -\text{tr} \Biggl\{& \left[\mathbf{m}-\text{E}_q\left(\mathbf{m}_0\right)\right] \left[\mathbf{m}-\text{E}_q\left(\mathbf{m}_0\right)\right]' \frac{\left(I_T \bigotimes V^{q,k}\bigotimes V^{q,l}\right)'}{\partial V^{q,l}_{i,j}} \\
    &\text{E}^*\left(\boldsymbol{\phi},\delta^{q,l}, \delta^{q,k}\right)\frac{\partial \left(I_T \bigotimes V^{q,k}\bigotimes V^{q,l}\right)}{\partial V^{q,l}_{i,j}} \Biggl\} \\
    -\text{tr} \Biggl\{ &M \frac{\left(I_T \bigotimes V^{q,k}\bigotimes V^{q,l}\right)'}{\partial V^{q,l}_{i,j}} \text{E}^*\left(\boldsymbol{\phi},\delta^{q,l}, \delta^{q,k}\right)\frac{\partial \left(I_T \bigotimes V^{q,k}\bigotimes V^{q,l}\right)}{\partial V^{q,l}_{i,j}} \Biggl\}- \frac{\delta^{q,l}}{\theta^l}.
\end{align*}
Off-diagonal entries in the Hessian matrix are 
\begin{align*}
    \frac{\partial^2 \text{ELBO}}{\partial \delta^{q,l}\partial V^{q,l}_{i,j}} = -\text{tr} \Biggl\{ & \frac{\text{E}^*\left(\boldsymbol{\phi},\delta^{q,l}, \delta^{q,k}\right)}{\partial \delta^{q,l}} \left[\mathbf{m}-\text{E}_q\left(\mathbf{m}_0\right)\right] \left[\mathbf{m}-\text{E}_q\left(\mathbf{m}_0\right)\right]'  \\
    & \left(I_T \bigotimes V^{q,k}\bigotimes V^{q,l}\right)'\frac{\partial \left(I_T \bigotimes V^{q,k}\bigotimes V^{q,l}\right)}{\partial V^{q,l}_{i,j}}  \Biggl\} \\
    -\text{tr} \Biggl\{ & \frac{\text{E}^*\left(\boldsymbol{\phi},\delta^{q,l}, \delta^{q,k}\right)}{\partial \delta^{q,l}}M \left(I_T \bigotimes V^{q,k}\bigotimes V^{q,l}\right)'\frac{\partial \left(I_T \bigotimes V^{q,k}\bigotimes V^{q,l}\right)}{\partial V^{q,l}_{i,j}} \Biggl\}- \frac{V^{q,l}_{i,j}}{\theta^l},
\end{align*} with $i\leq j$, and 
\begin{align*}
    \frac{\partial^2 \text{ELBO}}{\partial V^{q,l}_{i_1,j_1}\partial V^{q,l}_{i_2,j_2}} = -\text{tr} \Biggl\{& \left[\mathbf{m}-\text{E}_q\left(\mathbf{m}_0\right)\right] \left[\mathbf{m}-\text{E}_q\left(\mathbf{m}_0\right)\right]' \frac{\partial \left(I_T \bigotimes V^{q,k}\bigotimes V^{q,l}\right)'}{\partial V^{q,l}_{i_2,j_2}} \\
    &\text{E}^*\left(\boldsymbol{\phi},\delta^{q,l}, \delta^{q,k}\right)\frac{\partial \left(I_T \bigotimes V^{q,k}\bigotimes V^{q,l}\right)}{\partial V^{q,l}_{i_1,j_1}} \Biggl\} \\
    -\text{tr} \Biggl\{& \left[\mathbf{m}-\text{E}_q\left(\mathbf{m}_0\right)\right] \left[\mathbf{m}-\text{E}_q\left(\mathbf{m}_0\right)\right]' \left(I_T \bigotimes V^{q,k}\bigotimes V^{q,l}\right)' \\
    &\text{E}^*\left(\boldsymbol{\phi},\delta^{q,l}, \delta^{q,k}\right)\frac{\partial^2 \left(I_T \bigotimes V^{q,k}\bigotimes V^{q,l}\right)}{\partial V^{q,l}_{i_1,j_1}\partial V^{q,l}_{i_2,j_2}} \Biggl\} \\
    -\text{tr}\Biggl\{ &M \frac{\partial \left(I_T \bigotimes V^{q,k}\bigotimes V^{q,l}\right)'}{\partial V^{q,l}_{i_2,j_2}}\text{E}^*\left(\boldsymbol{\phi},\delta^{q,l}, \delta^{q,k}\right)\frac{\partial \left(I_T \bigotimes V^{q,k}\bigotimes V^{q,l}\right)}{\partial V^{q,l}_{i_1,j_1}} \Biggl\} \\
    -\text{tr}\Biggl\{ &M \left(I_T \bigotimes V^{q,k}\bigotimes V^{q,l}\right)'\text{E}^*\left(\boldsymbol{\phi},\delta^{q,l}, \delta^{q,k}\right)\frac{\partial^2 \left(I_T \bigotimes V^{q,k}\bigotimes V^{q,l}\right)}{\partial V^{q,l}_{i_1,j_1}\partial V^{q,l}_{i_2,j_2}} \Biggl\}
\end{align*}
for $i_1\leq j_1, i_2\leq j_2, i_1\neq i_2,j_1\neq j_2$.

By definition $\delta^{q,l}>L-1$ and the diagonal entries are real positive numbers. The constrained optimization problem is again solved using projected method with backtracking line search algorithm. The Newton's method has the merit to converge faster than linear rate to the optimal solution when the current value is in the neighborhood of the solution, however, when the current value is far away from the optimal point, the performance is less satisfying. Therefore, we perform both Newton's method and Cauchy's method in each iteration. The Cauchy's method simply uses the gradient to construct the updating rule and it has superior convergence radius compared to Newton's method.

\subsection{optimizing with respect to \texorpdfstring{$\delta^{q,k}$}{deltak} and \texorpdfstring{$V^{q,k}$}{Vk}}
\label{subsection:deltak_Vk}
The procedure to update $\delta^{q,k}$ and $V^{q,k}$ is essentially identical to the one outlined in Section \ref{subsection:deltal_Vl}. The optimization is performed using a hybrid of Cauchy and Newton's methods while the constraints are satisfied with projected method with backtracking line search. Below, we formulate first order conditions and the Hessian matrix.

\begin{align*}
    \frac{\partial \text{ELBO}}{\partial \delta^{q,k}} = -\frac{1}{2}\text{tr} \Biggl\{&\left(I_T \bigotimes V^{q,k}\bigotimes V^{q,l}\right) \left[\mathbf{m}-\text{E}_q\left(\mathbf{m}_0\right)\right] \left[\mathbf{m}-\text{E}_q\left(\mathbf{m}_0\right)\right]' \\
    & \left(I_T \bigotimes V^{q,k}\bigotimes V^{q,l}\right)' \frac{\partial \text{E}^*\left(\boldsymbol{\phi},\delta^{q,l}, \delta^{q,k}\right)}{\partial \delta^{q,k}} \Biggl\} \\
    -\frac{1}{2}\text{tr}\Biggl\{& \left(I_T \bigotimes V^{q,k}\bigotimes V^{q,l}\right)M\left(I_T \bigotimes V^{q,k}\bigotimes V^{q,l}\right)'\frac{\partial \text{E}^*\left(\boldsymbol{\phi},\delta^{q,l}, \delta^{q,k}\right)}{\partial \delta^{q,k}} \Biggl\} \\
    +\frac{LT}{4}&\sum_{k=1}^K \psi_1\left(\frac{\delta^{q,k}-k+1}{2}\right) + \frac{\delta^k - \delta^{q,k}}{4}\sum_{k=1}^K\psi_1\left(\frac{\delta^{q,k}-k+1}{2}\right)-\frac{1}{2\theta^k}\text{tr}\left(D^{q,k}\right) + \frac{K}{2}
\end{align*}
for degree of freedom $\delta^{q,k}$,
\begin{align*}
    \frac{\partial \text{ELBO}}{\partial V^{q,k}_{i,i}} = -\text{tr} \Biggl\{ & \left[\mathbf{m}-\text{E}_q\left(\mathbf{m}_0\right)\right] \left[\mathbf{m}-\text{E}_q\left(\mathbf{m}_0\right)\right]' \left(I_T \bigotimes V^{q,k}\bigotimes V^{q,k}\right)' \\
&\text{E}^*\left(\boldsymbol{\phi},\delta^{q,l}, \delta^{q,k}\right)\frac{\partial \left(I_T \bigotimes V^{q,k}\bigotimes V^{q,l}\right)}{\partial V^{q,l}_{i,i}} \Biggl\} \\
    -\text{tr} \Biggl\{ &M \left(I_T \bigotimes V^{q,k}\bigotimes V^{q,l}\right)'\text{E}^*\left(\boldsymbol{\phi},\delta^{q,l}, \delta^{q,k}\right)\frac{\partial \left(I_T \bigotimes V^{q,k}\bigotimes V^{q,l}\right)}{\partial V^{q,k}_{i,i}} \Biggl\} 
    + \frac{LT+\delta^k}{V^{q,k}_{i,i}} - \frac{\delta^{q,k}V^{q,k}_{i,i}}{\theta^k}
\end{align*}
for diagonal elements $V^{q,k}_{i,i}$, and
\begin{align*}
    \frac{\partial \text{ELBO}}{\partial V^{q,k}_{i,j}} = -\text{tr} \Biggl\{& \left[\mathbf{m}-\text{E}_q\left(\mathbf{m}_0\right)\right] \left[\mathbf{m}-\text{E}_q\left(\mathbf{m}_0\right)\right]' \left(I_T \bigotimes V^{q,k}\bigotimes V^{q,l}\right)' \\
    &\text{E}^*\left(\boldsymbol{\phi},\delta^{q,l}, \delta^{q,k}\right)\frac{\partial \left(I_T \bigotimes V^{q,k}\bigotimes V^{q,l}\right)}{\partial V^{q,k}_{i,j}} \Biggl\} \\
    -\text{tr}\Biggl\{ &M \left(I_T \bigotimes V^{q,k}\bigotimes V^{q,l}\right)'\text{E}^*\left(\boldsymbol{\phi},\delta^{q,l}, \delta^{q,k}\right)\frac{\partial \left(I_T \bigotimes V^{q,k}\bigotimes V^{q,l}\right)}{\partial V^{q,k}_{i,j}} \Biggl\}-\frac{\delta^{q,k}V^{q,k}_{i,j}}{\theta^k},
\end{align*}
for off-diagonal elements $V^{q,k}_{i,j}$ with $i<j$.
In the Hessian matrix, the second derivatives are
\begin{align*}
    \frac{\partial^2 \text{ELBO}}{\left(\partial \delta^{q,k}\right)^2} = &\frac{LT}{8}\sum_{k=1}^K \psi_2\left(\frac{\delta^{q,k}-k+1}{2}\right) - \frac{1}{4}\sum_{k=1}^K \psi_1\left(\frac{\delta^{q,k}-k+1}{2}\right)+\frac{\delta^k - \delta^{q,k}}{8}\sum_{k=1}^K \psi_2\left(\frac{\delta^{q,k}-k+1}{2}\right),
\end{align*}
\begin{align*}
    \frac{\partial^2 \text{ELBO}}{\left(\partial V^{q,k}_{i,i}\right)^2} = -\text{tr} \Biggl\{& \left[\mathbf{m}-\text{E}_q\left(\mathbf{m}_0\right)\right] \left[\mathbf{m}-\text{E}_q\left(\mathbf{m}_0\right)\right]' \frac{\left(I_T \bigotimes V^{q,k}\bigotimes V^{q,l}\right)'}{\partial V^{q,k}_{i,i}} \\
    &\text{E}^*\left(\boldsymbol{\phi},\delta^{q,l}, \delta^{q,k}\right)\frac{\partial \left(I_T \bigotimes V^{q,k}\bigotimes V^{q,l}\right)}{\partial V^{q,k}_{i,i}} \Biggl\} \\
    -\text{tr} \Biggl\{ &M \frac{\left(I_T \bigotimes V^{q,k}\bigotimes V^{q,l}\right)'}{\partial V^{q,k}_{i,i}} \text{E}^*\left(\boldsymbol{\phi},\delta^{q,l}, \delta^{q,k}\right)\frac{\partial \left(I_T \bigotimes V^{q,k}\bigotimes V^{q,l}\right)}{\partial V^{q,k}_{i,i}} \Biggl\}-\frac{LT+\delta^k}{\left(V^{q,k}_{i,i}\right)^2} - \frac{\delta^{q,k}}{\theta^k},
\end{align*} and
\begin{align*}
    \frac{\partial^2 \text{ELBO}}{\left(\partial V^{q,k}_{i,j}\right)^2} = -\text{tr} \Biggl\{& \left[\mathbf{m}-\text{E}_q\left(\mathbf{m}_0\right)\right] \left[\mathbf{m}-\text{E}_q\left(\mathbf{m}_0\right)\right]' \frac{\left(I_T \bigotimes V^{q,k}\bigotimes V^{q,l}\right)'}{\partial V^{q,k}_{i,j}} \\
    &\text{E}^*\left(\boldsymbol{\phi},\delta^{q,l}, \delta^{q,k}\right)\frac{\partial \left(I_T \bigotimes V^{q,k}\bigotimes V^{q,l}\right)}{\partial V^{q,k}_{i,j}} \Biggl\} \\
    -\text{tr} \Biggl\{ &M \frac{\left(I_T \bigotimes V^{q,k}\bigotimes V^{q,l}\right)'}{\partial V^{q,k}_{i,j}} \text{E}^*\left(\boldsymbol{\phi},\delta^{q,l}, \delta^{q,k}\right)\frac{\partial \left(I_T \bigotimes V^{q,k}\bigotimes V^{q,l}\right)}{\partial V^{q,k}_{i,j}} \Biggl\}- \frac{\delta^{q,k}}{\theta^k}.
\end{align*}
Off-diagonal entries in the Hessian matrix are 
\begin{align*}
    \frac{\partial^2 \text{ELBO}}{\partial \delta^{q,k}\partial V^{q,k}_{i,j}} = -\text{tr} \Biggl\{ & \frac{\text{E}^*\left(\boldsymbol{\phi},\delta^{q,l}, \delta^{q,k}\right)}{\partial \delta^{q,k}} \left[\mathbf{m}-\text{E}_q\left(\mathbf{m}_0\right)\right] \left[\mathbf{m}-\text{E}_q\left(\mathbf{m}_0\right)\right]'  \\
    & \left(I_T \bigotimes V^{q,k}\bigotimes V^{q,l}\right)'\frac{\partial \left(I_T \bigotimes V^{q,k}\bigotimes V^{q,l}\right)}{\partial V^{q,k}_{i,j}}  \Biggl\} \\
    -\text{tr} \Biggl\{ & \frac{\text{E}^*\left(\boldsymbol{\phi},\delta^{q,l}, \delta^{q,k}\right)}{\partial \delta^{q,k}}M \left(I_T \bigotimes V^{q,k}\bigotimes V^{q,l}\right)'\frac{\partial \left(I_T \bigotimes V^{q,k}\bigotimes V^{q,l}\right)}{\partial V^{q,k}_{i,j}} \Biggl\}- \frac{V^{q,k}_{i,j}}{\theta^k},
\end{align*} with $i\leq j$, and 
\begin{align*}
    \frac{\partial^2 \text{ELBO}}{\partial V^{q,k}_{i_1,j_1}\partial V^{q,k}_{i_2,j_2}} = -\text{tr} \Biggl\{& \left[\mathbf{m}-\text{E}_q\left(\mathbf{m}_0\right)\right] \left[\mathbf{m}-\text{E}_q\left(\mathbf{m}_0\right)\right]' \frac{\partial \left(I_T \bigotimes V^{q,k}\bigotimes V^{q,l}\right)'}{\partial V^{q,k}_{i_2,j_2}} \\
    &\text{E}^*\left(\boldsymbol{\phi},\delta^{q,l}, \delta^{q,k}\right)\frac{\partial \left(I_T \bigotimes V^{q,k}\bigotimes V^{q,l}\right)}{\partial V^{q,k}_{i_1,j_1}} \Biggl\} \\
    -\text{tr} \Biggl\{& \left[\mathbf{m}-\text{E}_q\left(\mathbf{m}_0\right)\right] \left[\mathbf{m}-\text{E}_q\left(\mathbf{m}_0\right)\right]' \left(I_T \bigotimes V^{q,k}\bigotimes V^{q,l}\right)' \\
    &\text{E}^*\left(\boldsymbol{\phi},\delta^{q,l}, \delta^{q,k}\right)\frac{\partial^2 \left(I_T \bigotimes V^{q,k}\bigotimes V^{q,l}\right)}{\partial V^{q,k}_{i_1,j_1}\partial V^{q,k}_{i_2,j_2}} \Biggl\} \\
    -\text{tr}\Biggl\{ &M \frac{\partial \left(I_T \bigotimes V^{q,k}\bigotimes V^{q,l}\right)'}{\partial V^{q,k}_{i_2,j_2}}\text{E}^*\left(\boldsymbol{\phi},\delta^{q,l}, \delta^{q,k}\right)\frac{\partial \left(I_T \bigotimes V^{q,k}\bigotimes V^{q,l}\right)}{\partial V^{q,k}_{i_1,j_1}} \Biggl\} \\
    -\text{tr}\Biggl\{ &M \left(I_T \bigotimes V^{q,k}\bigotimes V^{q,l}\right)'\text{E}^*\left(\boldsymbol{\phi},\delta^{q,l}, \delta^{q,k}\right)\frac{\partial^2 \left(I_T \bigotimes V^{q,k}\bigotimes V^{q,l}\right)}{\partial V^{q,k}_{i_1,j_1}\partial V^{q,k}_{i_2,j_2}} \Biggl\}
\end{align*}
for $i_1\leq j_1, i_2\leq j_2, i_1\neq i_2,j_1\neq j_2$.

\section{Italian mortality analysis}
\label{real data application}

To analyze non-linear relationship between mortality rates and covariates as well as evolving mortality trends of COVID-19 and other causes before and during the pandemic, we applied our method to the provisional monthly mortality data from Italy. The data, spanning from January 2015 to December 2020 (a total of $T=72$ months), includes declarations of 18 different causes of death reported by physicians for all deaths in Italy. These causes of death are listed in Table \ref{tb:death_causes}. Additionally, the death counts are compiled across 420 categories, derived from combining 10 age groups, 2 genders, and $L=21$ Italian regions, which together with mortality causes, gives 7,560 samples. In total, there are 544,320 observations. A detailed description of this mortality data is available at https://www.istat.it/it/archivio/240401. In our analysis, we set $K=17$ and exclude cause No.14 complications of pregnancy, childbirth and the puerperium. The reasons for discarding this category are the following. First, death counts are almost all 0's (30,165 out of 30,240); second, estimation with this category is unstable and the convergence rate of the CAVI algorithm drastically slows down; third, excluding one category does not affect the inference on other causes of death. Consequently, we examine $N=7,140$ time series $Y_{n,t}$ for $n=1,\dots,N, t=1,\dots,T$. In terms of non-Gaussian state space models, the dimension of observations is 7,140 while the dimension of latent states is 357. Due to high-dimensional structure of the data and the model, conventional sampling algorithms are not applicable, which motivates to approximate the posterior distribution in a variational approach.

\begin{table}[ht]
\centering
\caption{Causes of death according to the ICD-10.}
\label{tb:death_causes}
\begin{tabular}{ll}
\hline
1.&COVID-19 \\ \hline
2.&Some infectious and parasitic diseases \\ \hline
3.&Tumors \\ \hline
4.&Diseases of the blood and hematopoietic organs and \\
&some disorders of the immune system \\ \hline
5.&Endocrine, nutritional and metabolic diseases \\ \hline
6.&Psychic and behavioral disorders \\ \hline
7.&Diseases of the nervous system and sense organs \\ \hline
8.&Diseases of the circulatory system \\ \hline
9.&Diseases of the respiratory system \\ \hline
10.&Diseases of the digestive system \\ \hline
11.&Diseases of the skin and subcutaneous tissue \\ \hline
12.&Diseases of the musculoskeletal system and connective tissue \\ \hline
13.&Diseases of the genitourinary system \\ \hline
14.&Complications of pregnancy, childbirth and the puerperium \\ \hline
15.&Some morbid conditions that originate in the perinatal period \\ \hline
16.&Congenital malformations and chromosomal anomalies \\ \hline
17.&Symptoms, signs, abnormal results and ill-defined causes \\ \hline
18.&External causes of trauma and poisoning \\ \hline
\end{tabular}
\end{table}

\begin{figure}
    \centering
    \includegraphics[width=\linewidth]{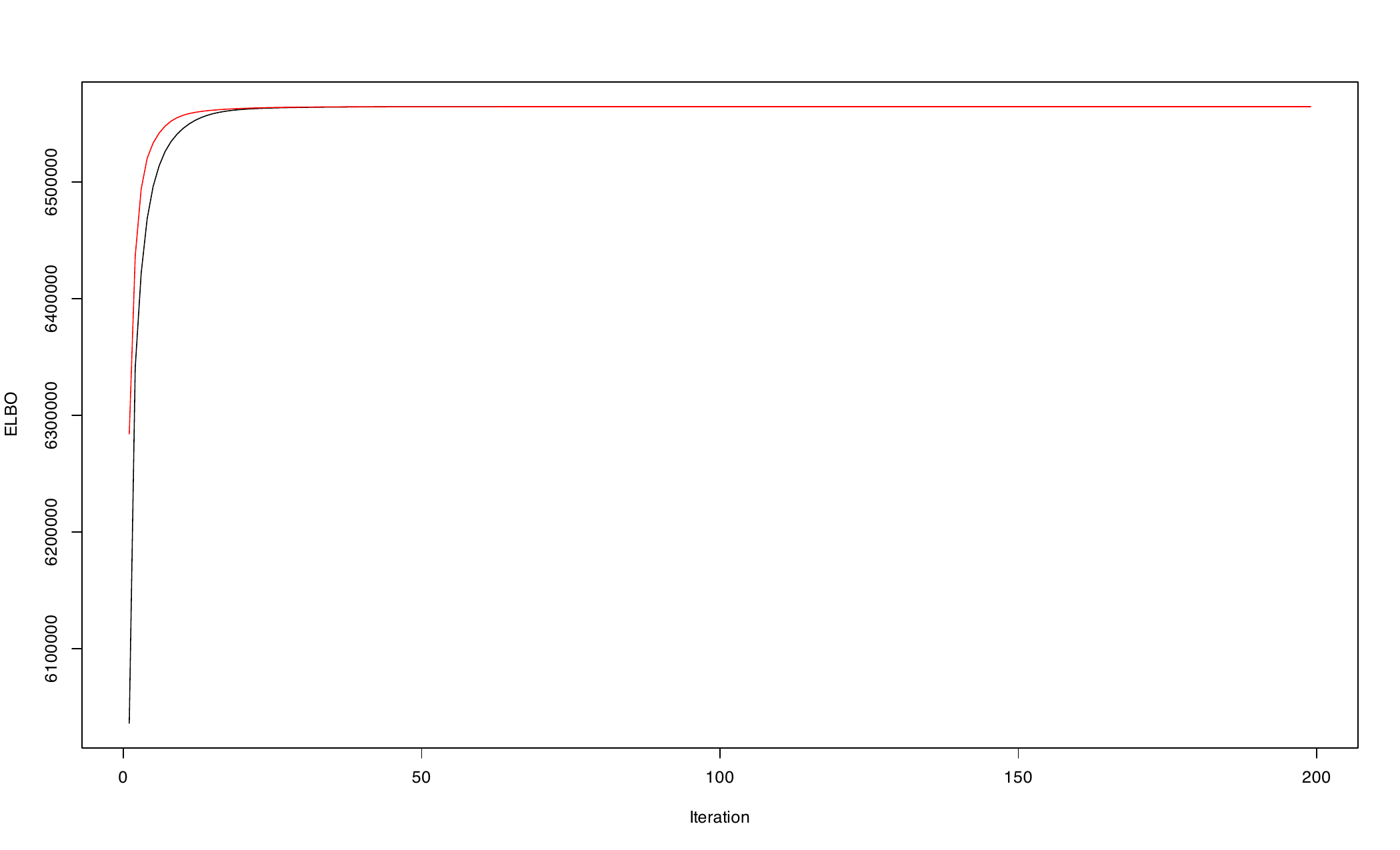}
    \caption{ELBO trajectories of the proposed CAVI algorithm. The first two ELBO values of each trajectory are dropped as they are placed distant from y-axis limits. The initalization gives ELBO values of $-2.0181\times 10^{10}$ and $-5.0756\times 10^{11}$ respectively.}
    \label{fig:elbo}
\end{figure}

We have briefly described covariates included in the model in Section \ref{Model} when formulating the GAM component. Now we elaborate more on one key variable, the Italian Stringency Index (ISI) developed by \citet{conteduca2022new}, analogous to the Oxford Stringency Index (OSI) by \citet{hale2021global}, which measures non-pharmaceutical interventions implemented in Italy to combat COVID-19. The index tracks both national and regional intervention intensity and it is ideal for studying mortality rate in regional level. We focus on non-linear interactions $f^{k,r}(k, r)$ between ISI and different causes of death, acknowledging the potential varying impacts of the pandemic on other mortality causes as highlighted in existing studies. As for the offsets $\epsilon_{n,t}$, we consider days per month, monthly aggregated COVID-19 cases by region, and regional population figures for other causes of death. Notably, for external causes like trauma and poisoning, we incorporate a mobility index from the Google COVID-19 Community Mobility Reports (Google LLC "Google COVID-19 Community Mobility Reports". https://www.google.com/covid19/mobility/), to model mortality rate changes per mobility unit. Parameters in the prior distributions are 
$\alpha_\lambda = 1, \beta_\lambda = 1000, \alpha_\phi = 10, \beta_\phi = 10,
\delta^k = K = 17, \delta^l = L, \theta^k = K-2 = 15, \theta^l = L-2 = 19, \sigma^2_\beta = 10,
\sigma^2_\mu = 1$. To check whether the proposed CAVI method achieves convergence, we start the algorithm from different initialization. The trajectories of ELBO are shown in Figure \ref{fig:elbo}; an optimal ELBO value of 6,564,738 is reached within reasonable steps despite that the initial values are randomly generated.

\begin{figure}[!]
\captionsetup[subfigure]{font=footnotesize}
     \centering
    \begin{subfigure}[t]{0.32\textwidth}
         \centering
         \includegraphics[width=\textwidth]{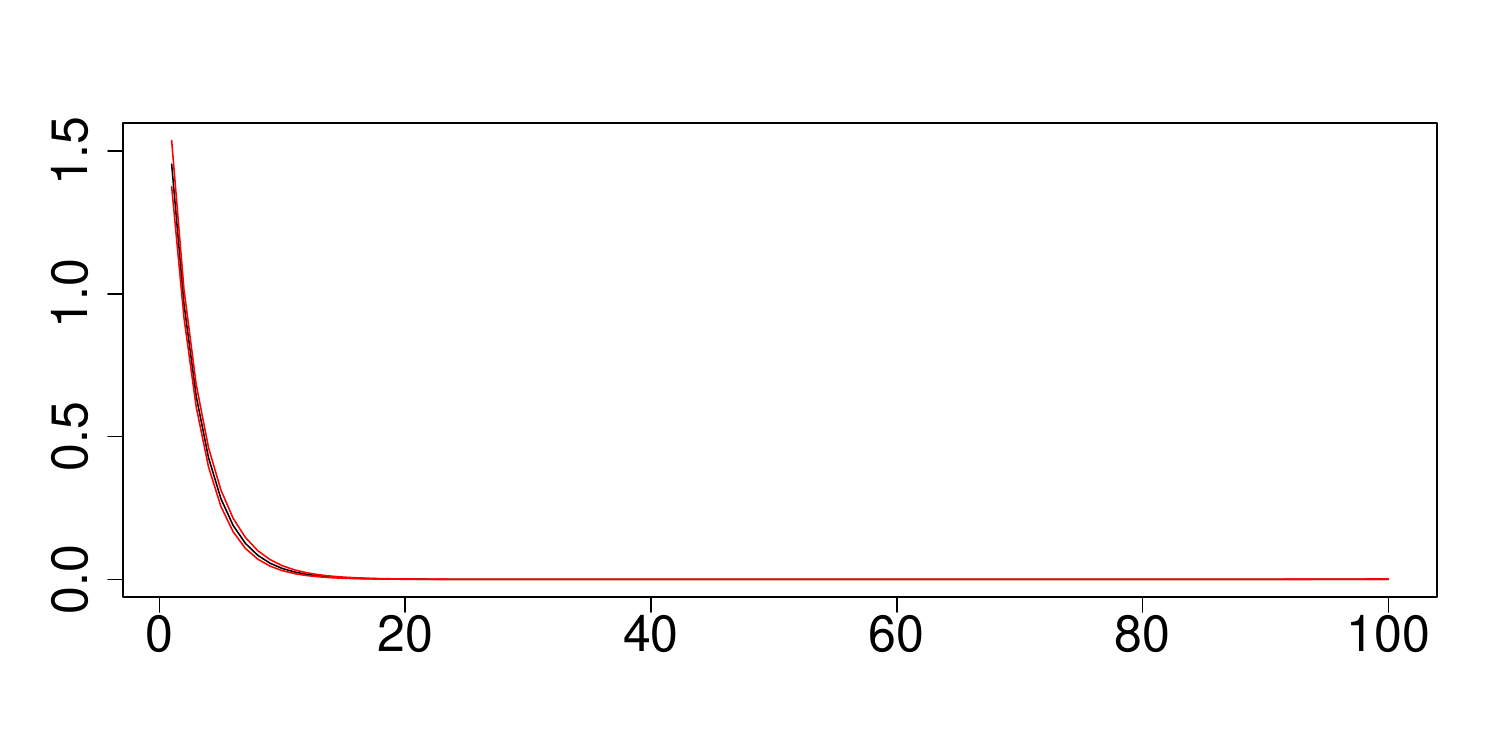}
         \caption{COVID-19}
         \label{subfig:isi_cause1}
     \end{subfigure}
    \begin{subfigure}[t]{0.32\textwidth}
         \centering
         \includegraphics[width=\textwidth]{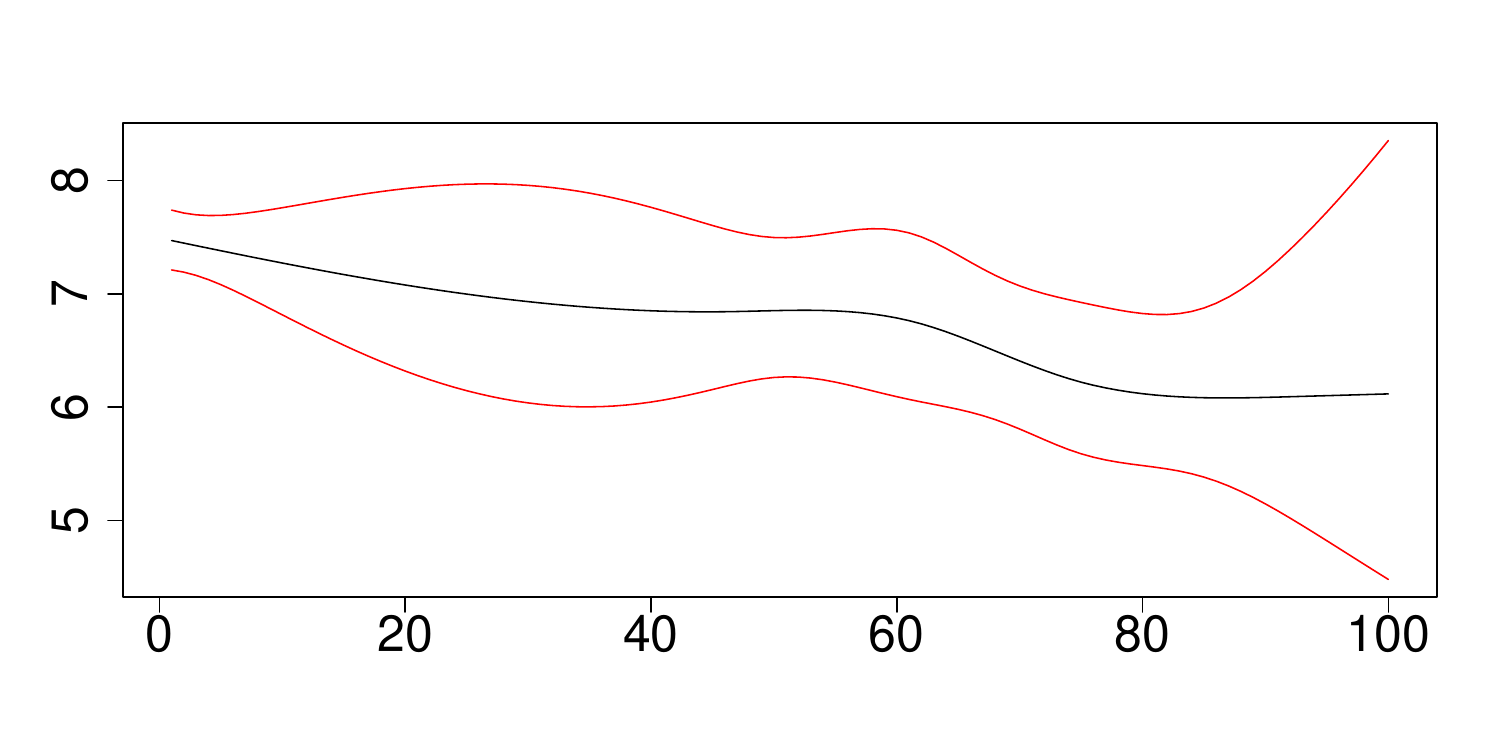}
         \caption{Some infectious and parasitic diseases}
         \label{subfig:isi_cause2}
     \end{subfigure}
    \begin{subfigure}[t]{0.32\textwidth}
         \centering
         \includegraphics[width=\textwidth]{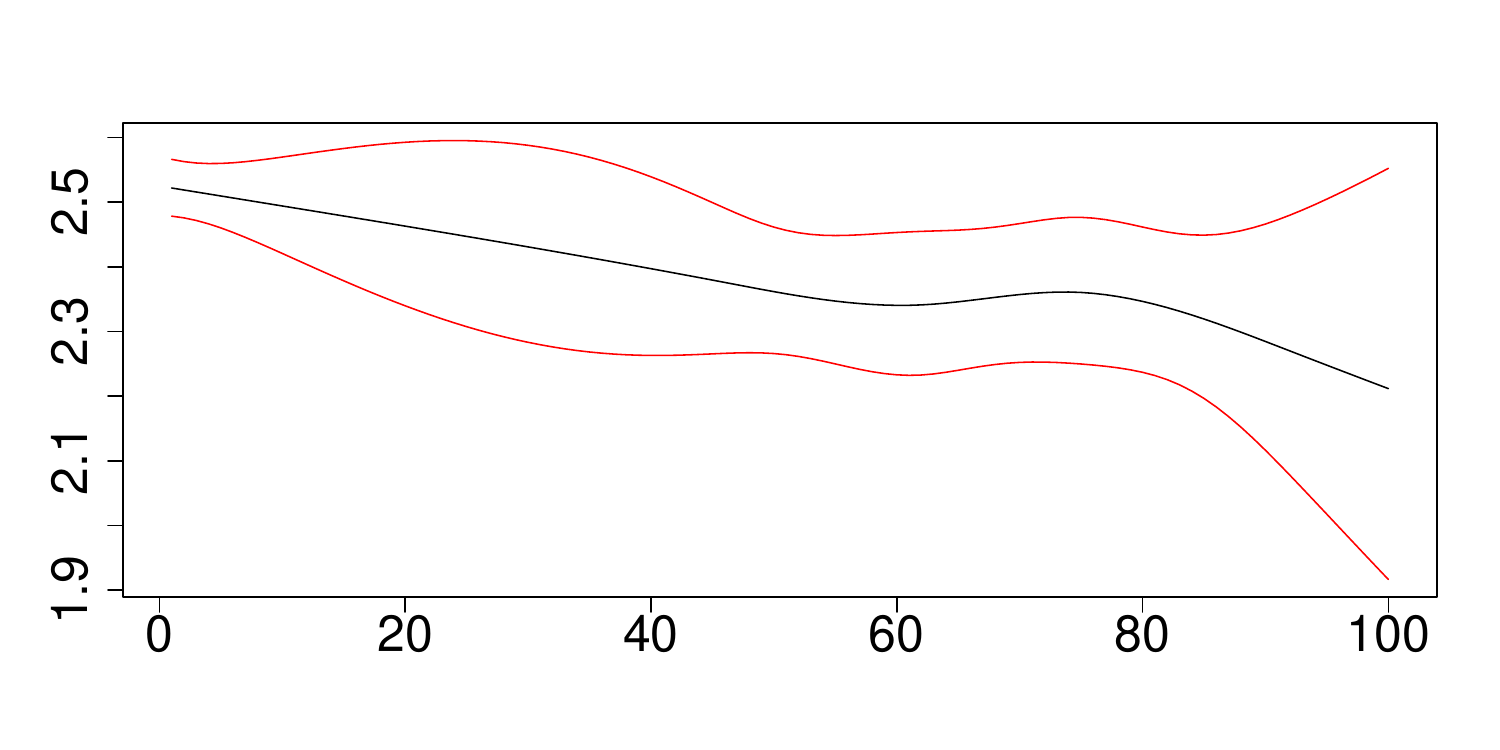}
         \caption{Tumors}
         \label{subfig:isi_cause3}
     \end{subfigure}
    
    \begin{subfigure}[t]{0.32\textwidth}
         \centering
         \includegraphics[width=\textwidth]{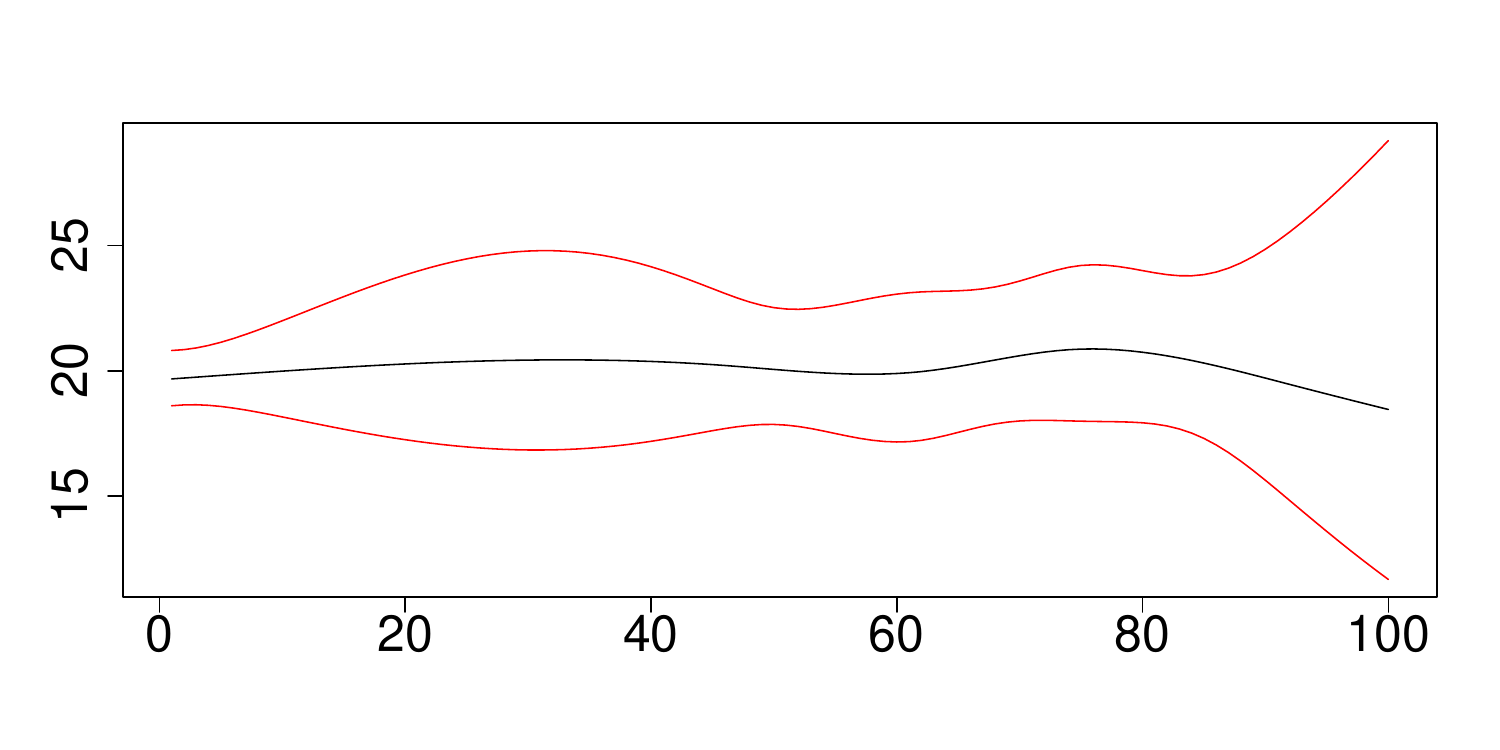}
         \caption{Diseases of the blood and hematopoietic organs and some disorders of the immune system}
         \label{subfig:isi_cause4}
     \end{subfigure}
    \begin{subfigure}[t]{0.32\textwidth}
         \centering
         \includegraphics[width=\textwidth]{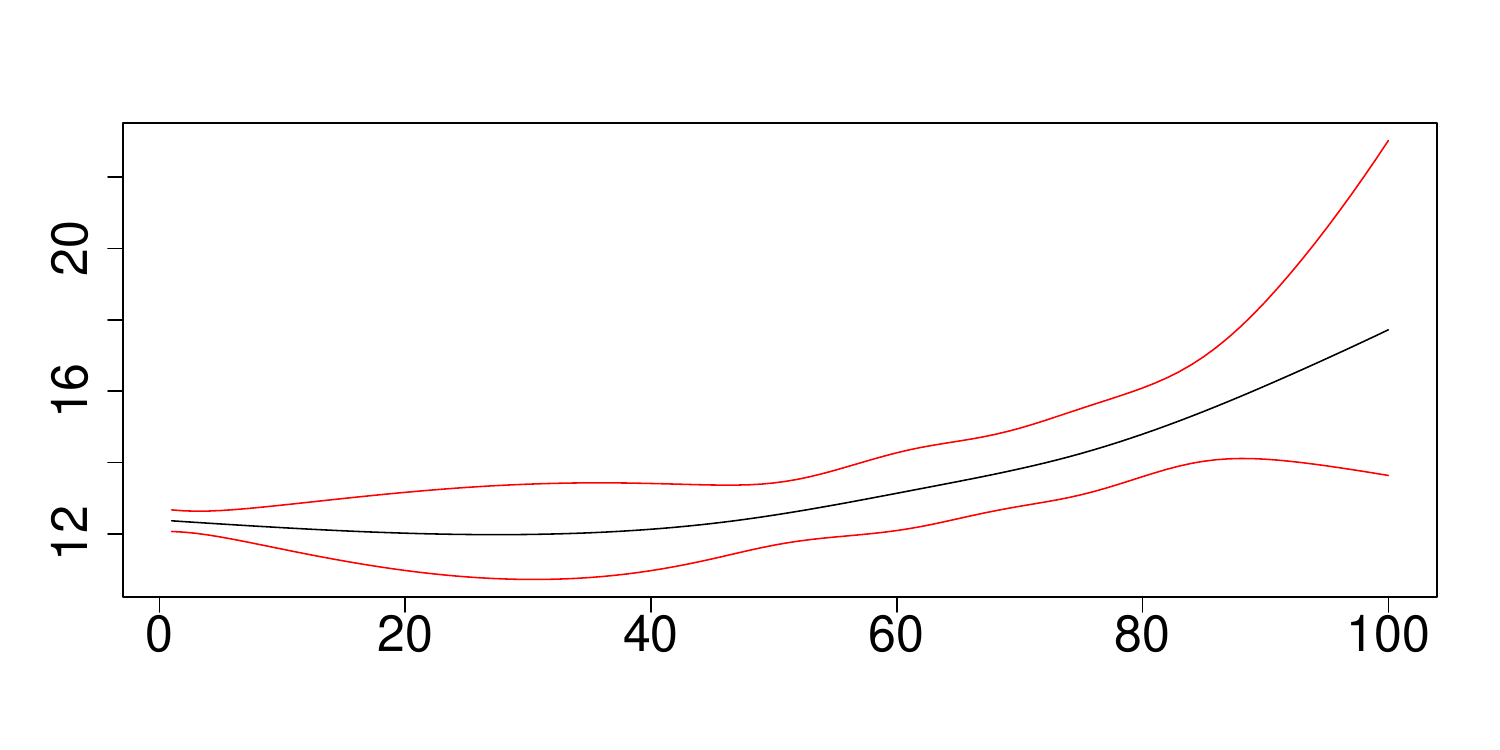}
         \caption{Endocrine, nutritional and metabolic diseases}
         \label{subfig:isi_cause5}
     \end{subfigure}
    \begin{subfigure}[t]{0.32\textwidth}
         \centering
         \includegraphics[width=\textwidth]{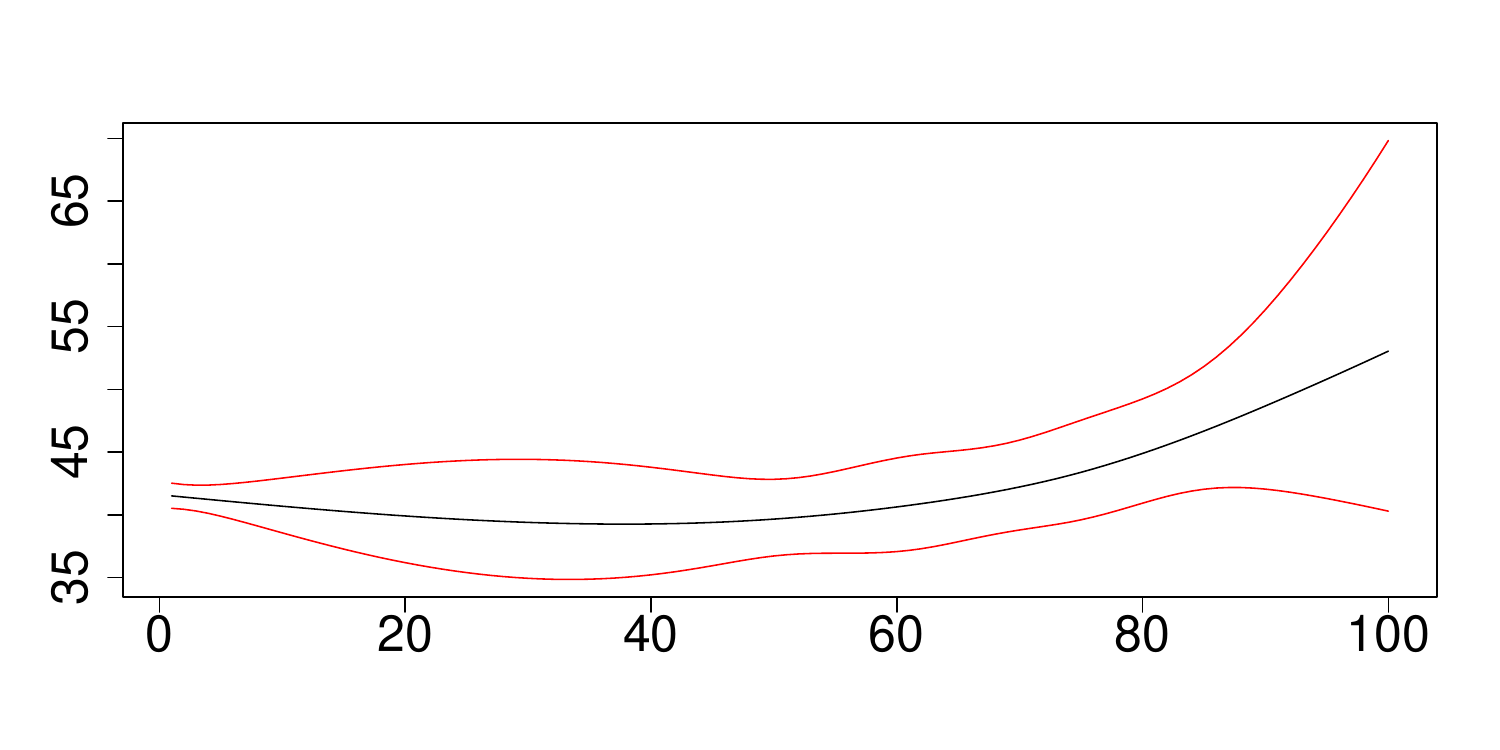}
         \caption{Psychic and behavioral disorders}
         \label{subfig:isi_cause6}
     \end{subfigure}

    \begin{subfigure}[t]{0.32\textwidth}
         \centering
         \includegraphics[width=\textwidth]{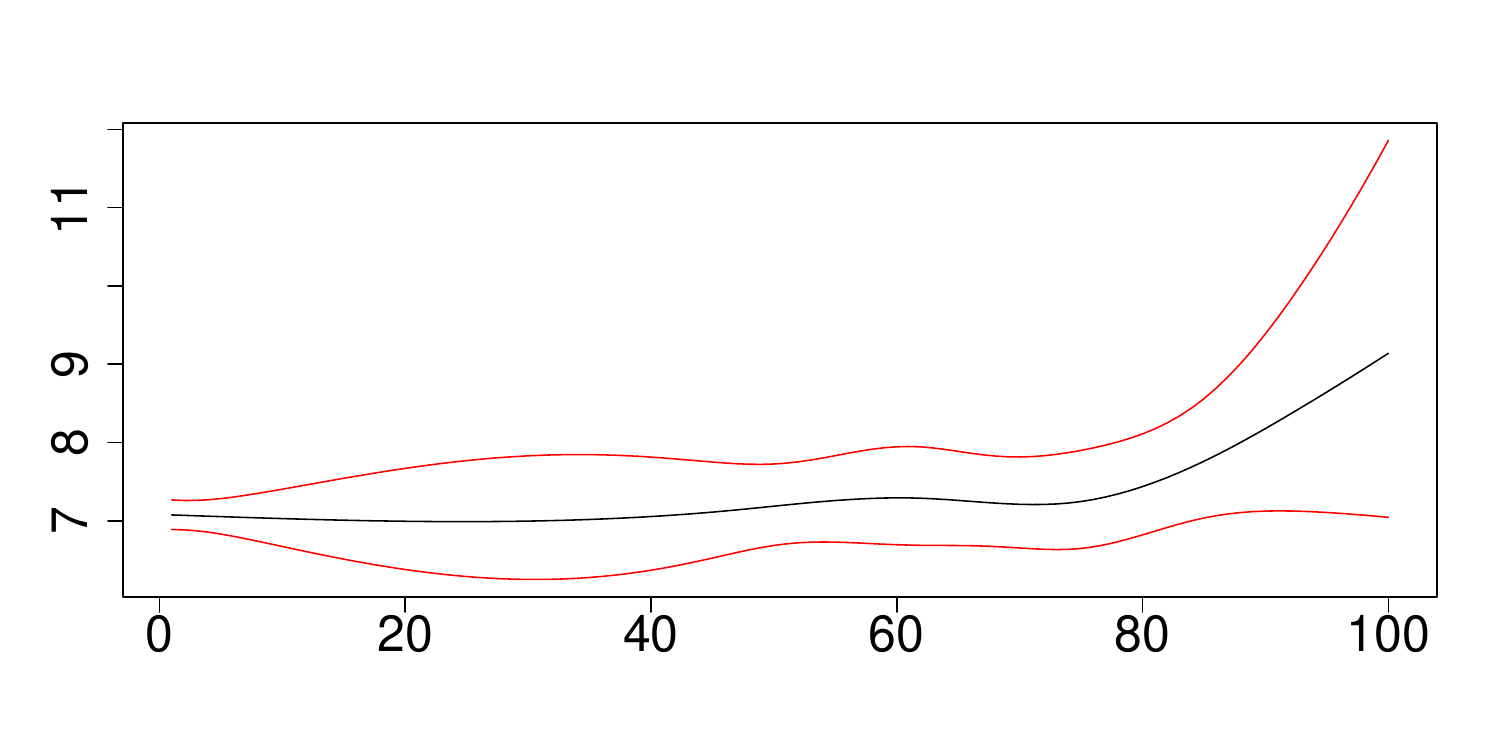}
         \caption{Diseases of the nervous system and sense organs}
         \label{subfig:isi_cause7}
     \end{subfigure}
    \begin{subfigure}[t]{0.32\textwidth}
         \centering
         \includegraphics[width=\textwidth]{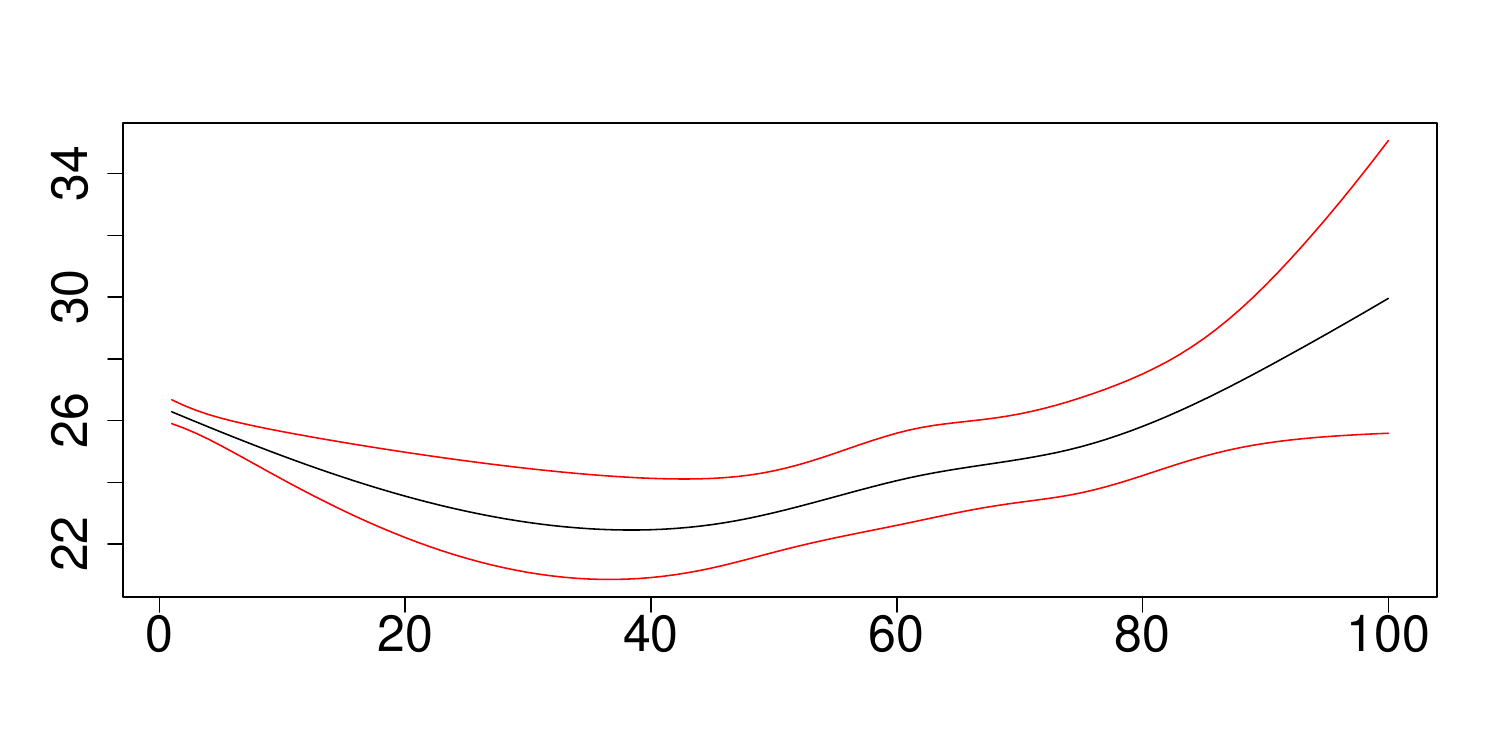}
         \caption{Diseases of the circulatory system}
         \label{subfig:isi_cause8}
     \end{subfigure}
    \begin{subfigure}[t]{0.32\textwidth}
         \centering
         \includegraphics[width=\textwidth]{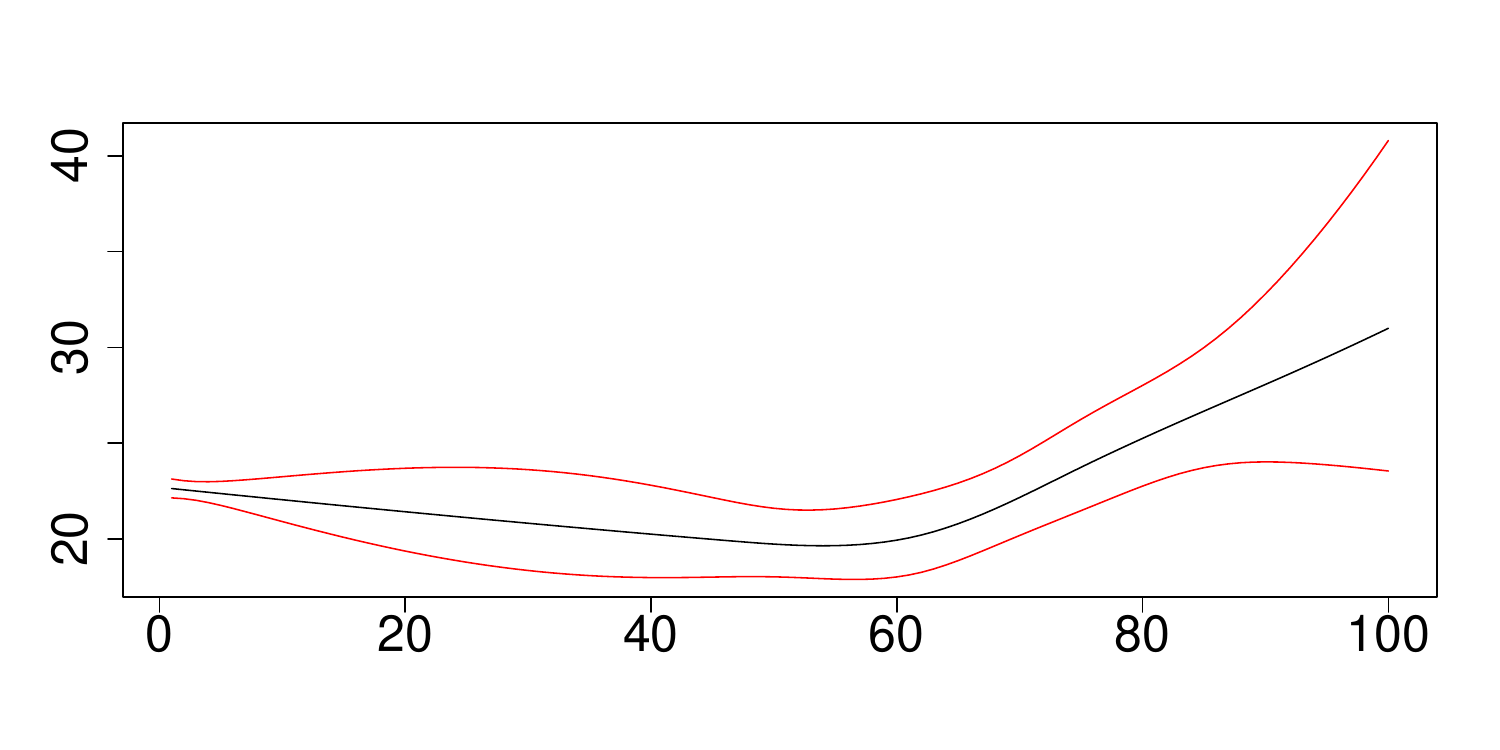}
         \caption{Diseases of the respiratory system}
         \label{subfig:isi_cause9}
     \end{subfigure}

         \begin{subfigure}[t]{0.32\textwidth}
         \centering
         \includegraphics[width=\textwidth]{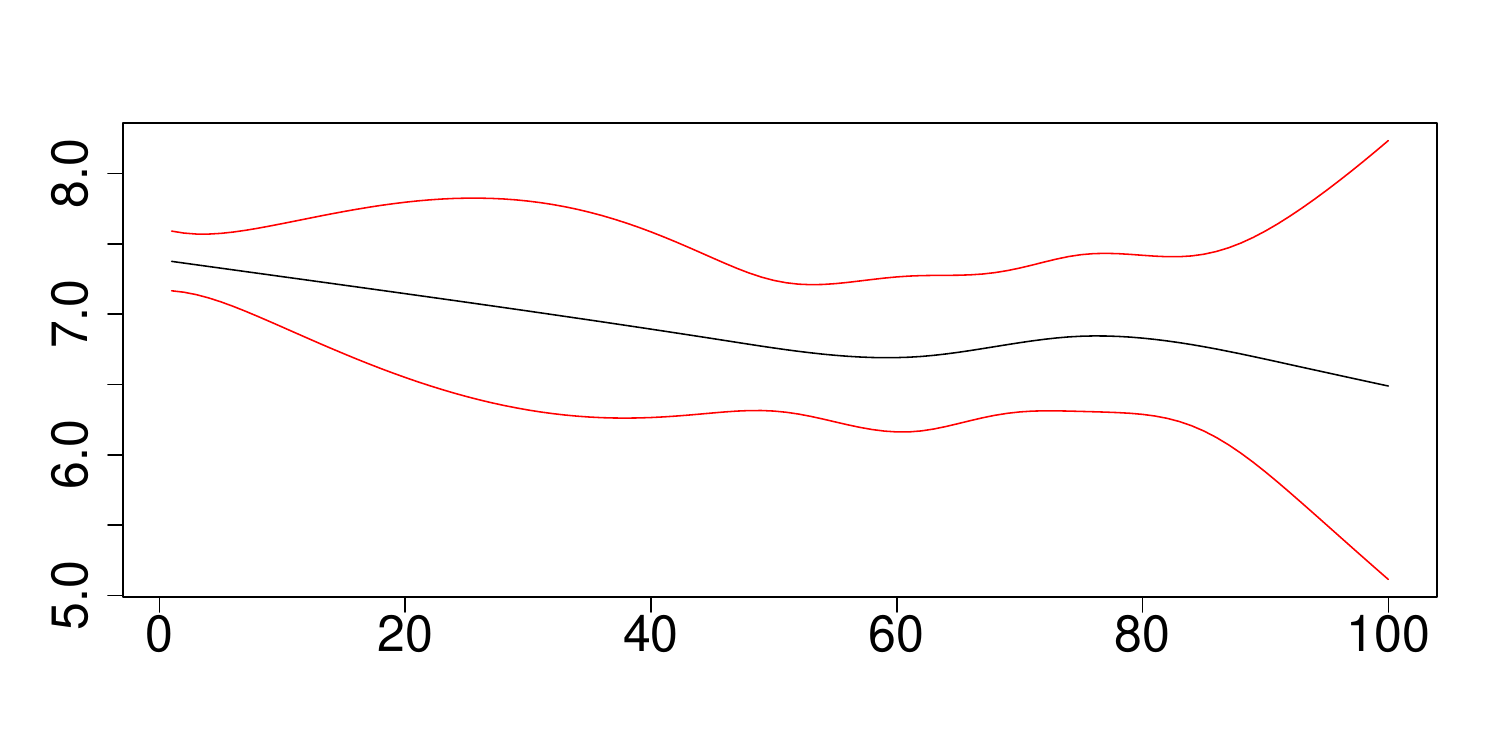}
         \caption{Diseases of the digestive system}
         \label{subfig:isi_cause10}
     \end{subfigure}
    \begin{subfigure}[t]{0.32\textwidth}
         \centering
         \includegraphics[width=\textwidth]{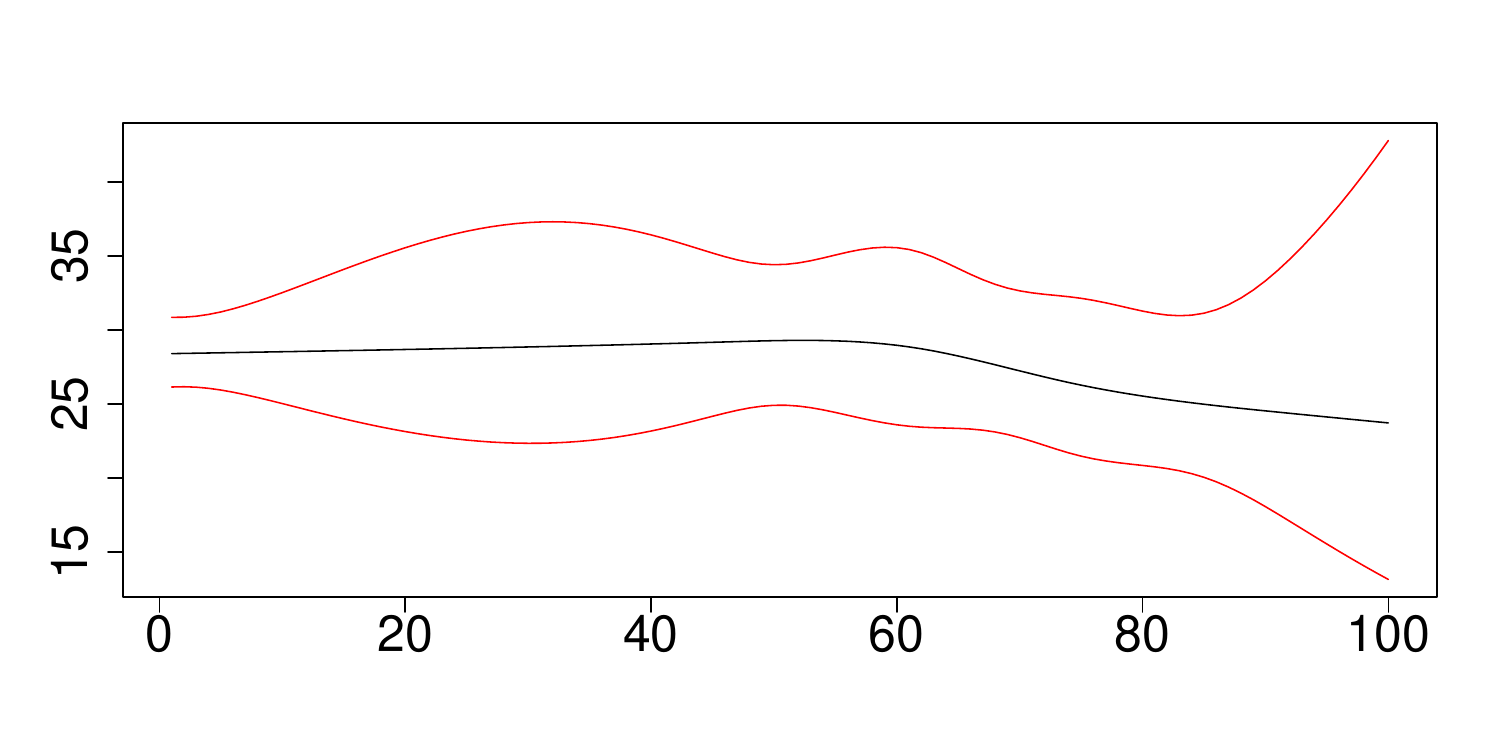}
         \caption{Diseases of the skin and subcutaneous tissue}
         \label{subfig:isi_cause11}
     \end{subfigure}
    \begin{subfigure}[t]{0.32\textwidth}
         \centering
         \includegraphics[width=\textwidth]{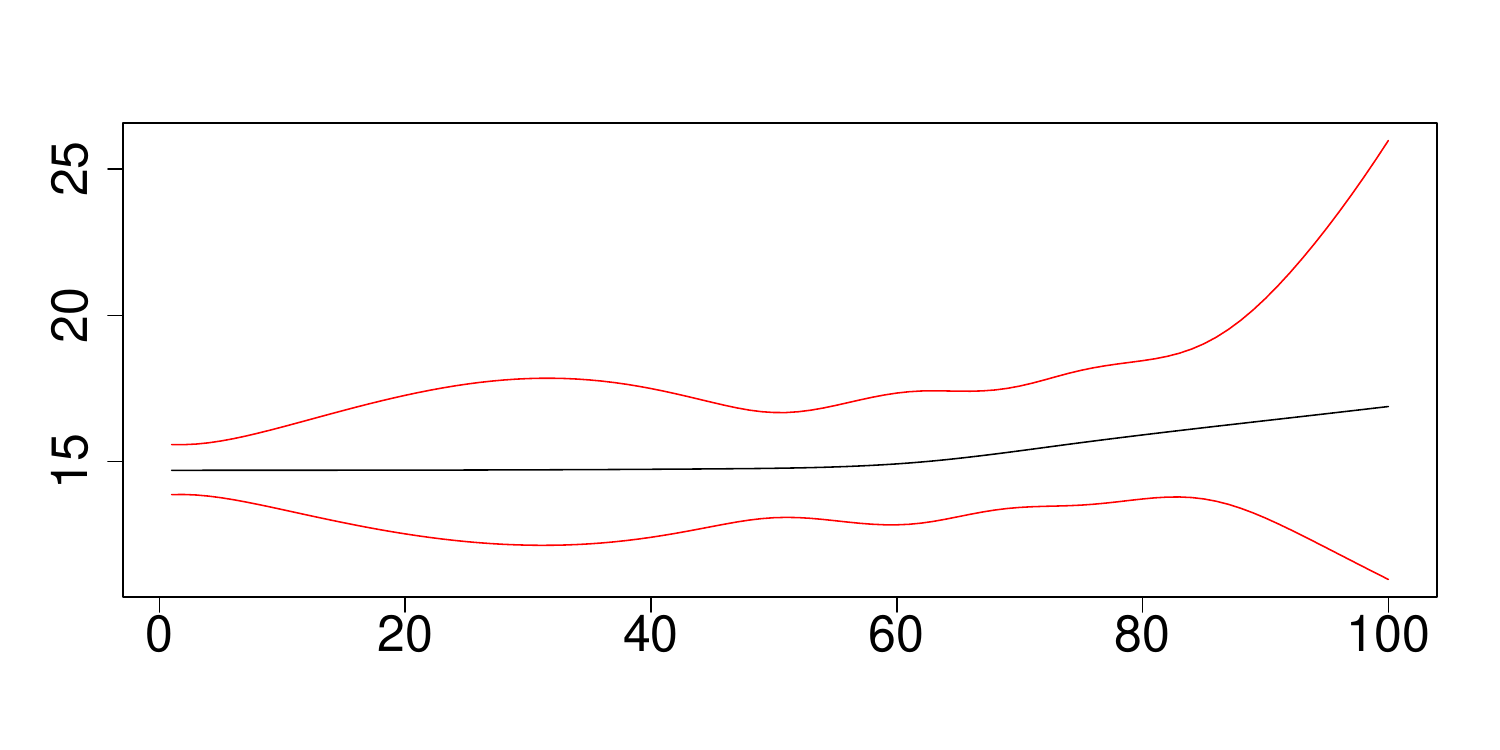}
         \caption{Diseases of the musculoskeletal system and connective tissue}
         \label{subfig:isi_cause12}
     \end{subfigure}

         \begin{subfigure}[t]{0.32\textwidth}
         \centering
         \includegraphics[width=\textwidth]{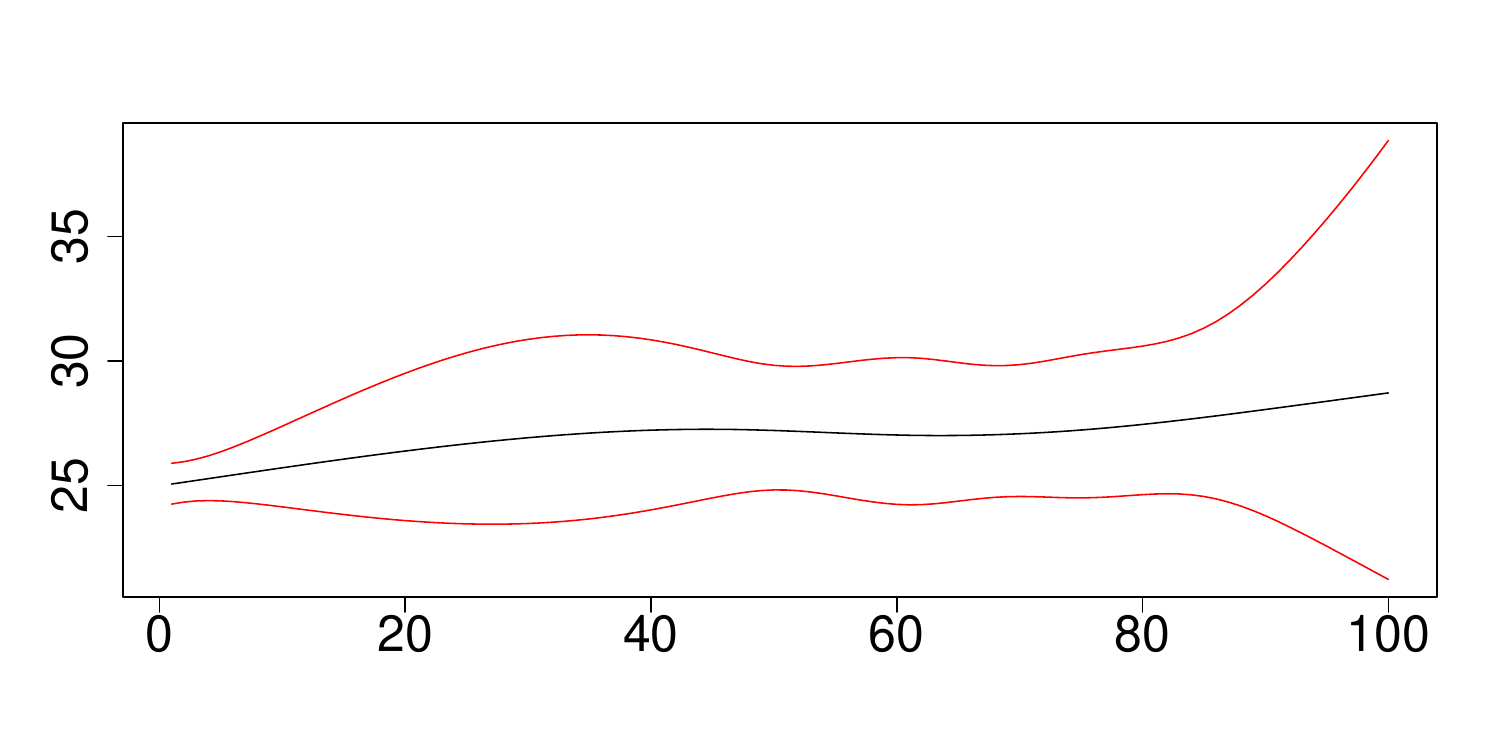}
         \caption{Diseases of the genitourinary system}
         \label{subfig:isi_cause13}
     \end{subfigure}
    \begin{subfigure}[t]{0.32\textwidth}
         \centering
         \includegraphics[width=\textwidth]{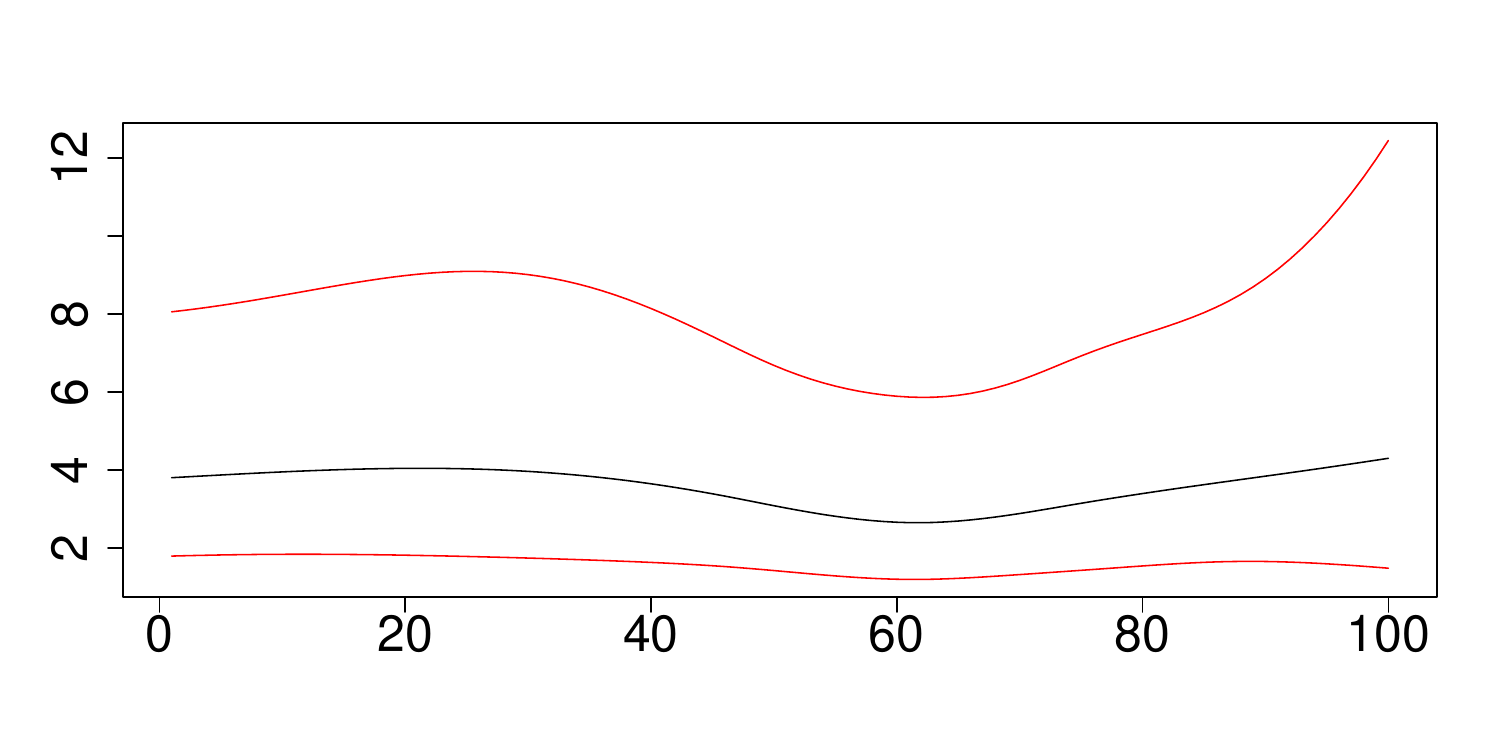}
         \caption{Some morbid conditions that originate in the perinatal period}
         \label{subfig:isi_cause15}
     \end{subfigure}
\begin{subfigure}[t]{0.32\textwidth}
         \centering
         \includegraphics[width=\textwidth]{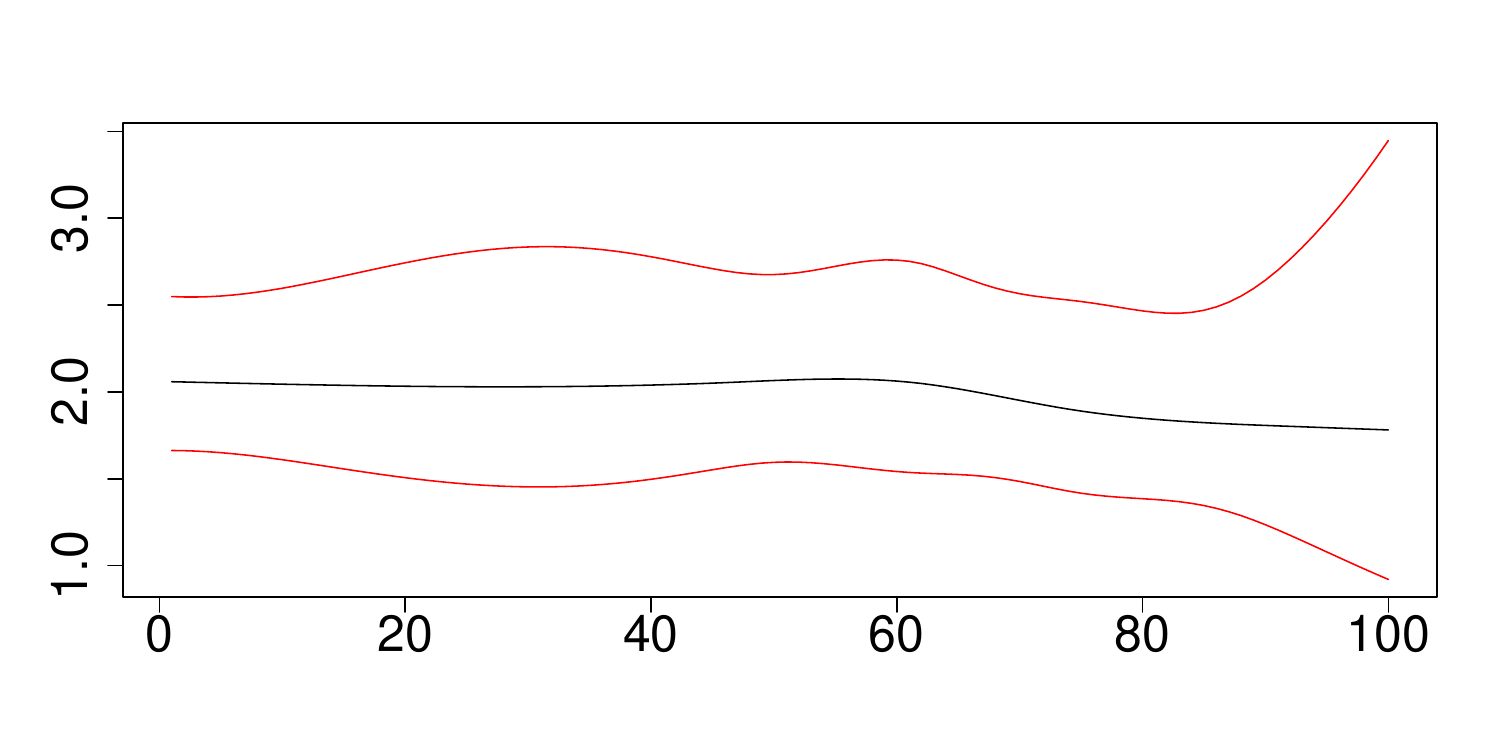}
         \caption{Congenital malformations and chromosomal anomalies}
         \label{subfig:isi_cause16}
     \end{subfigure}
     
    \begin{subfigure}[t]{0.32\textwidth}
         \centering
         \includegraphics[width=\textwidth]{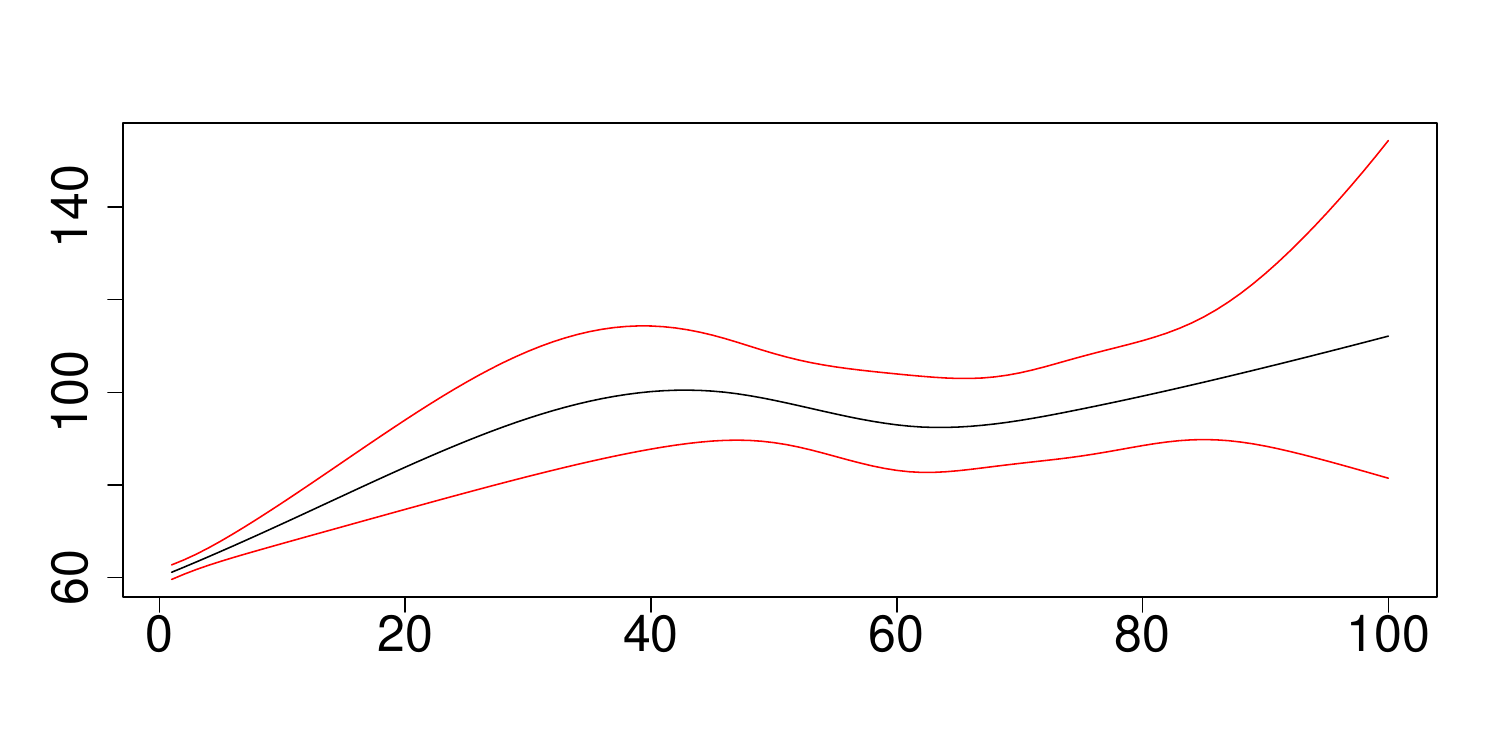}
         \caption{Symptoms, signs, abnormal results and ill-defined causes}
         \label{subfig:isi_cause17}
     \end{subfigure}
    \begin{subfigure}[t]{0.32\textwidth}
         \centering
         \includegraphics[width=\textwidth]{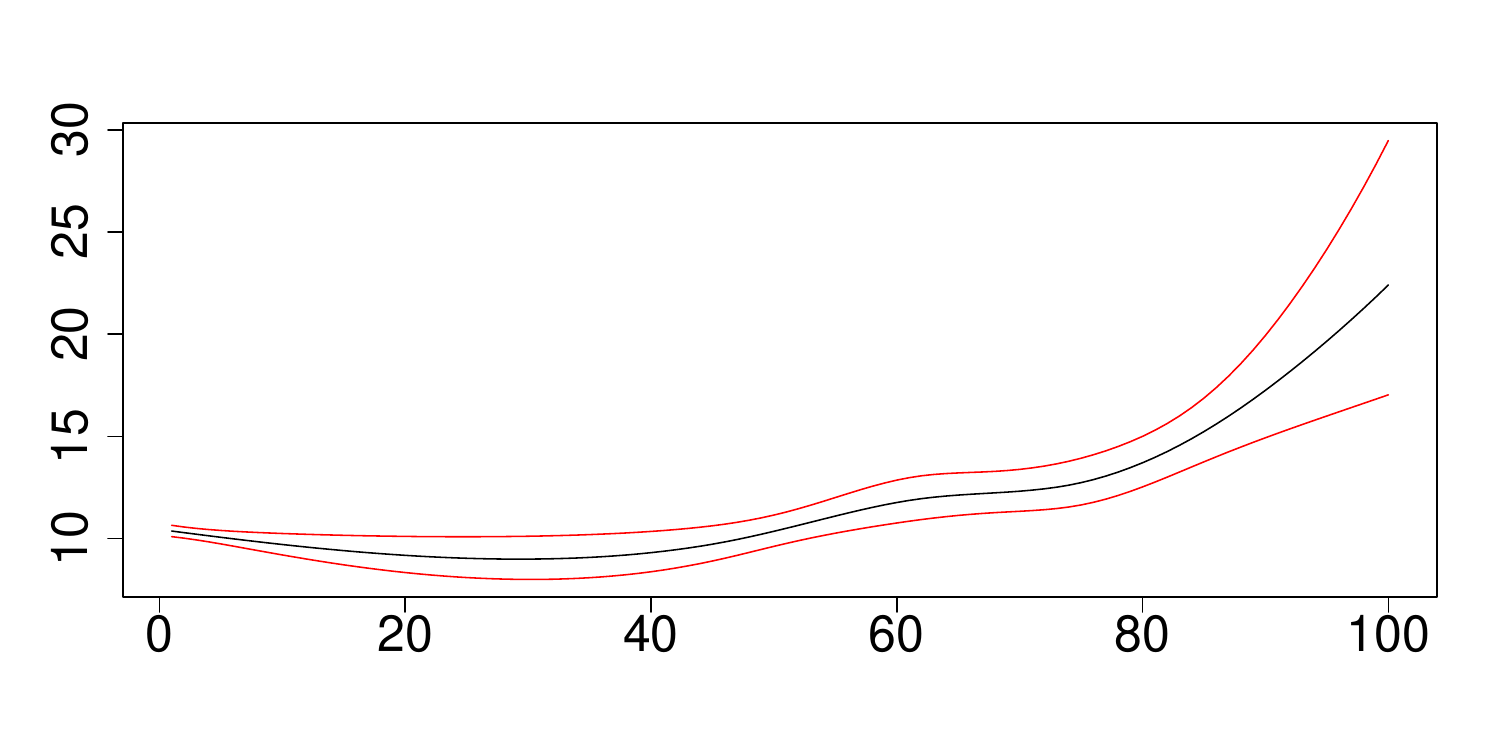}
         \caption{External causes of trauma and poisoning}
         \label{subfig:isi_cause18}
     \end{subfigure}
     
     \caption{Estimated mortality rates of 17 causes as a function of Italian Stringency Index (ISI). Black lines are estimated values and red lines are 95\% credible intervals.}
     \label{fig:isi_cause}
\end{figure}

We first discuss the non-linear relationship between mortality rates and ISI. Figure \ref{fig:isi_cause} shows that GAMs predict similar smoothing patterns for the following mortality causes, endocrine, nutritional and metabolic diseases in Figure \ref{subfig:isi_cause5}, psychic and behavioral disorders in Figure \ref{subfig:isi_cause6}, diseases of the nervous system and sense organs in Figure \ref{subfig:isi_cause7}, diseases of the circulatory system in Figure \ref{subfig:isi_cause8}, diseases of the respiratory system in Figure \ref{subfig:isi_cause9}, and external causes of trauma and poisoning in Figure \ref{subfig:isi_cause18}. When the Italian government implemented loose intervention measures, the mortality rates went down as the intervention became more strict until intervention reached certain median-to-low level (ISI between 20 and 40), the mortality rates took a reverse trend and went up as more strict policies were carried out. In general, early and mild interventions like social distancing and improved hygiene practices not only reduced the spread of COVID-19 but also other infectious diseases, which could benefit individuals with a wide range of health conditions. The initial stages of the pandemic also led to increased health awareness and changes to more regular lifestyle. However, these benefits subsided as healthcare resources were increasingly diverted to treat COVID-19 patients, individuals with other health conditions might have faced delayed or reduced access to necessary health care. More specifically, for some individuals with psychic and behavioral disorders, when the intervention measures were mild, staying at home might have initially reduced stressors such as workplace pressures or social anxiety, potentially benefiting mental health, however, the potential benefits were replaced by increased isolation and loneliness, heightened anxiety and stress as well as substance abuse when situation deteriorated and higher lockdown levels were enforced. As for diseases of the circulatory system and diseases of the respiratory system, initial reduction of mortality rates at low ISI levels could also be attributed to factors like reduction in air pollution, reduced exposure to allergens, less physical strain from commuting as a result of early intervention procedures. Lastly, we comment on external causes of trauma and poisoning. In the early stages of the pandemic, there was rising public vigilance and a focus on safety, which might have extended beyond COVID-19 precautions to general safety practices, potentially reducing accidents. Meanwhile, people were adapting to new routines which might have included safer practices at home, reducing the risk of domestic accidents or poisonings. When the stringency measures prolonged, people started to be impacted by mental health issues as well as social and economic hardship, which could potentially contribute to the upward trend.

Another common non-linear pattern of predicted mortality rates by ISI shared among some mortality causes is an overall downward trend. The mortality causes exhibiting such behaviors are some infectious and parasitic diseases in Figure \ref{subfig:isi_cause2}, tumors in Figure \ref{subfig:isi_cause3}, diseases of the digestive system in Figure \ref{subfig:isi_cause10}, diseases of the skin and sub-cutaneous tissue in Figure \ref{subfig:isi_cause11} and congenital malformations and chromosomal anomalies in Figure \ref{subfig:isi_cause16}. For these causes, benefits of social distancing due to high stringency levels dominates factors such as disrupted health care access, delayed diagnoses and treatments. We also note that credible intervals are wider when ISI is close to 100, suggesting distinct mortality patterns appearing in different regions and age groups.

\begin{figure}[t]
\captionsetup[subfigure]{font=footnotesize}
     \centering
    \begin{subfigure}[t]{0.32\textwidth}
         \centering
         \includegraphics[width=\textwidth]{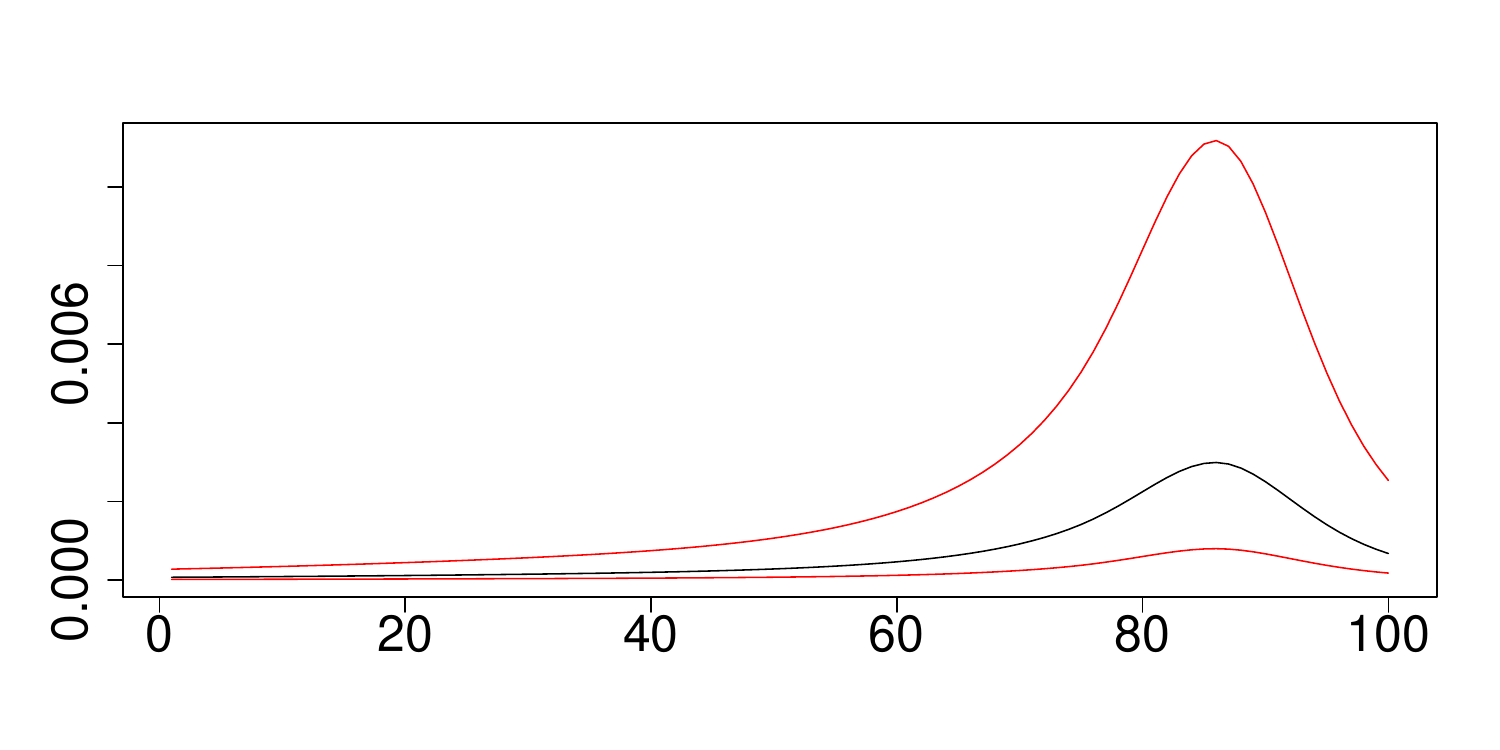}
         \caption{COVID-19}
         \label{subfig:age_cause1}
     \end{subfigure}
    \begin{subfigure}[t]{0.32\textwidth}
         \centering
         \includegraphics[width=\textwidth]{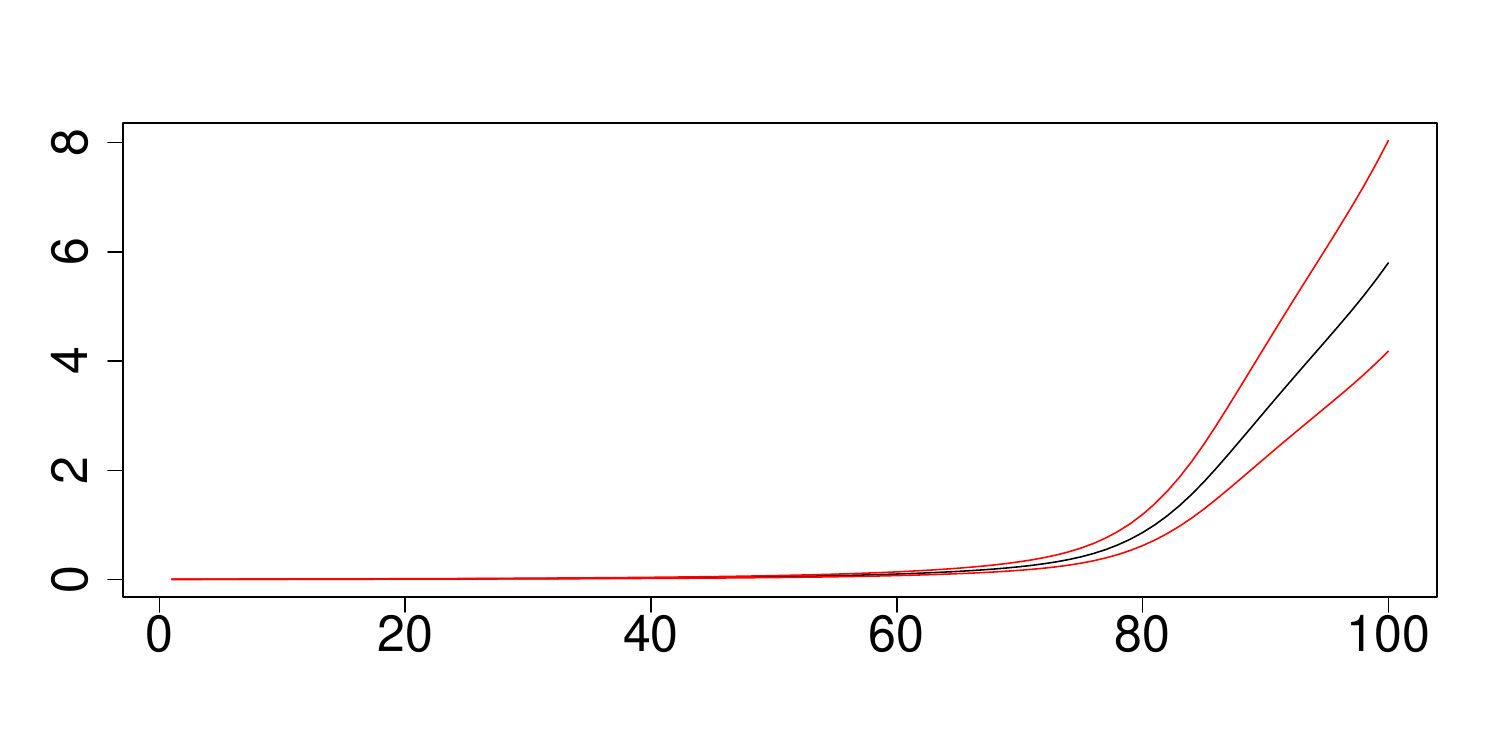}
         \caption{Some infectious and parasitic diseases}
         \label{subfig:age_cause2}
     \end{subfigure}
         \begin{subfigure}[t]{0.32\textwidth}
         \centering
         \includegraphics[width=\textwidth]{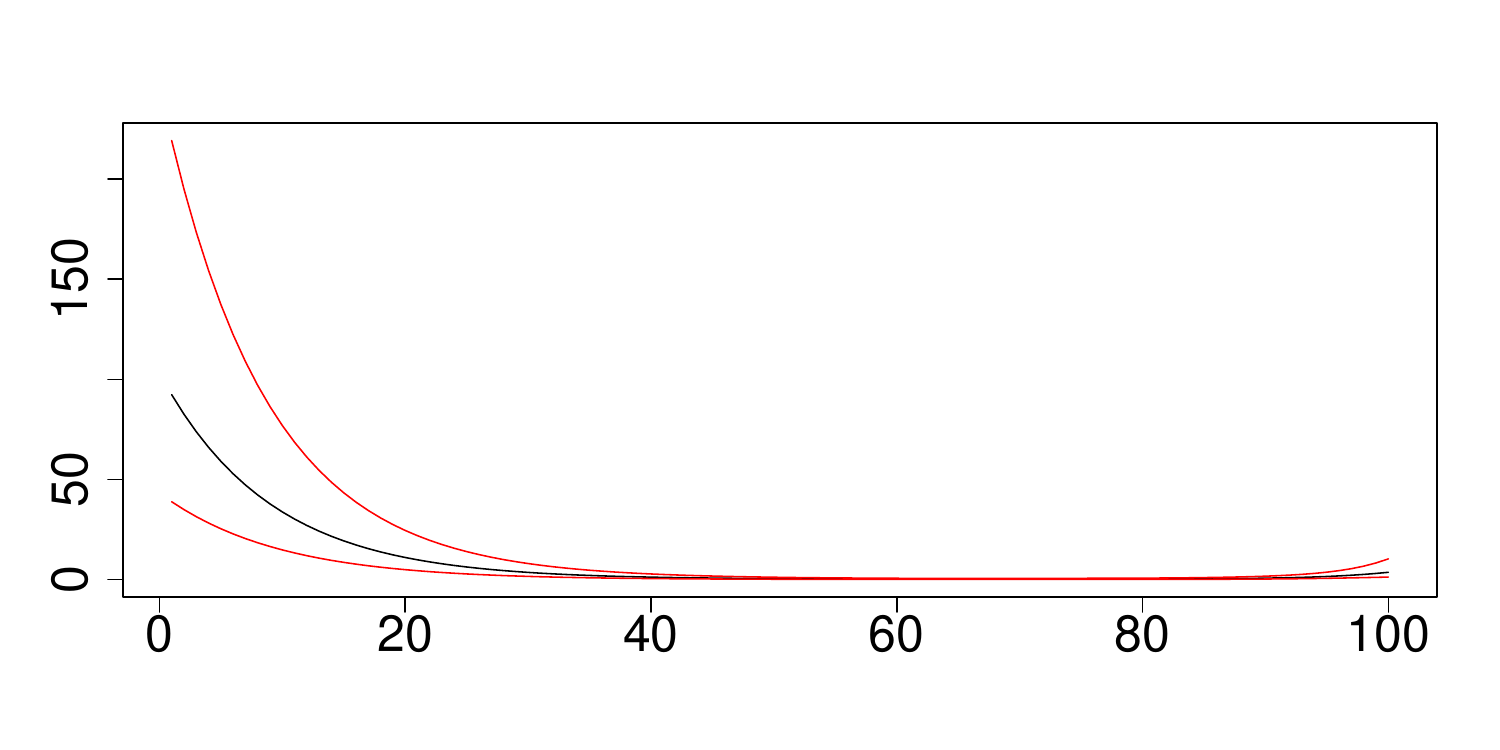}
         \caption{Some morbid conditions that originate in the perinatal period}
         \label{subfig:age_cause15}
     \end{subfigure}

     \caption{Estimated mortality rates of 3 selected causes as a function of age. Black lines are estimated values and red lines are 95\% credible intervals.}
     \label{fig:age_cause}
\end{figure}

We selected to display three representative mortality patterns between age and causes of death in Figure \ref{fig:age_cause}. Almost all mortality causes show the same feature that the mortality increases with age, similar to the pattern of some infectious and parasitic diseases shown in Figure \ref{subfig:age_cause2}. The mortality rate of COVID-19, however, peaked around slightly over 80 years old. Although there have been studies pointing out that age is a positive predictor of the morality rate \citep{bonanad2020effect, zhang2023risk}, their stratification usually sets 80 years old and above as a group. The age groups in the data set that we analyse are more refined with 4 separate age groups for age 80+. Our result offers new insights on the relationship between COVID-19 mortality and age.

As for temporal evolving $z_{n,t}^*$ in the Poisson rates, we focus on three aspects. The approximated posterior distribution of $\boldsymbol{\mu}$ indicates the average mortality levels of each cause of death in each region unexplained by covariates in the model. From Figure \ref{fig:mu_cause}, we can see that both tumors in Figure \ref{subfig:mu_cause3} and diseases of the circulatory system in Figure \ref{subfig:mu_cause8} contribute most to death counts whereas some morbid conditions that originate in the perinatal period shown in Figure \ref{subfig:mu_cause15} have the lowest mortality rate on average. Figure \ref{fig:mu_cause} also reveals certain geographical disparity. For instance, Piedmont, Liguria, Apulia and Abruzzo are the regions mostly affected by COVID-19 according to Figure \ref{subfig:mu_cause1}, however this conclusion is in contradiction with the observation that regions such as Lombardy, Veneto, and Campania experienced many cases and deaths. Recall that $\boldsymbol{\mu}$ measures unexplained mortality rate per 1,000 cases since we include COVID-19 cases in offset $\epsilon_{n,t}$, the discrepancy could be attributed to factors including healthcare system quality, timing of the outbreak, testing capacity and so on. For example, regions with less comprehensive testing may report fewer mild or asymptomatic cases, resulting in a higher mortality rate, as the total cases are underestimated. Nevertheless, to answer this question, we need  access to more covariates. Certain regional gap is more in line with conventional knowledge. Figure \ref{subfig:mu_cause5} shows that endocrine, nutritional and metabolic diseases are more deadly in the south as southern regions have higher rates of obesity and diabetes. Temperature is another risk factor behind the phenomenon.

\begin{figure}[!]
\captionsetup[subfigure]{font=footnotesize}
     \centering
    \begin{subfigure}[t]{0.32\textwidth}
         \centering
         \includegraphics[width=\textwidth]{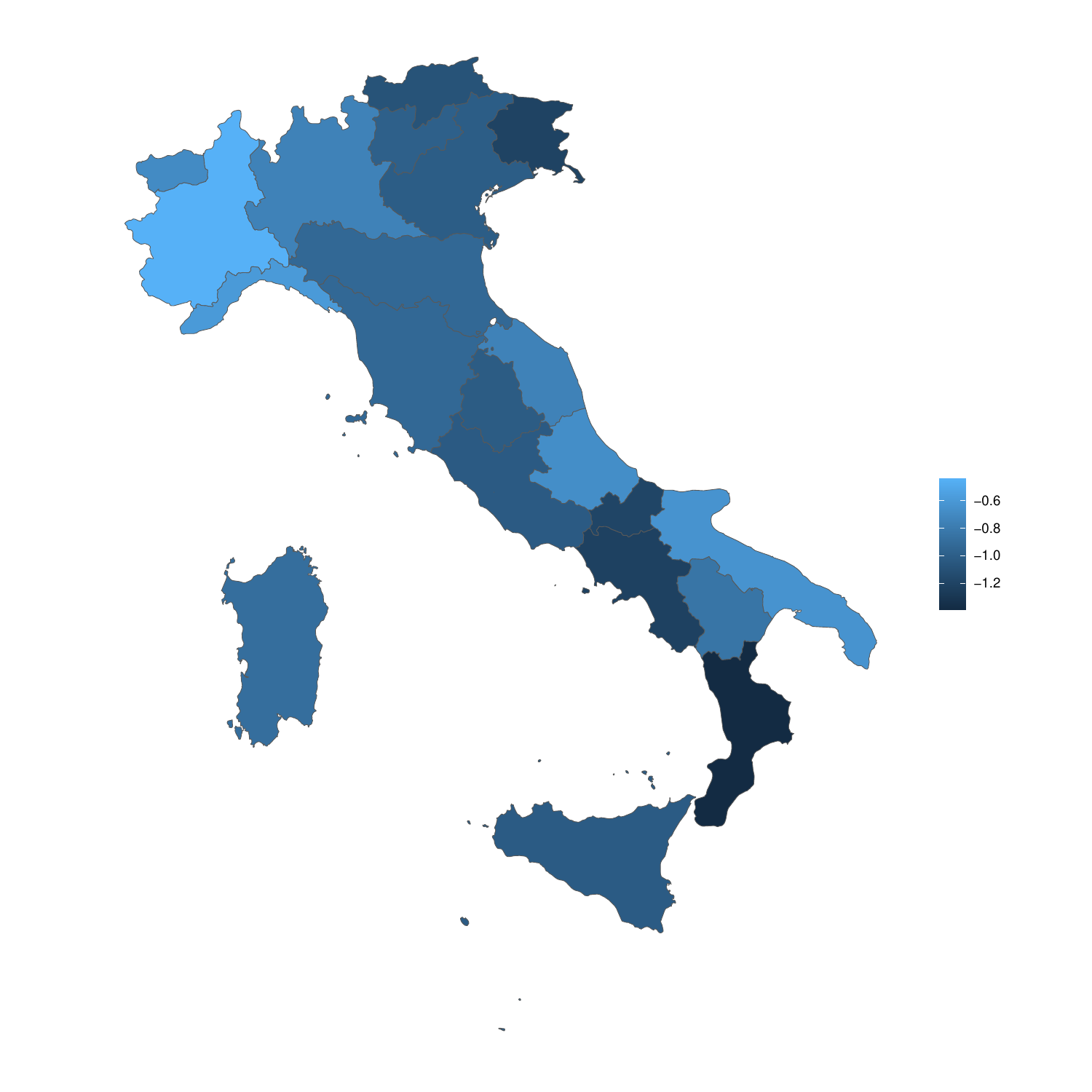}
         \caption{COVID-19}
         \label{subfig:mu_cause1}
     \end{subfigure}
    \begin{subfigure}[t]{0.32\textwidth}
         \centering
         \includegraphics[width=\textwidth]{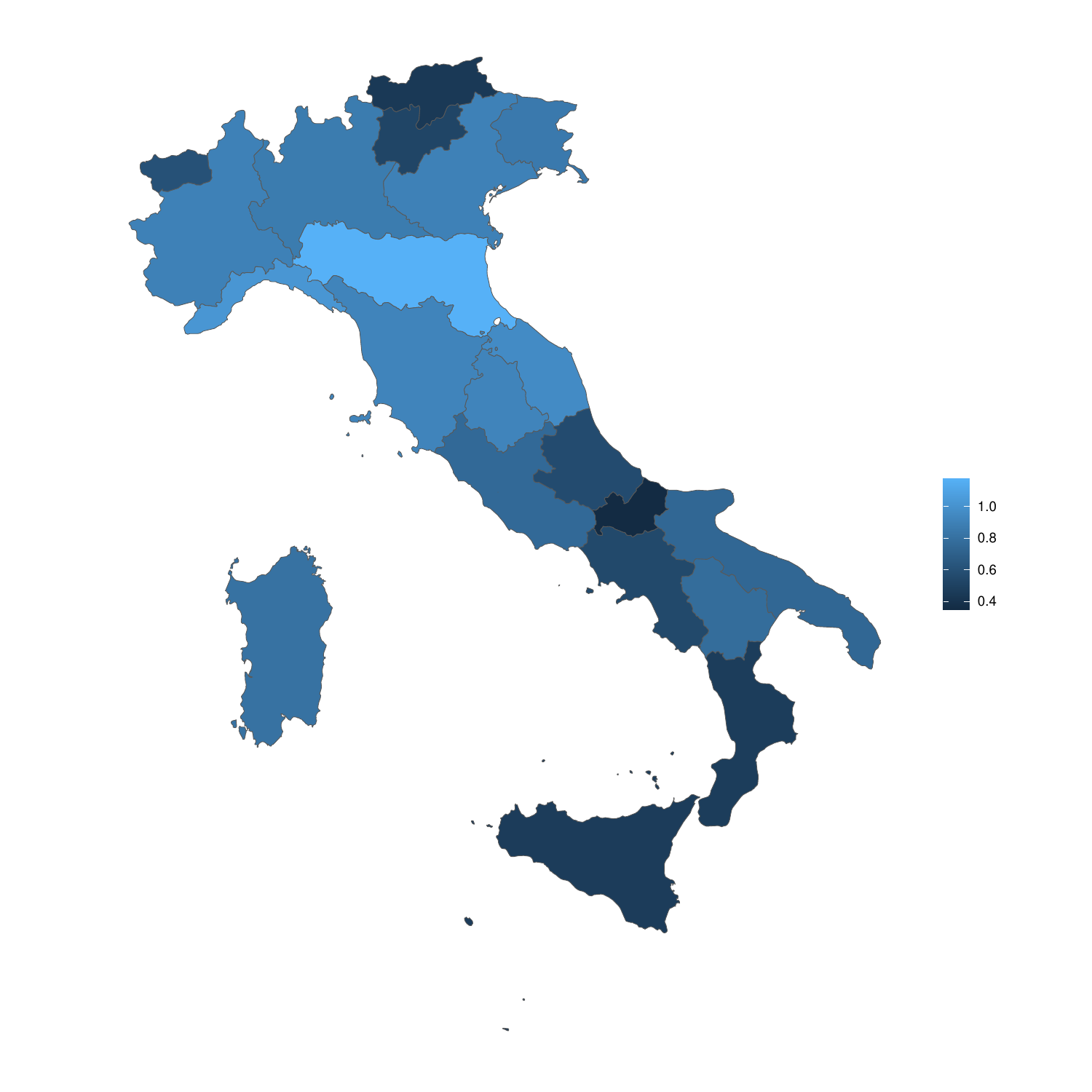}
         \caption{Some infectious and parasitic diseases}
         \label{subfig:mu_cause2}
     \end{subfigure}
    \begin{subfigure}[t]{0.32\textwidth}
         \centering
         \includegraphics[width=\textwidth]{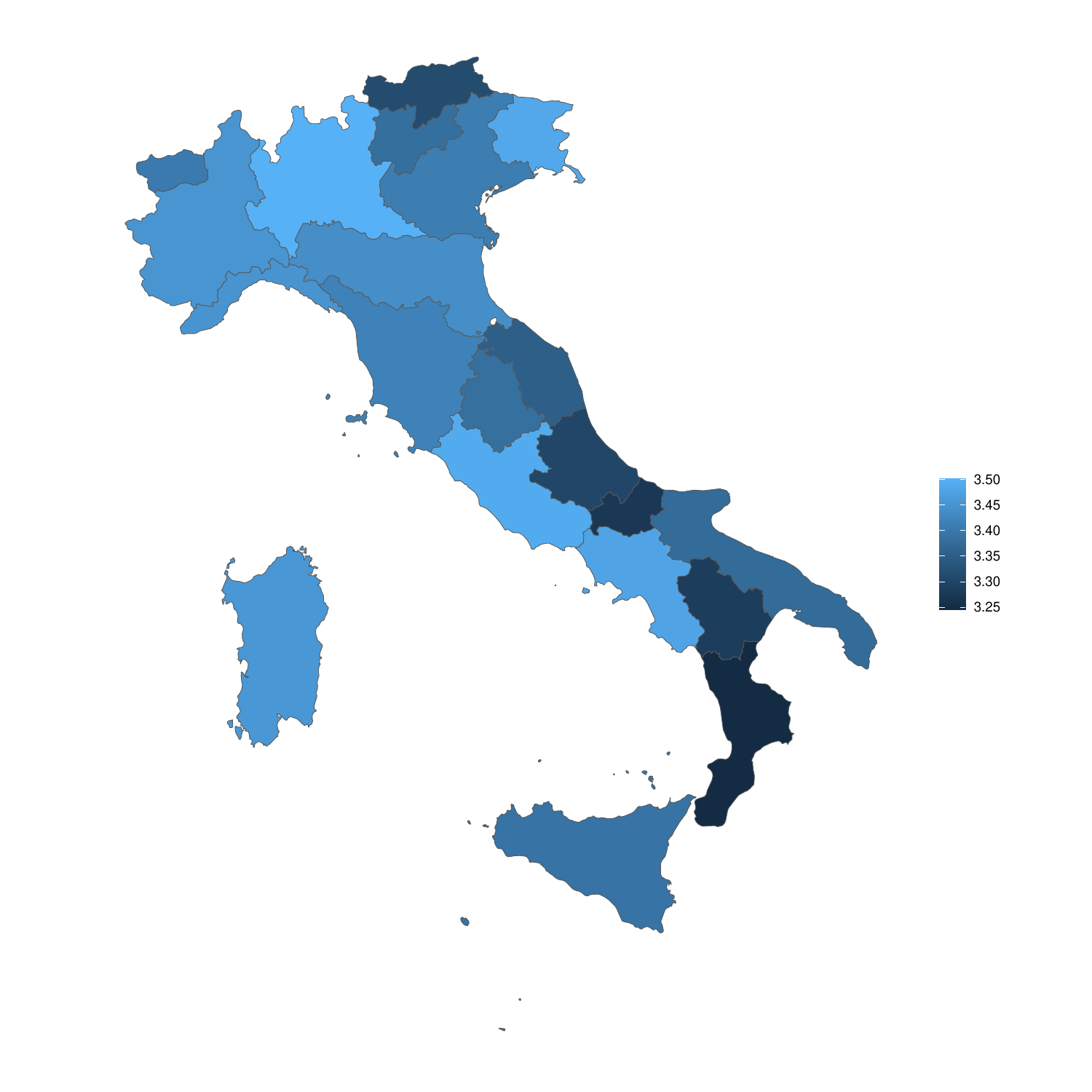}
         \caption{Tumors}
         \label{subfig:mu_cause3}
     \end{subfigure}
    
    \begin{subfigure}[t]{0.32\textwidth}
         \centering
         \includegraphics[width=\textwidth]{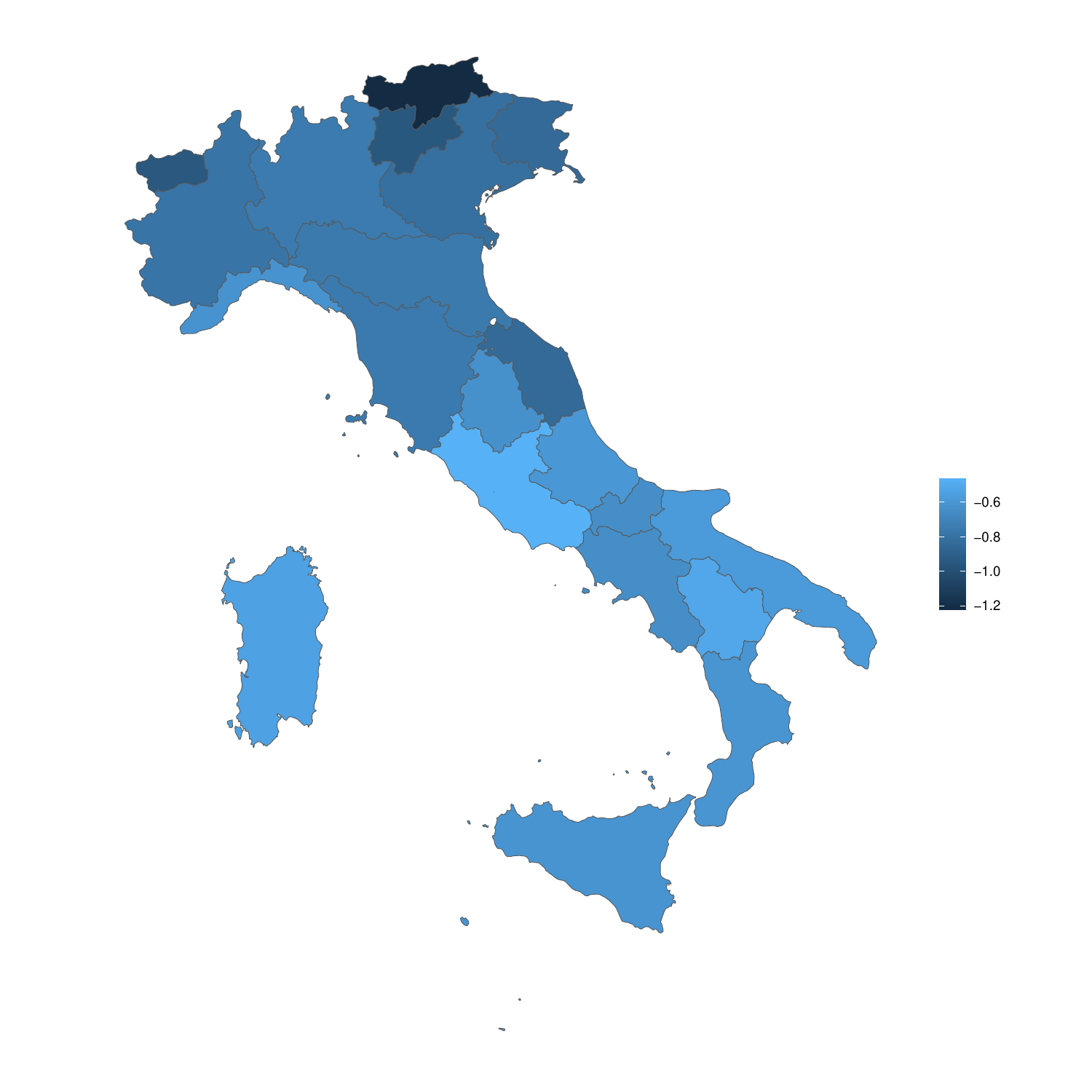}
         \caption{Diseases of the blood and hematopoietic organs and some disorders of the immune system}
         \label{subfig:mu_cause4}
     \end{subfigure}
    \begin{subfigure}[t]{0.32\textwidth}
         \centering
         \includegraphics[width=\textwidth]{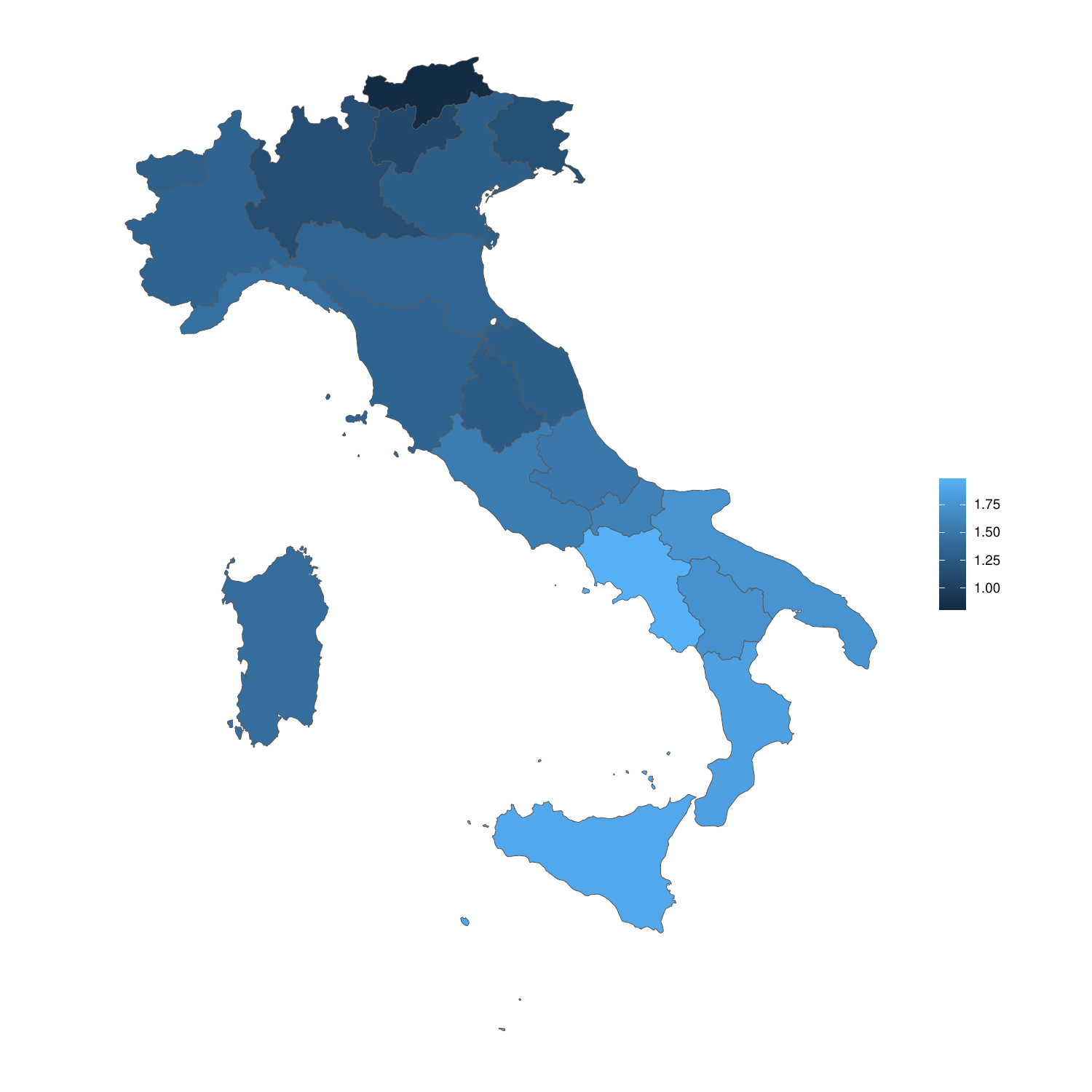}
         \caption{Endocrine, nutritional and metabolic diseases}
         \label{subfig:mu_cause5}
     \end{subfigure}
    \begin{subfigure}[t]{0.32\textwidth}
         \centering
         \includegraphics[width=\textwidth]{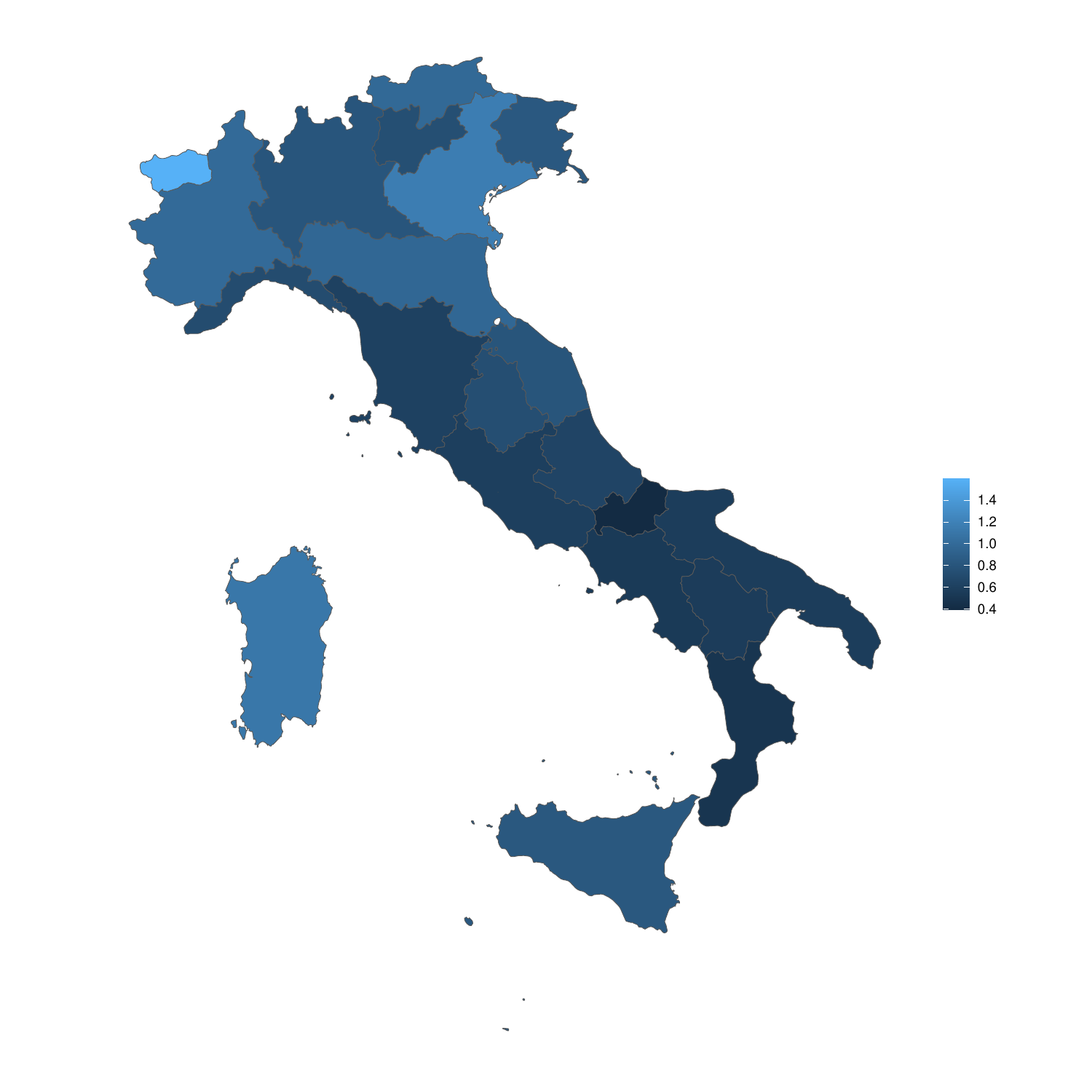}
         \caption{Psychic and behavioral disorders}
         \label{subfig:mu_cause6}
     \end{subfigure}

    \begin{subfigure}[t]{0.32\textwidth}
         \centering
         \includegraphics[width=\textwidth]{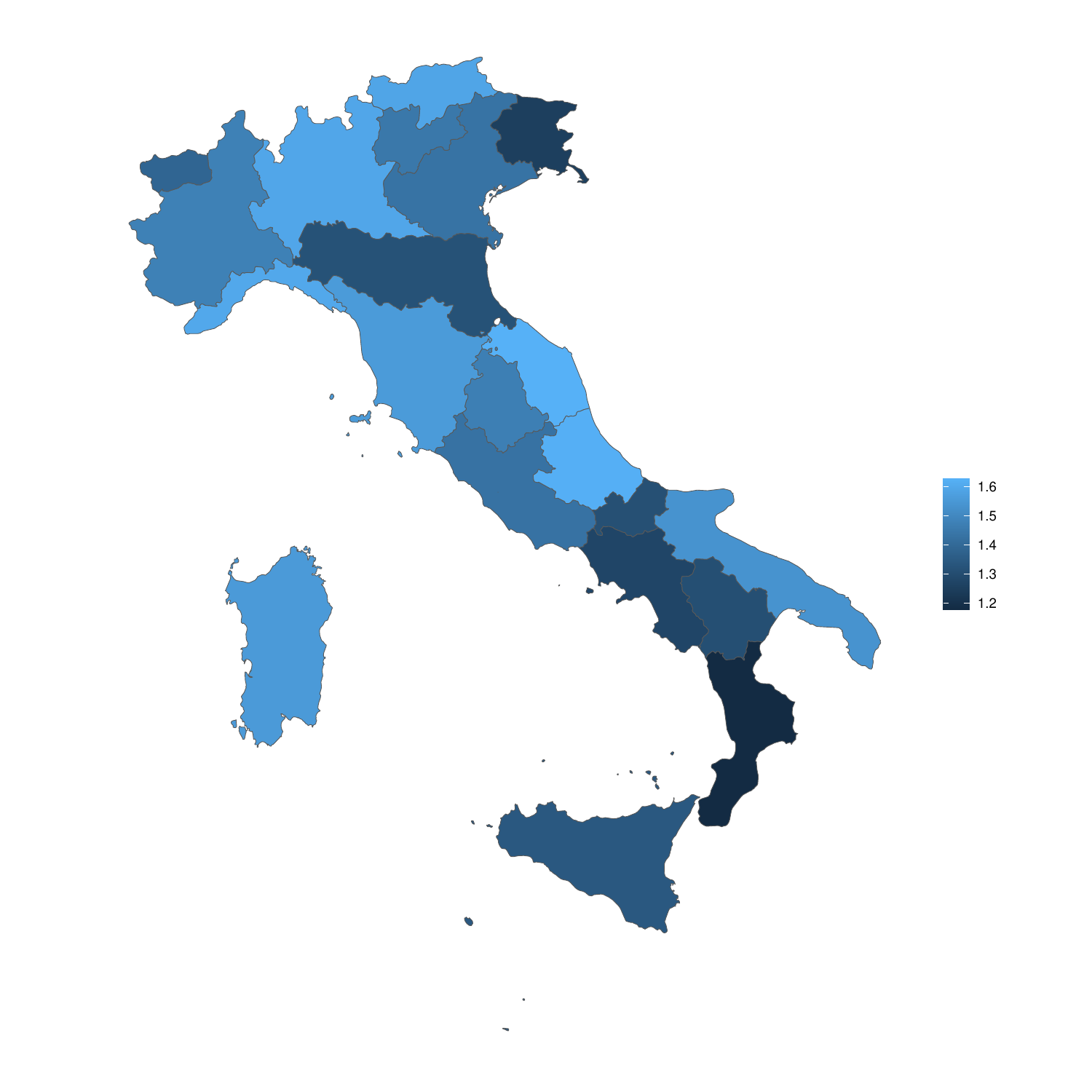}
         \caption{Diseases of the nervous system and sense organs}
         \label{subfig:mu_cause7}
     \end{subfigure}
    \begin{subfigure}[t]{0.32\textwidth}
         \centering
         \includegraphics[width=\textwidth]{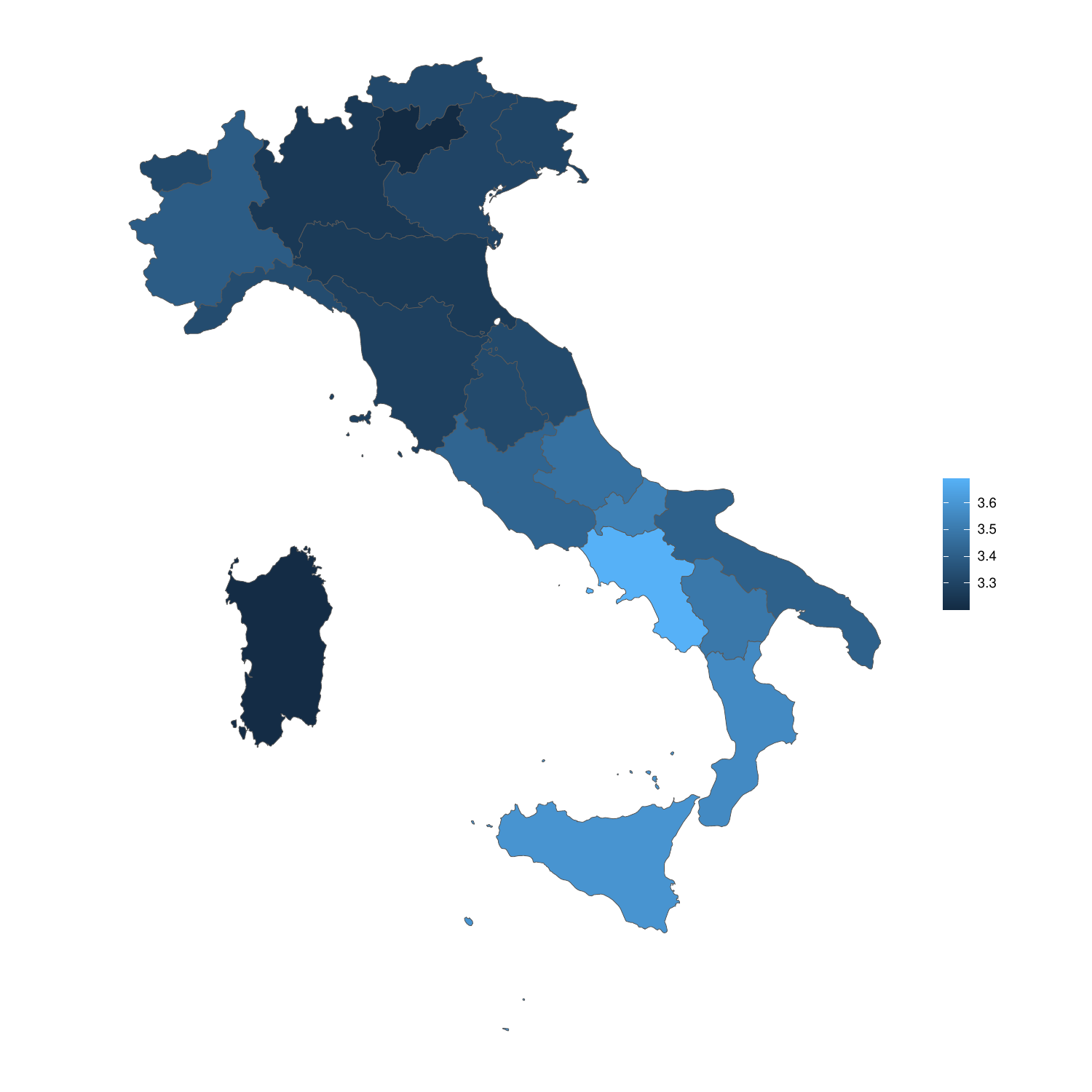}
         \caption{Diseases of the circulatory system}
         \label{subfig:mu_cause8}
     \end{subfigure}
    \begin{subfigure}[t]{0.32\textwidth}
         \centering
         \includegraphics[width=\textwidth]{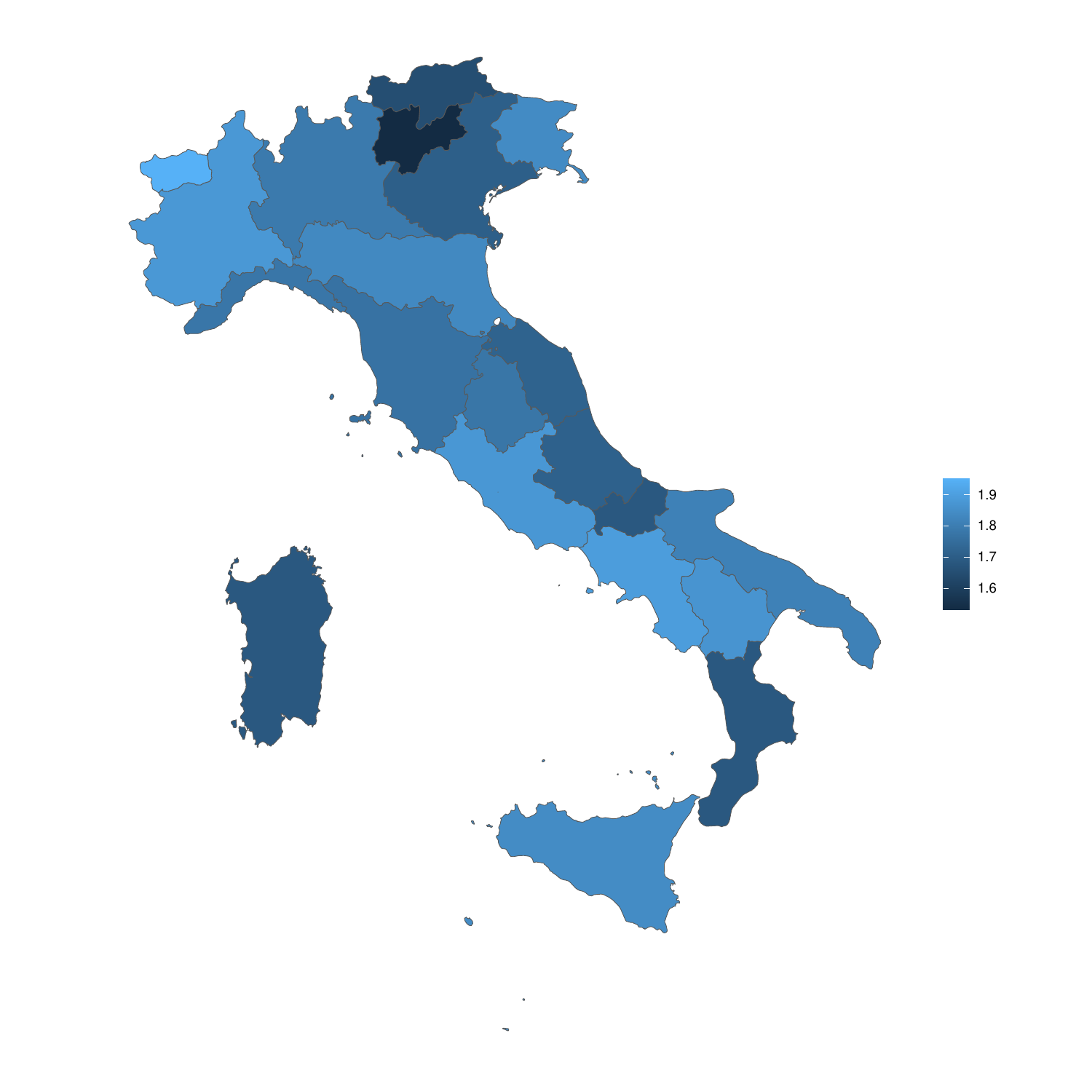}
         \caption{Diseases of the respiratory system}
         \label{subfig:mu_cause9}
     \end{subfigure}
     
     \caption{Estimated mortality means $\boldsymbol{\mu}$ of 17 causes in 21 Italian regions.}
\end{figure}

\begin{figure}[!]\ContinuedFloat
\captionsetup[subfigure]{font=footnotesize}
     \centering

         \begin{subfigure}[t]{0.32\textwidth}
         \centering
         \includegraphics[width=\textwidth]{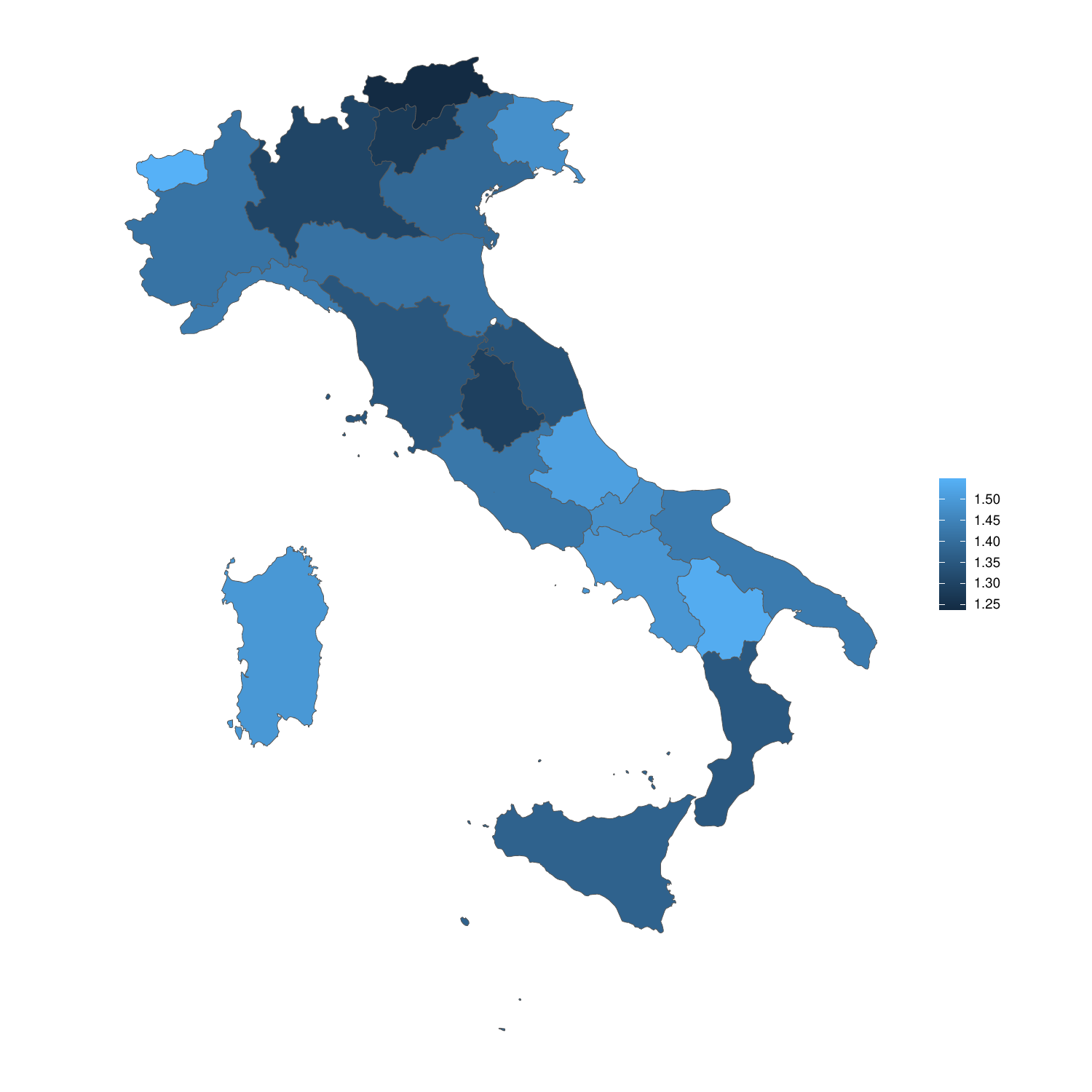}
         \caption{Diseases of the digestive system}
         \label{subfig:mu_cause10}
     \end{subfigure}
    \begin{subfigure}[t]{0.32\textwidth}
         \centering
         \includegraphics[width=\textwidth]{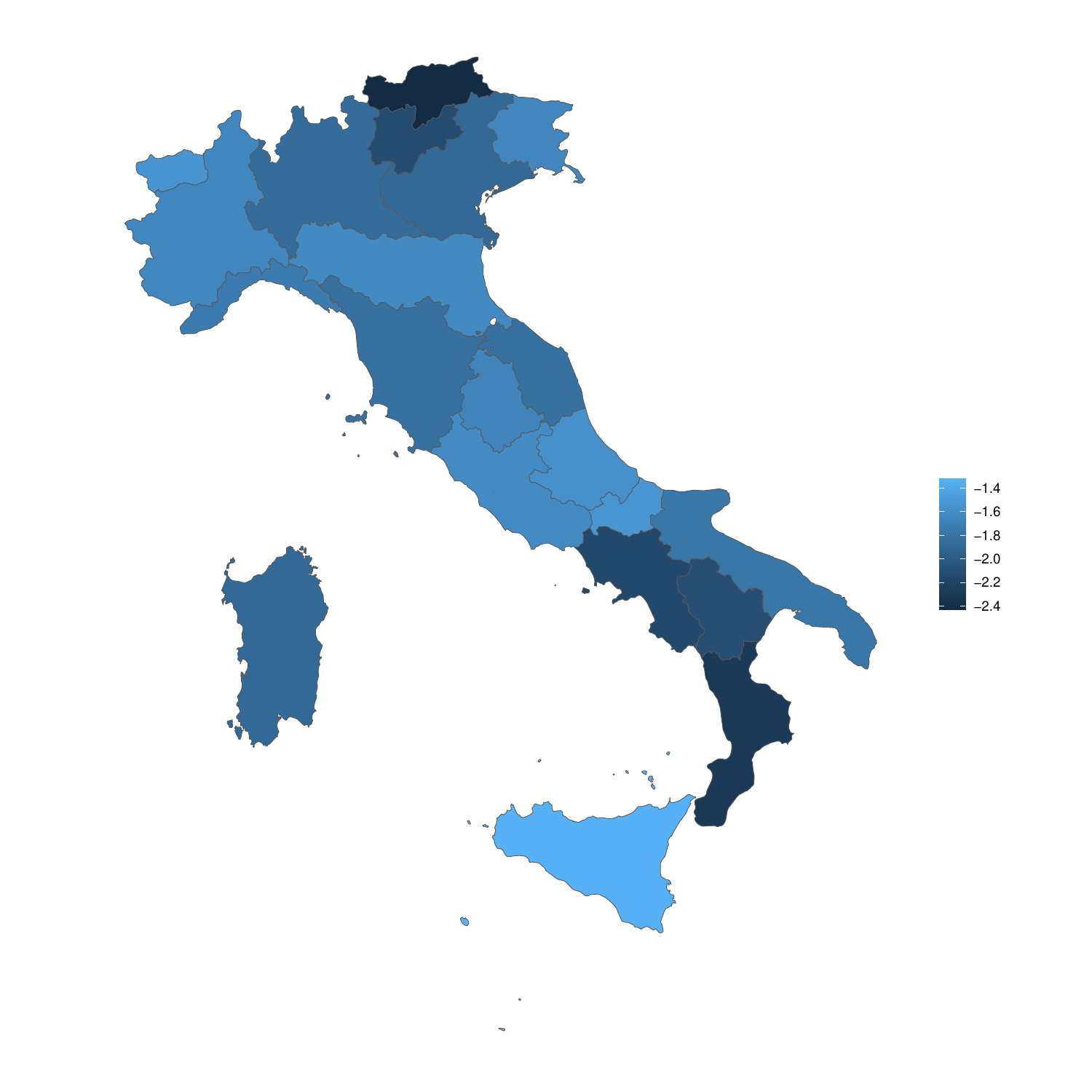}
         \caption{Diseases of the skin and subcutaneous tissue}
         \label{subfig:mu_cause11}
     \end{subfigure}
    \begin{subfigure}[t]{0.32\textwidth}
         \centering
         \includegraphics[width=\textwidth]{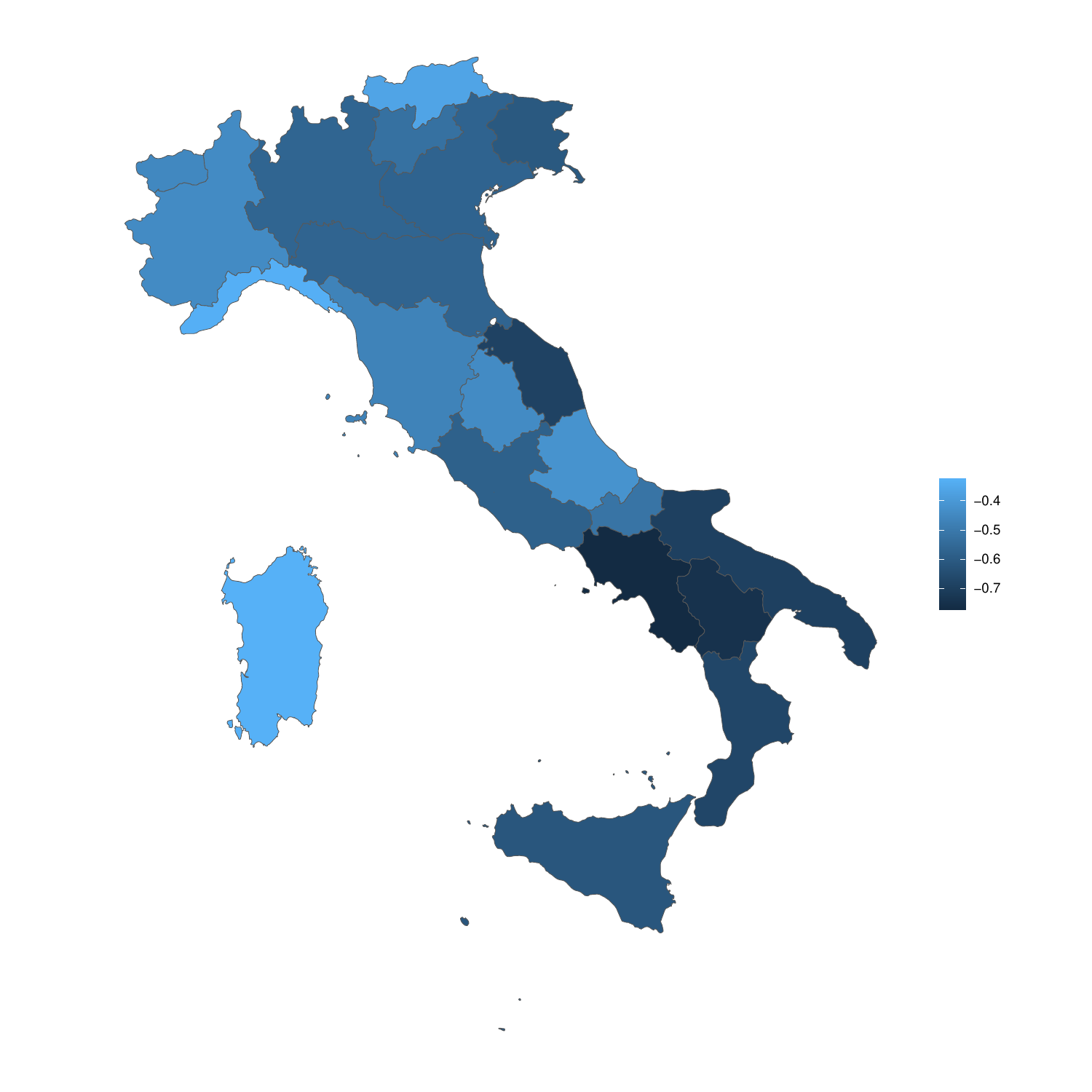}
         \caption{Diseases of the musculoskeletal system and connective tissue}
         \label{subfig:mu_cause12}
     \end{subfigure}

         \begin{subfigure}[t]{0.32\textwidth}
         \centering
         \includegraphics[width=\textwidth]{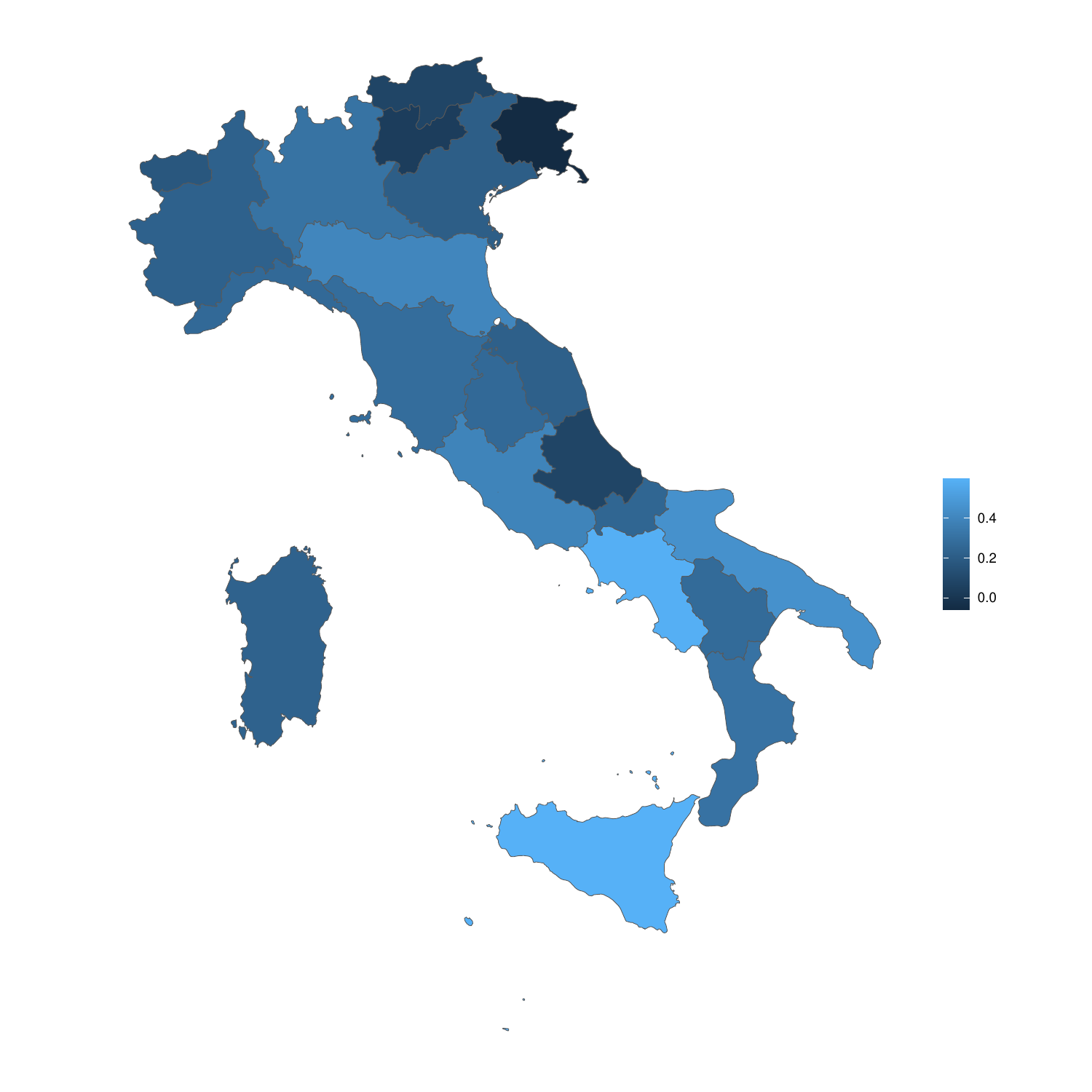}
         \caption{Diseases of the genitourinary system}
         \label{subfig:mu_cause13}
     \end{subfigure}
    \begin{subfigure}[t]{0.32\textwidth}
         \centering
         \includegraphics[width=\textwidth]{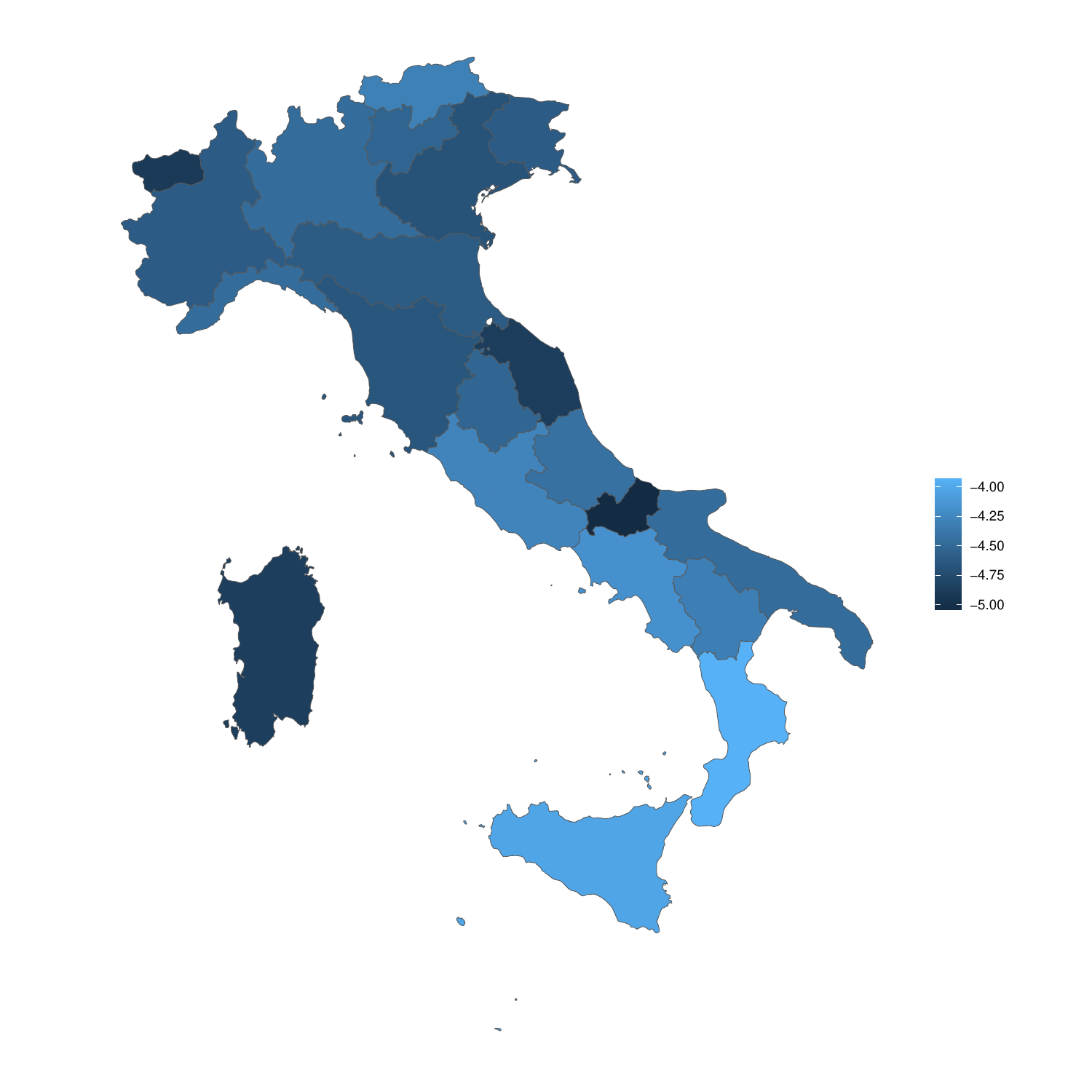}
         \caption{Some morbid conditions that originate in the perinatal period}
         \label{subfig:mu_cause15}
     \end{subfigure}
\begin{subfigure}[t]{0.32\textwidth}
         \centering
         \includegraphics[width=\textwidth]{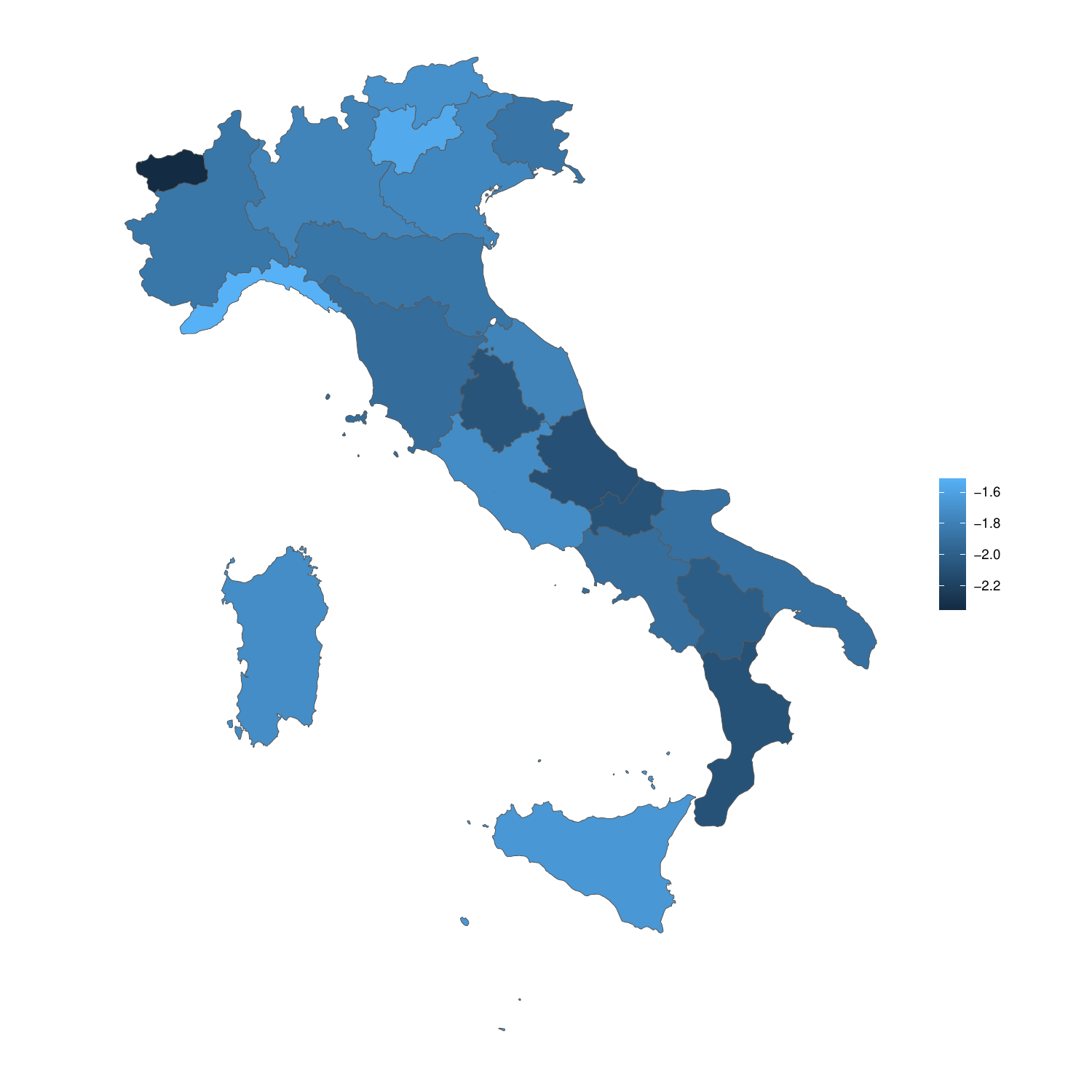}
         \caption{Congenital malformations and chromosomal anomalies}
         \label{subfig:mu_cause16}
     \end{subfigure}
     
    \begin{subfigure}[t]{0.32\textwidth}
         \centering
         \includegraphics[width=\textwidth]{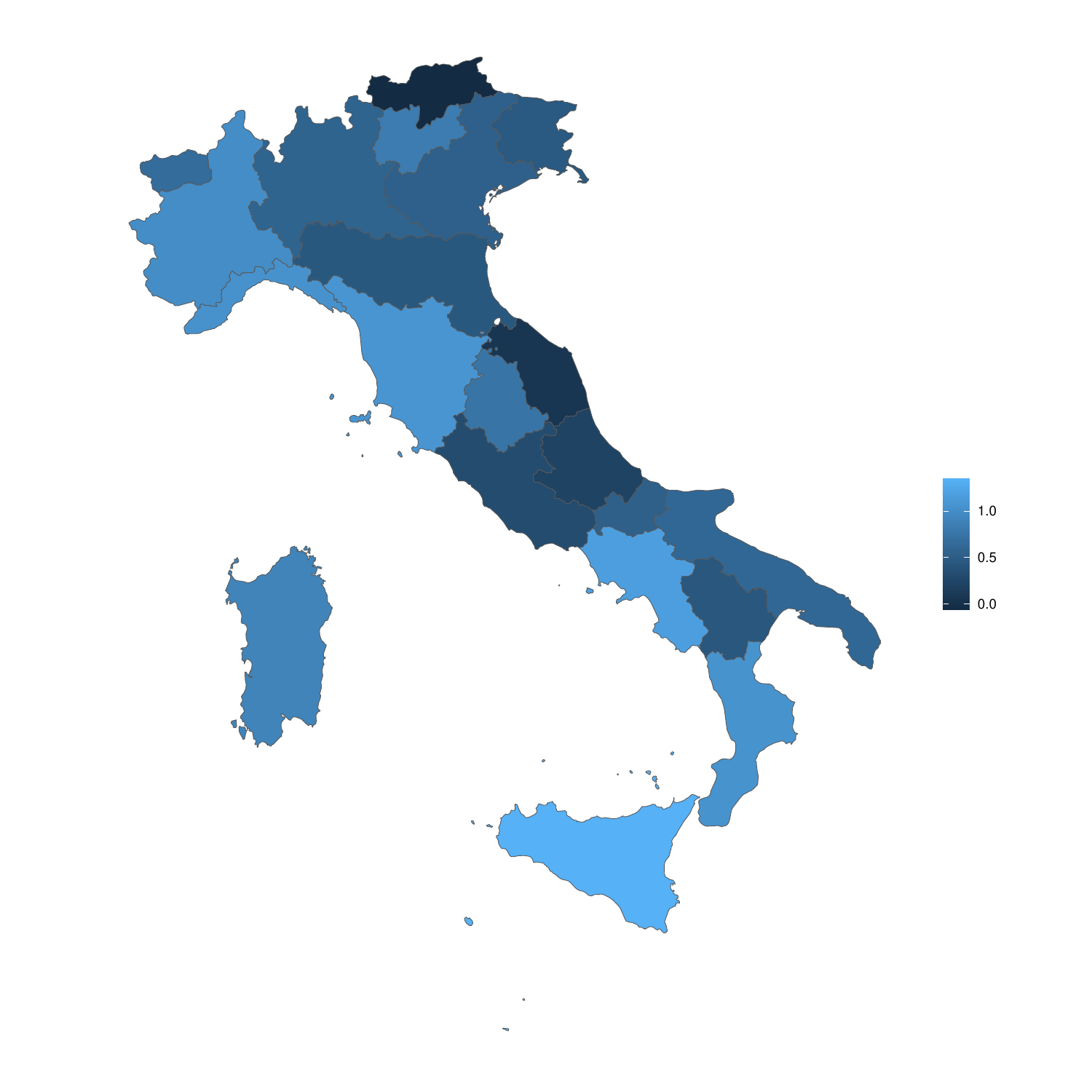}
         \caption{Symptoms, signs, abnormal results and ill-defined causes}
         \label{subfig:mu_cause17}
     \end{subfigure}
    \begin{subfigure}[t]{0.32\textwidth}
         \centering
         \includegraphics[width=\textwidth]{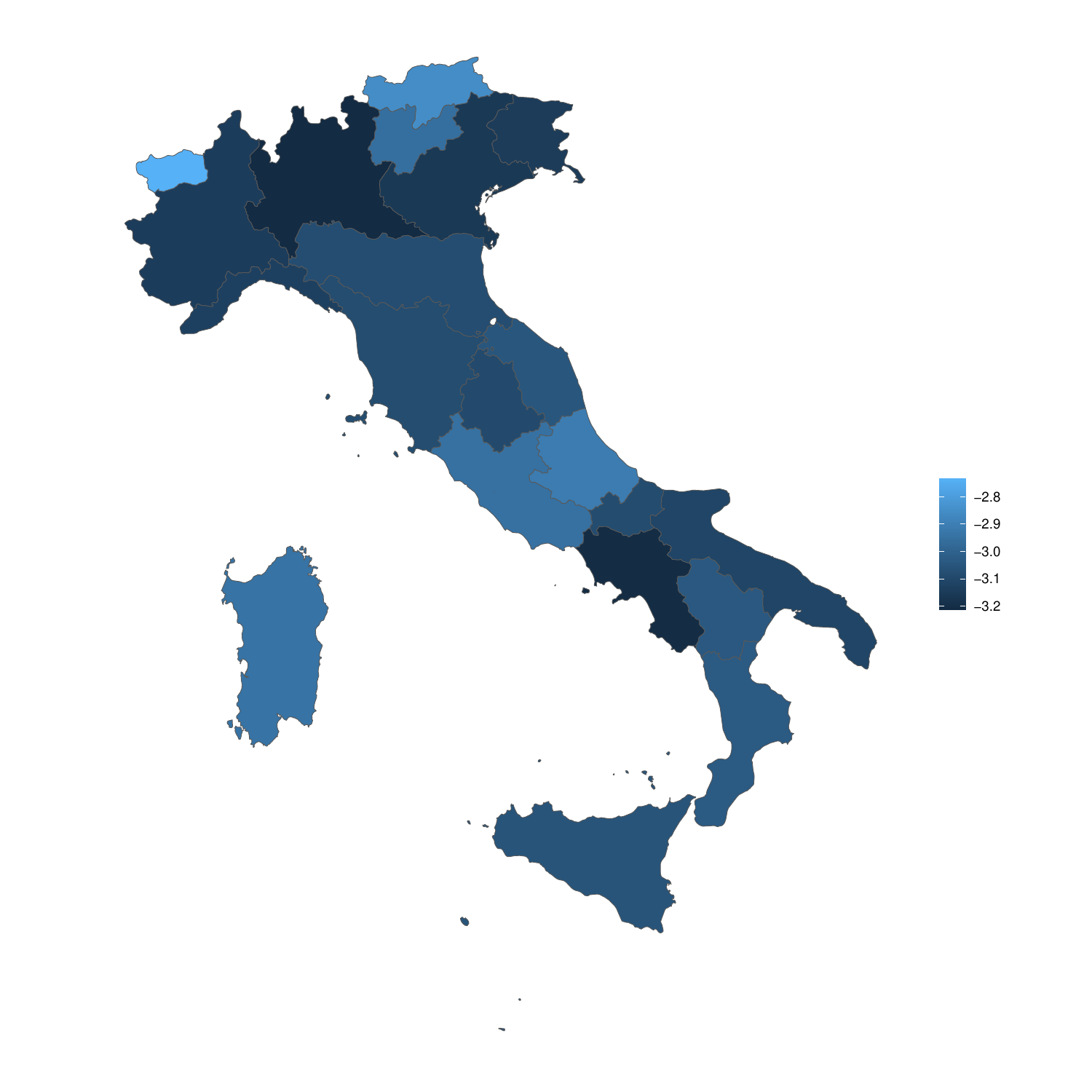}
         \caption{External causes of trauma and poisoning}
         \label{subfig:mu_cause18}
     \end{subfigure}
     
     \caption{Estimated mortality means $\boldsymbol{\mu}$ of 17 causes in 21 Italian regions.}
     \label{fig:mu_cause}
\end{figure}

\begin{figure}[t]
     \centering
         \begin{subfigure}[t]{0.49\textwidth}
         \centering
         \includegraphics[width=\textwidth]{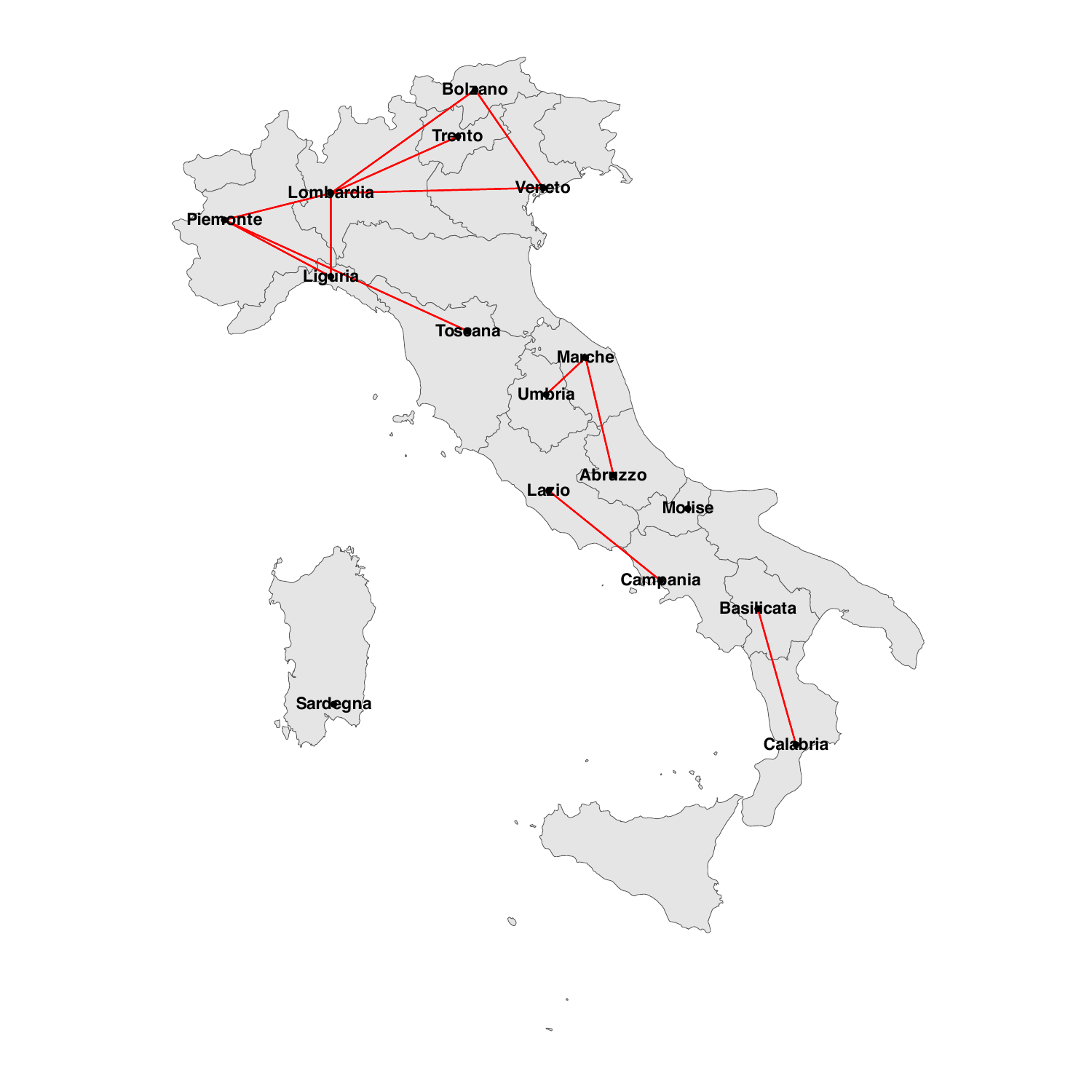}
         \caption{regional network from $D^{q,l}$}
         \label{subfig:ranef_region}
     \end{subfigure}
         \begin{subfigure}[t]{0.49\textwidth}
         \centering
         \includegraphics[width=\textwidth]{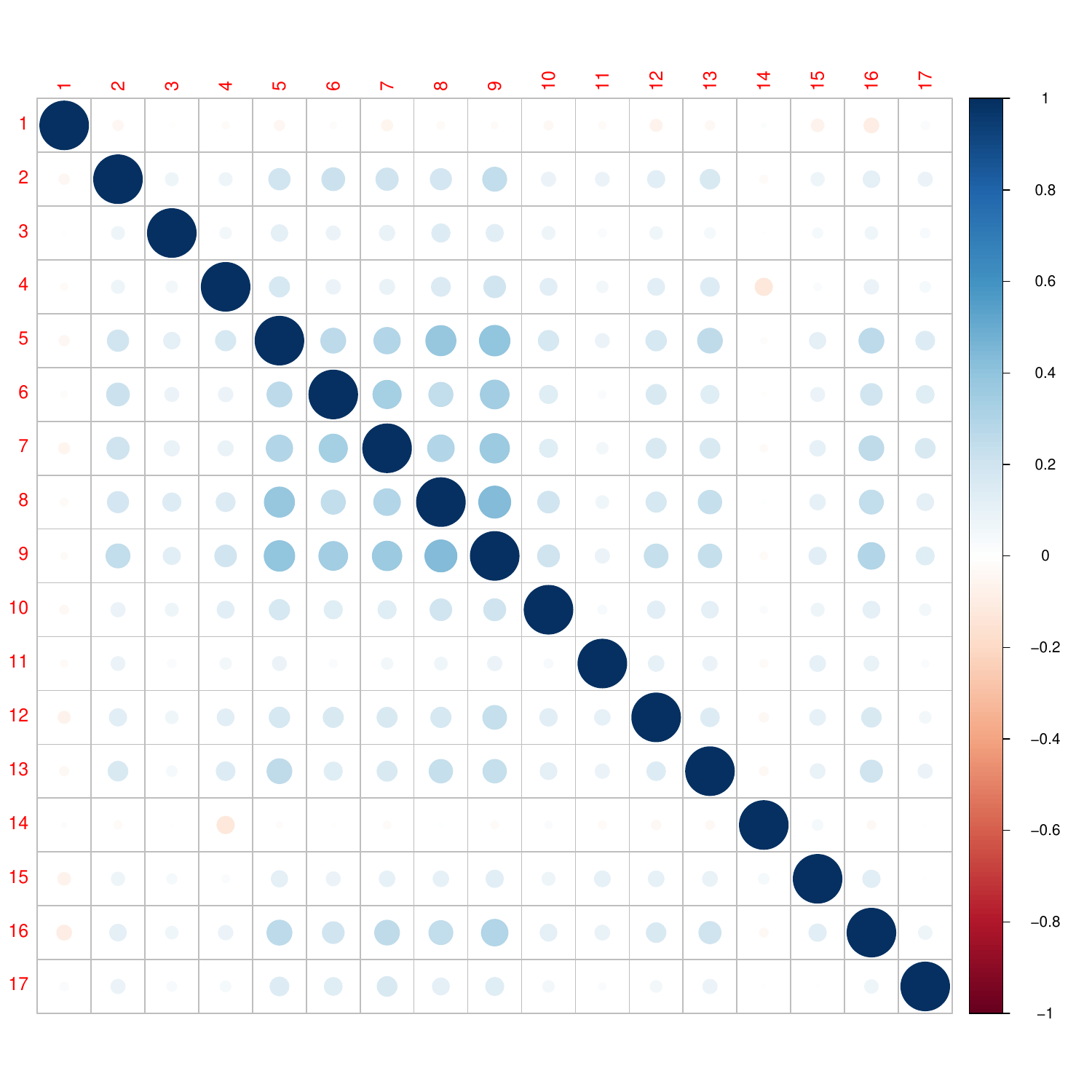}
         \caption{correlation matrix from $D^{q,k}$}
         \label{subfig:ranef_cause}
     \end{subfigure}
     \caption{Partial correlation and correlation derived from the optimal scale matrix of variational Wishart density.}
     \label{fig:ranef}
\end{figure}

We now discuss the other two aspects associated with $z^*_{n,t}$, the dependence structure captured by $(V^{q,l})'V^{q,l}$ for the scale matrix $D^{q,l}$ of the variational Wishart density and the correlation captured by $(V^{q,k})'V^{q,k}$ for the scale matrix $D^{q,k}$. Since $D^{q,l}$ contains spatial information that controls the dynamic of $z^*_{n,t}$, we compute the partial correlation matrix, that is the normalized inverse, of $D^{q,l}$ and then set the entries in the inverse matrix whose absolute value are below 0.1 to 0 while replace the remaining entries with 1. Figure \ref{subfig:ranef_region} maps the resulting adjacency matrix into Italy where red edges between two regions stand for 1's in the adjacency matrix, indicating a direct statistical relationship between them that is not explained by their relationships with other regions. There edges form four clusters; northern Italian regions are connected, and  Lombardy appears to be the central hub for disease spread with many edges connected to it, possibly due to higher mobility or being a major transportation center. The remaining three clusters lie in the center and the southern Italy. The behavior indicates that regions inside each cluster might experience similar trends in mortality rates due to migration patterns, economic ties, climate conditions, or population behaviors. On the other hand, the absence of an edge between two regions implies conditional independence, meaning that once the influence of all other regions is controlled for, there is no direct statistical relationship between these two regions. This is the case with most regions.

Lastly, we turn to the correlation structure among various mortality causes. Figure \ref{subfig:ranef_cause} shows strong positive correlation inside the following 5 causes of death, endocrine, nutritional and metabolic diseases, psychic and behavioral disorders, diseases of the nervous system and sense organs, diseases of the circulatory system, and diseases of the respiratory system. In terms of the relationship between COVID-19 and other causes of death, we only observe weak negative relationship between COVID-19 and symptoms, signs, abnormal results and ill-defined causes, which suggests that improved detection of COVID-19 death likely reduced the number of deaths classified under ill-defined causes. The lack of correlation between between COVID-19 and other causes of death indicates that after controlling government intervention intensity during the pandemic, COVID-19 itself does not have strong influence on moralities in other death categories. 

\section{Summary and Future Work}
\label{Summary and Future Work}

We develop an effiecient variational inference procedure for Bayesian dynamic GAMs where the dependent variable follows a Poisson distribution. The GAM component in the Poisson rate captures non-linear relationship between the outcome and covariates while the dynamics in the rate has a state space representation. The latent states in the model are assumed to be stationary with covariance matrix defined using a Kronecker product that disentangles one large covariance matrix into smaller covariance matrices. High-dimensional setting is considered, which prohibits the use of exact sampling algorithms, therefore we adopt variational algorithm to achieve fast convergence. The approach is applied to Italian mortality data to facilitate our understanding of mortality pattern changes during the COVID-19 outbreak.

Even though we concentrate on explanatory power of the model, it is potentially suitable for forecasting as the state space component models dynamics in mortality patterns. Another extension makes more sophisticated assumptions on the covariance matrix of the latent states. For instance, it is interesting to impose hierarchical G-Wishart distribution prior instead of Wishart prior on the precision matrices so that conditional dependence structure can be directly inferred from the model. Alternatively, the covariance matrix can be time-varying, allowing for more flexibility to account for changes in the dependence structure, especially before the COVID-19 pandemic and after the pandemic. We leave these discussion to future work.

\clearpage

\bibliographystyle{apalike}
\bibliography{ref}

\end{document}